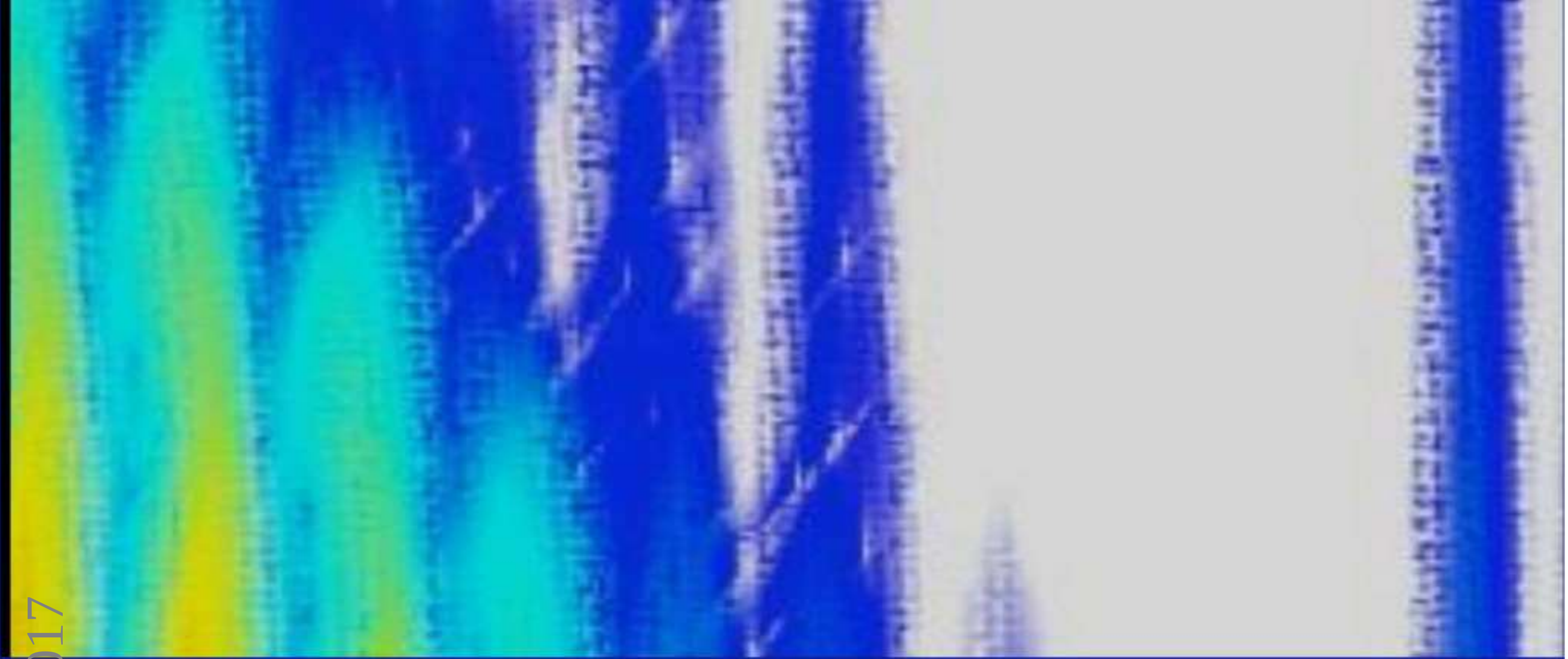

# Study of diatomic molecules under short intense laser pulses

Rui E. Ferreira da Silva

PhD Thesis
January 2016

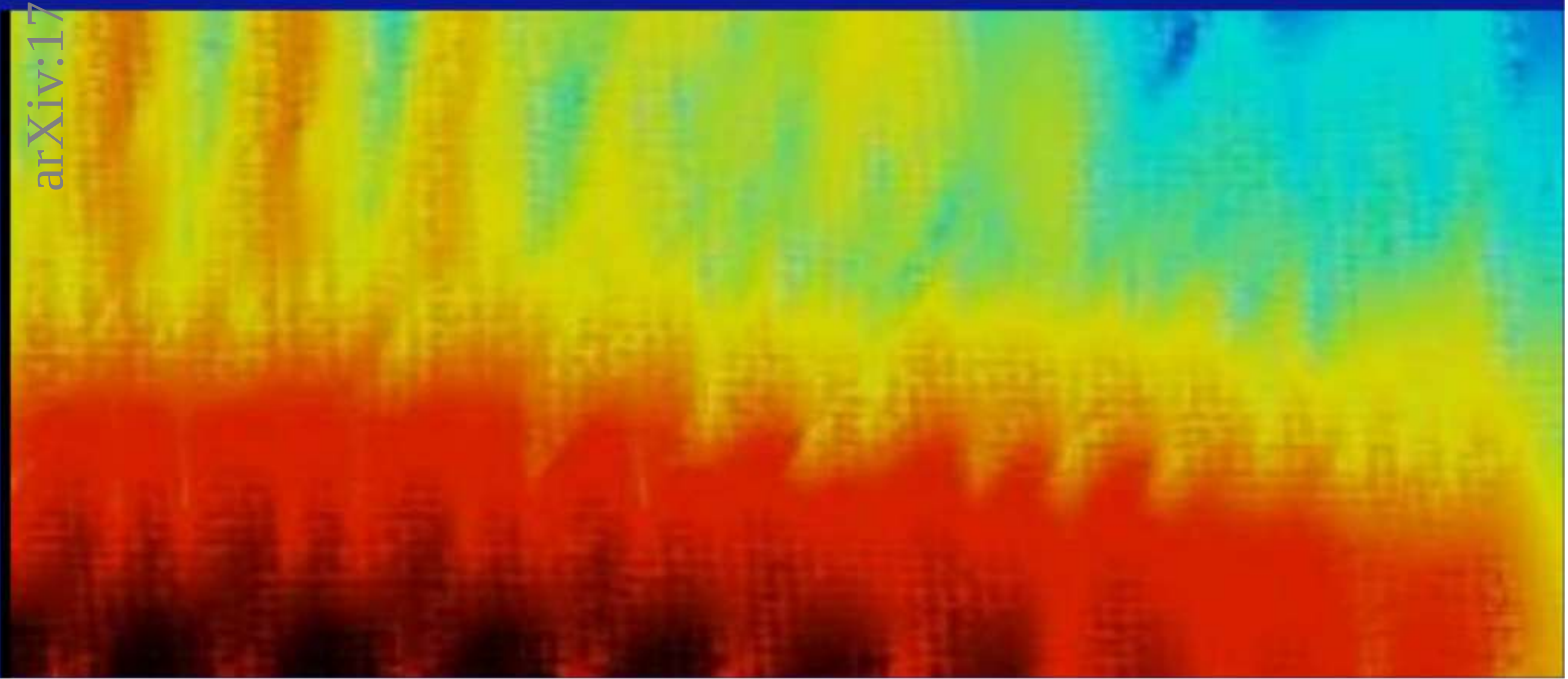

DOCTORAL THESIS

# STUDY OF DIATOMIC MOLECULES UNDER SHORT INTENSE LASER PULSES

UNIVERSIDAD AUTONOMA DE MADRID

RUI EMANUEL FERREIRA DA SILVA

Supervised by Fernando Martín and Paula Rivière

Madrid, January 2016

*Dedicado à memória do meu avô Ferreirinha e da minha avó Ferreirinha.*
*Dedicado à minha melhor amiga, confidente, namorada e mulher, Maitreyi.*

# ACKNOWLEDGEMENTS

First of all I want to thank Paula Rivière. It was her comprehension, knowledge and patience that guide me in the hard path of research. For the long hours discussing science, teaching me the marvelous world of Atomic Physics, to teach me a lot of programming, to be the most comprehensive and kind person that one can have as supervisor. Thank you Paula.

Then I would like to thank Fernando Martín for the opportunity of working in such research group. For the confidence that puts on me by offering me the chance of proceed my PhD in Madrid. Because with his fantastic physical intuition I was able to understand results that for me at the beginning were like chinese characters.

I also need to say a word to Henri Bachau. Thanks for the help in my stage at Bordeaux and for the discussions that made the Resolvent Technique easier. Thanks also for the rides to the airport. I really appreciate your disponibility for even the silliest question. I can not forget Fabrice Catoire. It was a pleasure to chat with someone so interesting and so brilliant. Thanks for introducing me to the wonderful world of the Resolvent Technique.

A thanks to Professor Manuel Yañez, Professora Otilia Mó and Professor Manuel Alcamí for the organization of the EMTCCM.

I would like to thanks to all my colleagues of the master for having great times with them. A special thanks to Alberto Muzas, Humberto, Juan Pablo and Ana for the everyday friendship.

A thanks to all professors, students and post-docs of the Chemistry Department.

Agradeço aos meus amigos e família porque são o meu porto de abrigo.

Ao meu companheiro português em Madrid, Bruno Amorim, obrigado pelos jogos de squash, pelos jogos na Play2 e por seres o grande amigo que encontrei em Madrid.

Agradeço aos meus pais, que me deram uma vida sem igual e que todos os dias se esforçaram para que eu e as minhas irmãs tivessemos a melhor vida possível. Porque sem eles não seria quem sou, e sem eles não estaria aqui.

Gracias Maitreyi, porque has sido lo mejor que me ha pasado en la vida.

## INSTITUTIONAL SUPPORT

With the support of the Erasmus Mundus programme of the European Union (FPA 2010-0147).

The research leading to these results has received funding from the European Union Seventh Framework Progrprograamme (FP7/2007-2013) under grant agreement n° 264951. Funded by XChem project (ERC grant agreement n° 290853).

With the support of Fundação para a Ciência e Tecnologia under grant number SFRH/BD/84053/2012.

Funded by Spanish government.

# CONTENTS

# LIST OF FIGURES

## ABBREVIATIONS

| | |
|---|---|
| ATI | Aboveabove threshold ionization |
| BO | Born-Oppenheimer |
| $CAK_e$ | Correlated angular and electronic kinetic energy spectrum |
| $CAK_N$ | Correlated angular and nuclear kinetic energy spectrum |
| CAP | Complex absorbing potential |
| CEP | Carrier-envelope phase |
| CKE | Correlated energy spectrum |
| COLTRIMS | Cold Target Recoil Ion Momentum Spectroscopy |
| EKE | Photoelectron-kinetic energy spectrum |
| EMF | Electromagnetic field |
| HHG | High Harmonic Generation |
| IR | Infrared |
| KER | Kinetic energy release |
| LOPT | Lowest order pertubation theory |
| NKE | Nuclear kinetic energy spectrum |
| NRQM | Non-relativistic Quantum Mechanics |
| QED | Quantum Electrodynamics |
| ROM | Resolvent operator method |
| SI | International System of Units |
| TDSE | Time-dependent Schrödinger equation |
| TOF | Time-of-flight |
| VMI | Velocity map imaging |
| XUV | Extreme ultraviolet |

# ABSTRACT

In this work we report *ab initio* calculations for the $H_2^+$ molecule interacting with ultrashort intense laser pulses. We analyse several observables that can, in principle, be available experimentally, in order to get a deeper understanding of the strong field molecular dynamics. In particular, we will focus our attention to the interplay between electronic and nuclear dynamics and how the two motions are correlated.

We have extended the Resolvent Technique to molecules and we have extracted the correlated energy spectra in different regimes of ionization: from the perturbative regime to the tunneling ionization regime. We have applied this new method, called resolvent operator method (ROM), to a 1+1D model of the $H_2^+$ molecule. We have applied this method to several photoionization regimes and we have compared with previous results in the literature. We show that the correlated spectra can provide us much more information than the integrated spectra on the electron or nuclear energy. We report the correlated energy spectrum from the multiphoton ionization to the tunneling ionization regime and how the sharing of the excess energy is different for the two regimes.

We have also applied the ROM to a 3D model of the $H_2^+$ molecule in which angular distributions can be obtained. With these angular correlated spectra we show that the contributions of electrons coming from absorption of different number of photons can be disentangled.

We have calculated the high harmonic spectra for a 3D model of the $H_2^+$ molecule and its isotopes. In these calculations we have observed that for sufficiently long pulses and for light molecules we observe even harmonic generation, a phenomenon that is not expected in a centrosymmetric molecule.

# RESUMEN

En este trabajo presentamos cálculos *ab initio* para la molécula de $H_2^+$ en la presencia de pulsos láser ultra cortos e intensos. Analizamos distintos observables que pueden, en principio, ser obtenidos experimentalmente, para obtener una mayor comprensión de la dinámica molecular en campos láser intensos. En particular, enfocaremos nuestra atención al intercambio entre las dinámicas electrónicas y nucleares y cómo los dos movimientos están correlacionados.

Hemos extendido la técnica del resolvente a moléculas y hemos extraído los espectros correlacionados en la energía en diferentes regímenes de ionización: desde el régimen perturbativo hasta la ionización por efecto túnel. Hemos aplicado este nuevo método, llamado método del operador resolvente (ROM), a un modelo 1+1D de la molécula de $H_2^+$. Hemos aplicado este método a distintos regímenes de fotoionización y hemos comparado nuestros resultados con resultados obtenidos anteriormente en la literatura. Hemos demostrado que los espectros correlacionados en energía nos dan más información que los espectros integrados en energía electrónica o energía nuclear. Enseñamos los espectros correlacionados en energía desde la ionización multifotónica hasta la ionización por túnel y cómo el reparto del exceso de energía es diferente en los dos regímenes.

También hemos aplicado el ROM a un modelo 3D de la molécula de $H_2^+$ donde las distribuciones angulares pueden ser obtenidas. Con estas distribuciones angulares hemos demostrado que podemos distinguir distintos canales de ionización que provienen de la absorción de distintos números de fotones.

Hemos calculado el espectro de armónicos para un modelo 3D de la molécula de $H_2^+$ y sus isótopos. En estos cálculos hemos observado que para pulsos suficientemente largos y para moléculas ligeras observamos la generación de armónicos pares, un fenómeno que no es trivial en una molécula con un centro de inversión.

# Part I

# INTRODUCTION

# 1

## INTRODUCTION

In 1878, Muybridge took pictures of a horse during a gallop with a temporal resolution enough to see, in detail, the real movement of the legs of the horse. It was the beginning of a revolution in photography, but also in science. For the first time, we were able to take pictures to see in detail the motion of fast processes. The motivation of this work is of the same nature as Muybridge's work. Instead of understanding processes at the human scale, both temporal and spatial, we want to go further on the comprehension of the dynamics of atoms and molecules. To do that, we must use new methods that are appropriate at the typical atomic scale. A difficulty arises from the fact that at the atomic and molecular scale the particles are no longer governed by the laws of Classical Mechanics, but Quantum Mechanics, since the typical action is of the order of $\hbar$.

It was back in the 1960's that a spot at the energy of the second harmonic was discovered. A crystal irradiated with an 800 nm intense laser was emitting light at 400 nm [4]. The field of non-linear optics was born. At the same time, it became possible to generate coherent monochromatic light: laser science was born. Over the following decades, the laser science went into shorter and shorter time scales (of the order of a few femtoseconds) and into more intense pulses that can create an electric field comparable to the electric field felt by a bound electron in an atom. This was the beginning of ultrafast optics.

As stated before, to see a process in detail we must capture images at that time scale. With this idea in mind, experiments were done using ultrafast lasers on molecules to probe the nuclear motion, [5], since the typical time scale for it (the femtosecond (fs) domain) was experimentally available. With Zewail's experiments a new field was opened, Femtochemistry. In this field people try to manipulate molecules, in order to track chemical reactions in real time and also to manipulate them. It has a huge technological potential in pharmaceutical and chemical industry.

In the 1980's, it was observed that atoms under an intense IR field irradiate photons with energy equal to $E = (2N + 1)\,\omega$, where $\omega$ is the frequency of the IR pulse [6]. These atomic spectra had some structure and its main feature was the cutoff at $I_p + 3.17U_p$, where $U_p = \frac{I}{4\omega^2}$ is the ponderomotive energy, that is the cycle averaged energy gain of a free electron in the electric field, $I$ is the intensity of the field and $\omega$ its frequency. This phenomenon is called High Harmonic generation (HHG). HHG was explained by Corkum [7] and Kulander *et al.* [8] with the so-called Three Step Model and latter by a quantum model [9] based on the Strong Field Approximation. One of the properties of the field irradiated by the atom was the low duration of the pulse, of the order of hundreds of attoseconds. This is the typical timescale of the electronic motion. The scientific community had then the chance to produce laser pulses that were able to probe the electronic motion in atoms and molecules. The field of Attophysics was created [10].

In several recent experiments it was demonstrated that electronic motion can be controlled in an attosecond time scale using strong fs driving laser pulses [11]. This motion can also be used to measure directly the electric field of light [12], to produce XUV laser pulses with a few hundreds of attoseconds [13], and even to image electronic orbitals [14]. The importance and technological potential of this new field in physics and chemistry, which allows one to access the internal dynamics of atoms and molecules, is growing every day.

New experimental techniques have been developed with this objective of visualizing and manipulating electronic dynamics in an attosecond time scale, by extracting the energy spectrum of particles ejected upon photoionization. Among these we find time-of-flight (TOF) techniques, velocity map imaging (VMI) [15] and Cold Target Recoil Ion Momentum Spectroscopy (COLTRIMS) [16]. These techniques can be combined with theoretical calculations to reveal the structural and dynamical information about molecules.

The aim of this work will be the computation and analysis of the several observables in molecular strong field ionization. To do that, we will solve the Time Dependent Schrödinger equation (TDSE) and analyze the energy of the fragments or the harmonic spectra generated in order to get a deeper insight in the dynamics of the molecule when it is subject to a very strong laser field. This work will study the molecular dynamics by studying two different and complementary observables: i) on one hand we will study the energy of particles ejected upon photoionization; ii) on the other hand we will analyse the radiation emitted by a molecule interacting with a strong IR laser field, the HHG spectra.

The simplest molecule in Nature is the $H_2^+$ molecular ion. Its simplicity allows us to solve numerically the TDSE in an almost exact way. For this reason, the $H_2^+$ system is one of the most studied molecular systems in strong field physics [2, 17–24] and will be the system studied in this work.

In the present work, we apply theoretical tools to study molecular processes under the presence of ultrashort pulses. To do so we solve the TDSE of an $H_2^+$ molecule interacting with a strong laser field. To solve the TDSE, we discretize the partial differential equations that our system obeys and solve them in a numerical grid, in the so-called grid method [25].

The grid method presents several advantages for solving numerically the TDSE. One of the major advantages is its simplicity: we only need to write down the partial differential equations that describe the dynamics of our system and propagate the partial differential equation. Another advantage is the fact that the grid method can describe even processes where the wavefunction is completely distorted from its initial wavefunction during the propagation. This allow us to describe all the processes in which we have a strong interaction between the laser pulse and our molecular system. Therefore, the numerical exploration of the field of Strong Field Physics is mostly done using the grid method.

However, the grid method also presents several challenges. From the computational point of view, the grid method is very expensive and it becomes prohibitive as long as one increases the size of the system. If we have a $D$-dimensional problem and we describe each spatial dimension with $N$ points, our wavefunction will be described by $N^D$ points. As $D$ increases, the computational effort increases enormously.

Another difficulty in the grid method is the analysis of the results. At the end of the propagation of the TDSE, we will have a wavefunction that is defined at each spatial point of our numerical grid. However, to extract information from this wavefunction, we can project it on eigenfunctions of the observable that we are interested in. In the grid method, the calculation of eigenfunctions of the Hamiltonian can be a very hard task and it must be avoided. In this work, we will present a way to extract the correlated electronic and nuclear kinetic energy spectrum in molecular strong field ionization by means of the Resolvent technique [26–29].

This work emphasizes the importance of looking at correlated spectra in order to have a better picture of the molecular dynamics in strong field ionization. These correlated spectra can provide more information than the integrated electronic or nuclear kinetic energy spectra. By looking at these integrated spectra we are loosing information on the correlation between electronic and nuclear dynamics.

The HHG spectra will be also calculated and will be used as a complementary observable that allow us to trace the dynamics of molecular strong field ionization. In particular, we will show the influence of the nuclear dynamics in the HHG.

The starting point of this thesis is the description of the theoretical methods that were used during this work. In Chapter 2, we will describe the dynamical equations that describe a homonuclear diatomic molecule interacting with a laser pulse. In Chapter 3, we will describe the theoretical methods to evaluate the energy spectrum in a grid calculation. The molecular resolvent operator method (ROM) is introduced in this chapter. The theory related to the ROM was published in [28, 29]. In Chapter 4, we describe qualitatively the Keldysh theory of ionization [30] and the two different regimes of strong field ionization are presented: multiphoton ionization and tunneling ionization. In Chapter 5, we present the three step model that is a cornerstone in the understanding of the HHG. The way to calculate the HHG spectra is deduced from classical and quantum considerations.

In Chapter 6, we present the results of the application of the ROM to the 1+1D model of the $H_2^+$ molecule. Several calculations were performed and cover frequencies that go from the XUV to the IR. We show the correlated kinetic-energy (CKE) spectra for each pulse and we show that the correlated spectra can provide us more information than the integrated spectra. These results were published in [28].

In Chapter 8, we present the results of the application of the ROM to the 3D model of the $H_2^+$ molecule. Several calculations were performed in the sub-fs regime for different frequencies. We present the CKE and correlated angular and nuclear kinetic-energy ($CAK_N$) spectra for each pulse and we show that the angular correlated spectra can be useful to disentangle the different ionization channels in the photoionization of $H_2^+$. These results were published in [29].

In Chapter 7, we present a theoretical study of the transition between the multiphoton ionization regime and the tunneling regime. The study of the CKE spectra show us that the way electron and nuclei share the excess energy is different if we are in the multiphoton or tunneling regime. These results were published in [31].

In Chapter 9, we present a theoretical study of the HHG in different isotopes of the $H_2^+$ molecule. We study the influence of the nuclear dynamics on the HHG spectra and show that even harmonic generation appears for light molecules and long pulses. Even harmonic generation is attributed to electron localization. These results are being prepared for publication.

Atomic units (a.u.) are used throughout the thesis unless stated otherwise.

# Part II

# THEORY

# 2

# LASER-MATTER INTERACTION

> *"Physical concepts are free creations of the human mind, and are not, however it may seem, uniquely determined by the external world. In our endeavor to understand reality we are somewhat like a man trying to understand the mechanism of a closed watch. He sees the face and the moving hands, even hears its ticking, but he has no way of opening the case. If he is ingenious he may form some picture of a mechanism which could be responsible for all the things he observes, but he may never be quite sure his picture is the only one which could explain his observations. He will never be able to compare his picture with the real mechanism and he cannot even imagine the possibility or the meaning of such a comparison. But he certainly believes that, as his knowledge increases, his picture of reality will become simpler and simpler and will explain a wider and wider range of his sensuous impressions. He may also believe in the existence of the ideal limit of knowledge and that it is approached by the human mind. He may call this ideal limit the objective truth."*
>
> Albert Einstein and Leopold Infield

Our goal is the study of molecules interacting with intense ultrashort laser pulses. In physics, we must always be aware of the approximations that we are using to be conscious of the validity of our results. The interaction of an electromagnetic field (EMF) and a molecule is completely described by the laws of Quantum Electrodynamics (QED) [32]. In the regime that we are working, the velocity of the particles is small compared with the velocity of light, $v \ll c$, and the density of photons in the laser field is very high. Based on these two assumptions, it is reasonable to work in the framework of classical electromagnetism for the electromagnetic field and to treat the massive particles (electrons and nuclei) with non-relativistic quantum mechanics (NRQM). This is often known as the semi-classical approximation.

The semi-classical approximation ignores the fact that the EMF is a quantum field and treat the EMF with the Maxwell Equations. This is a good approximation, since a typical laser field has a very high density of photons making the gain or loss of a photon negligible. Within the semi-classical approximation we can describe both the absorption and the induced emission of a photon by a molecule. Since we are neglecting the fact that the EMF is a quantum field we cannot describe processes such as spontaneous emission, where an excited state of matter decays to the groundstate by emitting one photon.

To have a global idea of the regime in which we are working, we can use the *cube of physics* [33] (see Fig. 2.1.1). The three fundamental constants are $c$, $\hbar$ and $G$. The exact description would be at coordinates $(1,1,0)$ (QED) and we are going to work on coordinates $(1/2,1,0)$. This is the case because we are working on the assumption that the massive particles behave as non-relativistic particles but light, the EMF, is intrinsically a relativistic phenomena.

In this chapter we will describe the dynamical equations for the EMF, the dynamical equations for the particles and we will introduce several approximations that are going to be used throughout the text.

## 2.1 MAXWELL EQUATIONS[1]

The Maxwell equations are a set of differential equations that unifies both electricity and magnetism, and they read (in SI units):

$$\nabla \cdot \boldsymbol{E}(\boldsymbol{r},t) = \rho(\boldsymbol{r},t)\,\varepsilon_0^{-1} \tag{2.1.1}$$
$$\nabla \cdot \boldsymbol{B}(\boldsymbol{r},t) = 0 \tag{2.1.2}$$
$$\nabla \times \boldsymbol{E}(\boldsymbol{r},t) = -\partial_t \boldsymbol{B}(\boldsymbol{r},t) \tag{2.1.3}$$
$$\nabla \times \boldsymbol{B}(\boldsymbol{r},t) = c^{-2}\partial_t \boldsymbol{E}(\boldsymbol{r},t) + \varepsilon_0^{-1}c^{-2}\boldsymbol{j}(\boldsymbol{r},t), \tag{2.1.4}$$

where $\boldsymbol{E}$ is the electric field, $\rho$ is the charge density, $\varepsilon_0$ is the vacuum permittivity, $\boldsymbol{B}$ is the magnetic field, $c$ is the speed of light and $\boldsymbol{j}$ is the current density.

In this way we can define the vector potential and the scalar potential and obtain both the electric field and the magnetic field from

$$\boldsymbol{B} = \nabla \times \boldsymbol{A} \tag{2.1.5}$$
$$\boldsymbol{E} = -\partial_t \boldsymbol{A} - \nabla U, \tag{2.1.6}$$

where $\boldsymbol{A}$ is the potential vector and $U$ is the scalar potential.

1 This section is in SI units

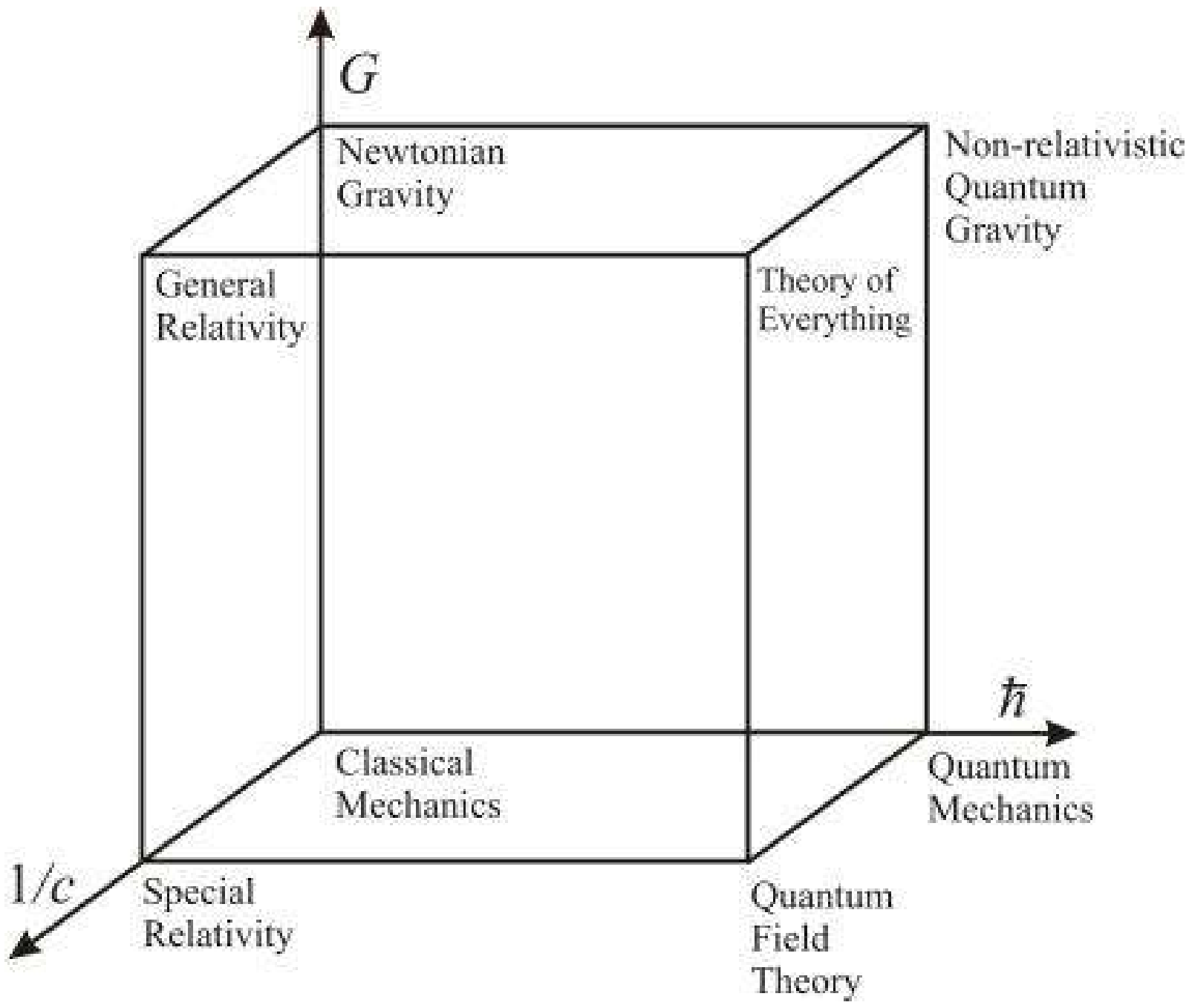

Figure 2.1.1.: *Cube of physics.*

Defining in this way the magnetic and electric fields, Eqs. (2.1.2) and (2.1.3) are satisfied by definition. Replacing the previous relationships in the other two Maxwell equations, we obtain the dynamical equations for the vector and scalar potentials

$$\Delta U = -\frac{1}{\varepsilon_0}\rho - \nabla \cdot \partial_t \boldsymbol{A}, \quad (2.1.7)$$

$$\left(\frac{1}{c^2}\partial_t^2 - \Delta\right)\boldsymbol{A} = \varepsilon_0^{-1}c^{-2}\boldsymbol{j} - \nabla\left(\nabla \cdot \boldsymbol{A} + \frac{1}{c^2}\partial_t U\right). \quad (2.1.8)$$

There are infinite ways to define the same pair of electric and magnetic fields from the potentials. The transformation

$$\boldsymbol{A}' = \boldsymbol{A} + \nabla F \quad (2.1.9)$$

$$U' = U - \partial_t F, \quad (2.1.10)$$

being $F$ a scalar function, does not change the electric and magnetic fields. This is called a *gauge transformation*. Since we have such a freedom to choose the potentials, we can impose conditions to fix $\boldsymbol{A}$ and $U$.

In this work, we will not solve the coupled Maxwell-Schrödinger equations [19] and so we are neglecting the effect of the charges and its current on the EMF. We will assume that the external EMF is imposed and we will only solve the Schrödinger equation coupling the matter to an external EMF.

### 2.1.1 *Coulomb gauge*

We will introduce the Coulomb gauge. This gauge imposes that the $\boldsymbol{A}$ vector field is a transverse field, that is $\nabla \cdot \boldsymbol{A} = 0$. In this gauge, Eq. (2.1.7) takes the form of the Poisson equation, $\nabla^2 U = -\rho/\varepsilon_0$ which is well known in electrostatics. Fixing that $U(|\boldsymbol{r}| = \infty, t) = 0$, we can show [32] that the scalar potential is just the Coulomb potential,

$$U_C(\boldsymbol{r}, t) = \frac{1}{4\pi\varepsilon_0} \int d^3r' \frac{\rho(\boldsymbol{r}', t)}{|\boldsymbol{r} - \boldsymbol{r}'|}. \tag{2.1.11}$$

### 2.1.2 *Göppert-Mayer gauge*

Starting from the Coulomb gauge, i.e. $\boldsymbol{A}$ and $U$ refer to the vector and scalar potential in the Coulomb gauge, we define the scalar function

$$F(\boldsymbol{r}, t) = -(\boldsymbol{r} - \boldsymbol{r}_0) . \boldsymbol{A}(\boldsymbol{r}_0, t) .$$

The new potentials are now

$$\boldsymbol{A}'(\boldsymbol{r}, t) = \boldsymbol{A}(\boldsymbol{r}, t) - \boldsymbol{A}(\boldsymbol{r}_0, t) \tag{2.1.12}$$

$$U'(\boldsymbol{r}, t) = U_C(\boldsymbol{r}, t) + (\boldsymbol{r} - \boldsymbol{r}_0) . \partial_t \boldsymbol{A}(\boldsymbol{r}_0, t) . \tag{2.1.13}$$

We will return to this gauge when we discuss the interaction of a system of particles with the EMF.

## 2.2 PARTICLES IN AN ELECTROMAGNETIC FIELD[2]

The starting point of any problem in Quantum Mechanics is the Hamiltonian. This is the operator that will dictate the time-evolution of the system. The non-relativistic Hamiltonian for a system of charged particles coupled to an external field is [32]

2 This section is in SI units

$$\begin{aligned} \hat{H}(t) &= \sum_i \frac{1}{2m_i}\left(\hat{\boldsymbol{p}}_i - q_i \boldsymbol{A}(\hat{\boldsymbol{r}}_i, t)\right)^2 + \sum_i q_i U(\hat{\boldsymbol{r}}_i, t) \qquad (2.2.1)\\ &= \sum_i \frac{1}{2m_i}\hat{\boldsymbol{p}}_i^2 - \sum_i \frac{q_i}{2m_i}\left(\hat{\boldsymbol{p}}_i \cdot \boldsymbol{A}(\hat{\boldsymbol{r}}_i, t) + \boldsymbol{A}(\hat{\boldsymbol{r}}_i, t)\cdot \hat{\boldsymbol{p}}_i\right)\\ &+ \sum_i \frac{q_i^2}{2m_i}\boldsymbol{A}^2(\hat{\boldsymbol{r}}_i, t) + \sum_i q_i U(\hat{\boldsymbol{r}}_i, t) \qquad (2.2.2) \end{aligned}$$

where $m_i$ and $q_i$ are the mass and the charge of each particle. The operators $\hat{\boldsymbol{r}}_i$ and $\hat{\boldsymbol{p}}_i$ are the position and momentum operators of each particle obeying the commutation relationship $\left[\hat{\boldsymbol{r}}_i, \hat{\boldsymbol{p}}_j\right] = i\hbar\delta_{ij}$.

The dynamical equation that is imposed in NRQM is the Time-dependent Schrödinger equation (TDSE)

$$i\hbar\partial_t \left|\Psi(t)\right\rangle = \hat{H}(t)\left|\Psi(t)\right\rangle . \qquad (2.2.3)$$

We will discuss in this work several ways to solve this equation and also ways to extract information from this dynamical equation.

### 2.2.1 *Gauge invariance of the TDSE*

We can solve the TDSE in the gauge that we prefer. A gauge transformation will only transform the wavefunction by a unitary transformation [32]. In detail if the wavefunction described in the first representation (with the electromagnetic potentials $\boldsymbol{A}$ and $U$) is $\Psi(..., \boldsymbol{r}_i, ..., t)$, after performing a gauge transformation with the scalar function $F$ and solving the TDSE we can show [32] that the wavefunction on the new representation (with the electromagnetic potentials $\boldsymbol{A}' = \boldsymbol{A} + \nabla F$ and $U' = U - \partial_t F$) is given by the unitary transformation

$$\hat{\boldsymbol{T}} = \exp\left(\frac{i}{\hbar}\sum_i q_i F(\hat{\boldsymbol{r}}_i, t)\right) \qquad (2.2.4)$$

and the new wavefunction is $\Psi'(..., \boldsymbol{r}_i, ..., t) = \hat{\boldsymbol{T}}\Psi'(..., \boldsymbol{r}_i, ..., t)$.

### 2.2.2 *Long-wavelength approximation (or dipole approximation)*

In the matter-light interactions studied in this work, the wavelength of the light, $\lambda$, is usually very large compared to the dimensions of the system under study. For example, for a Ti:sapphire laser the typical wavelength is 800 nm, much larger than the dimensions of atomic hydrogen ($a_0 = 0.0529177$ nm). Under these conditions, the amplitude of the external field is practically constant over the spatial extent of the molecule and the vector potential can be replaced by the vector potential at the center of the molecule $\boldsymbol{r}_0$, and

$$\boldsymbol{A}\left(\boldsymbol{r},t\right) = \boldsymbol{A}\left(\boldsymbol{r}_0,t\right), \tag{2.2.5}$$

this is the so-called long-wavelength approximation or dipole approximation. Making this approximation, we can see that the magnetic field $\boldsymbol{B}$ vanishes. This is only valid when the velocity of our particles is small compared to $c$, otherwise the magnetic term on the Lorentz force becomes important. Working on the Coulomb gauge, Eq. (2.2.1) takes the following form

$$\begin{aligned}\hat{H}\left(t\right) &= \sum_i \frac{1}{2m_i}\left(\hat{\boldsymbol{p}}_i - q_i\boldsymbol{A}\left(\boldsymbol{r}_0,t\right)\right)^2 + \sum_i q_i U_C\left(\hat{\boldsymbol{r}}_i\right) & (2.2.6)\\ &= \hat{H}_0 + \hat{H}_I\left(t\right). & (2.2.7)\end{aligned}$$

where the interaction Hamiltonian is denoted by $\hat{H}_I$ and the unperturbed Hamiltonian is denoted by $\hat{H}_0$.

In the Coulomb gauge we have that $\hat{\boldsymbol{p}}$ and $\boldsymbol{A}$ commute, since the divergence of the vector potential is zero. Assuming the dipole approximation, the interaction Hamiltonian will no longer depend on the position operator of the particles. Also the quadratic term in the vector potential is just a time-dependent scalar in the Hamiltonian and it will not couple different states [34]. Therefore, we can state the following final form for the interaction Hamiltonian in the long-wavelength approximation

$$\hat{H}_I^V\left(t\right) = -\sum_i \frac{q_i}{m_i}\hat{\boldsymbol{p}}_i.\boldsymbol{A}\left(\boldsymbol{r}_0,t\right). \tag{2.2.8}$$

The interaction term written like this is usually refereed in the literature of strong-field physics as the *velocity-gauge*.

By making the gauge transformation (Göppert-Mayer gauge) stated in Section 2.1.2, the unperturbed Hamiltonian, $\hat{H}_0$, is unchanged. The new electromagnetic potentials are

$$A' = 0 \tag{2.2.9}$$
$$U' = U + (\boldsymbol{r} - \boldsymbol{r}_0) \cdot \partial_t \boldsymbol{A}(\boldsymbol{r}_0, t). \tag{2.2.10}$$

The electric field of an electromagnetic wave in the Coulomb gauge is just $\boldsymbol{E}(\boldsymbol{r}, t) = -\partial_t \boldsymbol{A}(\boldsymbol{r}, t)$ [34]. Making use of this, the interaction Hamiltonian in the *length gauge* is

$$\hat{H}_I^L(t) = -\sum_i q_i (\hat{\boldsymbol{r}}_i - \boldsymbol{r}_0) . \boldsymbol{E}(\boldsymbol{r}_0, t). \tag{2.2.11}$$

The two gauges presented in this subsection are the most common ones in strong field physics. They are formally equivalent, since we know that QM is gauge invariant. However in numerical calculations only when the exact limit is achieved the two gauges coincide. Several comparisons were presented in the literature [35, 36]. Usually the length gauge is preferred for calculations where the wavefunction is expanded over a spatial grid because in this way the interaction Hamiltonian is a diagonal term in the total Hamiltonian. In this work we will always use the *length gauge*, unless otherwise stated.

### 2.2.3 *Laser pulse*

In this subsection, the mathematical form of the laser pulse is given taking into account that we are working under the dipole approximation, i.e., the electric laser field does not have an explicit dependence on the spatial coordinates. In this work we will restrict our study to laser pulses where the electric field is linearly polarized along the internuclear axis. The external laser field is described by the product of an envelope function, $f(t)$, and a sin function with frequency $\omega$ as

$$E(t) = E_0 f(t) \sin(\omega t + \delta) \tag{2.2.12}$$

where $E_0$ is related to the peak intensity of the laser pulse and in atomic units is expressed as

$$E_0\,(\text{a.u.}) = \sqrt{\frac{I\,(\text{W}/\text{cm}^2)}{I_0}} \tag{2.2.13}$$
$$I_0 = 3.51 \times 10^{16} \text{W}/\text{cm}^2. \tag{2.2.14}$$

The envelope function can take several forms but we will use the $\cos^2$ envelope

$$f(t) = \begin{cases} \cos^2\left(\frac{\pi t}{T}\right) & |t| \leq \frac{T}{2} \\ 0 & |t| > \frac{T}{2} \end{cases} \tag{2.2.15}$$

where $T$ is the total duration of the laser pulse. The relation between the pulse duration and the number of optical cycles, $n_{cycles}$, contained in the laser pulse is simply given by

$$T = \frac{2\pi}{\omega} n_{cycles}. \tag{2.2.16}$$

With these equations we have established all the dynamical equations that we will solve in this work. One of our major goals is to solve the following TDSE

$$i\hbar\partial_t \Psi(..., \boldsymbol{r}_i, ..., t) = \left(\hat{H}_I^L(t) + \hat{H}_0\right)\Psi(..., \boldsymbol{r}_i, ..., t) \tag{2.2.17}$$

knowing the wavefunction at the initial time $t_0$.

## 2.3 DIATOMIC MOLECULE INTERACTING WITH A LASER FIELD

### 2.3.1 *Separation of center-of-mass motion*[3]

Consider an $n$-electron diatomic molecule with nuclei of masses $M_a$ and $M_b$ and charges $ea$ and $eb$, being $e$ the modulus of the electron charge [37]. The position of the particles in the laboratory frame are $\boldsymbol{R}_a$, $\boldsymbol{R}_b$ for the nuclei position and $\boldsymbol{r}_{i'}$ for the position of the $i^{th}$ electron. We consider that the electric field is polarized along the $z$ axis. The Hamiltonian of this system in the dipole approximation and in length gauge is

$$\hat{H}(t) = -\frac{\hbar^2}{2}\left\{\frac{1}{M_a}\nabla_a^2 + \frac{1}{M_b}\nabla_b^2 + \frac{1}{m}\sum_{i=1}^{n}\nabla_{ei'}^2\right\} + V_C + \hat{H}_I^L(t) \tag{2.3.1}$$

$$\hat{H}_I^L(t) = -eE_z(t)\left[az_a + bz_b - \sum_{i=1}^{n} z_{ei'}\right] \tag{2.3.2}$$

where $m$ is the mass of the electron, $V_C$ is the Coulomb interaction potential and $E_z(t)$ is the electric field. We want to separate the center-of-mass motion and in order to do that we introduce new coordinates. We will introduce the center-

3 This subsection is in SI units

of-mass coordinate, $\boldsymbol{R}_{CM}$, a relative nuclear coordinate, $\boldsymbol{R}_n$, and the distance of each electron to the center of mass of the two nuclei, $\boldsymbol{r}_i$. The new coordinates can be expressed as

$$\boldsymbol{R}_{CM} = \frac{1}{M}\left(M_a\boldsymbol{R}_a + M_b\boldsymbol{R}_b + \sum_{i=1}^{n} m\boldsymbol{r}_{i'}\right), \quad (2.3.3)$$

$$\boldsymbol{R} = \boldsymbol{R}_a - \boldsymbol{R}_b, \quad (2.3.4)$$

$$\boldsymbol{r}_i = \boldsymbol{r}_{i'} - \frac{1}{M_a + M_b}\left(M_a\boldsymbol{R}_a + M_b\boldsymbol{R}_b\right). \quad (2.3.5)$$

The total Hamiltonian is now written in these new coordinates as

$$\hat{H}(t) = \hat{H}_{CM}(t) + \hat{H}_{internal}(t) \quad (2.3.6)$$

$$\hat{H}_{CM}(t) = -\frac{\hbar^2}{2}\left(\frac{1}{M_a + M_b + nm}\nabla^2_{CM}\right) - eE_z(t)(a + b - n)z_{CM} \quad (2.3.7)$$

$$\begin{aligned}\hat{H}_{internal}(t) = & -\frac{\hbar^2}{2}\left\{\frac{1}{M_n}\nabla^2_n + \frac{1}{M_a + M_b}\sum_{i=1}^{n}\sum_{j\neq i}\nabla_i \cdot \nabla_j\right\} \\ & - \frac{\hbar^2}{2}\left\{\frac{1}{\mu_e}\sum_{i=1}^{n}\nabla^2_i\right\} \\ & + V_C - eE_z(t)\left[\frac{aM_b - bM_a}{M_a + M_b}\right]z_n \\ & + eE_z(t)\left[1 + \frac{(a + b - n)m}{M_a + M_b + nm}\right]\sum_{i=1}^{n} z_i \end{aligned} \quad (2.3.8)$$

where $M_n = M_aM_b/\left(M_a + M_b\right)$ and $\mu_e = m\left(M_a + M_b\right)/\left(M_a + M_b + m\right)$. The center-of-mass motion was separated from the internal coordinates and describes the motion of a particle $\left(M_a + M_b + nm\right)$ moving in an electrostatic field with charge $e\left(a + b - n\right)$. We are interested in the processes that involve the internal degrees of freedom so we will only solve the dynamics for the internal Hamiltonian.

### 2.3.2 *Hamiltonian for the $H_2^+$ molecule and its isotopes*

In this work we will focus our attention to the simplest molecule in nature, i.e., $H_2^+$ and its isotopes ($D_2^+$ and $T_2^+$) [38]. For these cases we have that $n = 1$, $a = b = 1$ and $M_a = M_b = M_X$ where $M_X$ is standing for the mass of the nucleus of hydrogen, deuterium or tritium. Simplifying, the Hamiltonian for the internal degrees of freedom reads (in atomic units)

$$\begin{aligned} \hat{H}_{internal}(t) &= -\frac{1}{2}\left\{\frac{2}{M_X}\nabla_n^2 + \frac{1}{\mu_e}\nabla^2\right\} + V_C \\ &+ E_z(t)\left[1 + \frac{1}{2M_X+1}\right]z, \end{aligned} \tag{2.3.9}$$

$$\hat{H}_{internal}(t) = \hat{H}_0 + \hat{H}_I^L(t), \tag{2.3.10}$$

$$\hat{H}_I^L(t) = E_z(t)\left[1 + \frac{1}{2M_X+1}\right]z, \tag{2.3.11}$$

$$\mu_e = 2M_X/(2M_X+1), \tag{2.3.12}$$

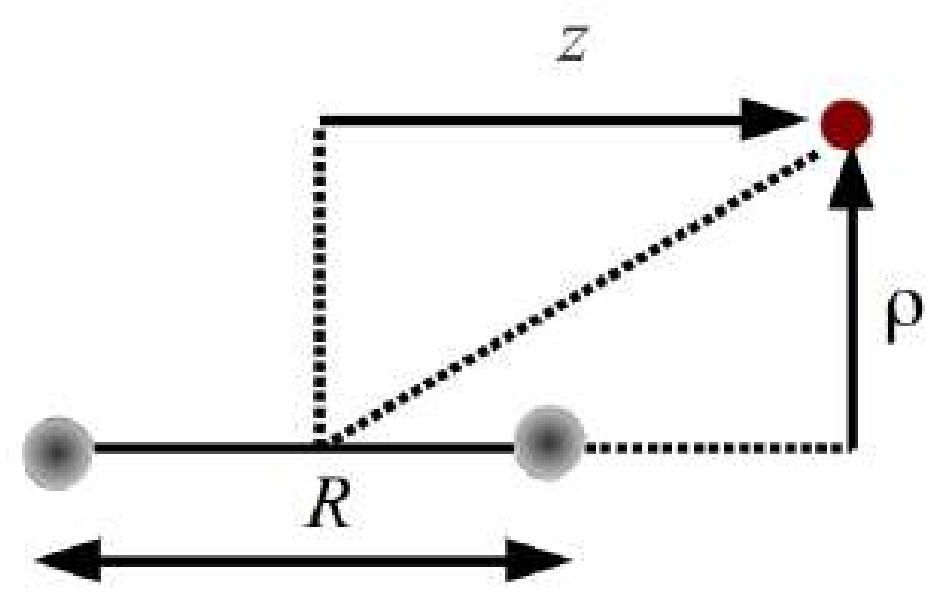

Figure 2.3.1.: Coordinate system used to describe the $H_2^+$ system

We will drop the internal label to the Hamiltonian and from now on $\hat{H}_{internal}(t)$ is going to be refereed just as $\hat{H}(t)$.

Neglecting all rotational effects fixes the orientation of the internuclear axis (see Fig. 2.3.1). We will express the electron position, $\vec{r}$, in cylindrical coordinates $(\rho, z, \phi)$ with the origin in the middle of our molecular axis, $z$ parallel to the internuclear axis and making $R$ the vibrational coordinate. We will study laser fields that are polarized along the $z$ axis and we start from states with $m = 0$, being $m$ the quantum number associated to the $z$ component of the angular momentum ($\hat{L}_z$). In this case we can take advantage of the cylindri-cal symmetry of the problem and the term $\frac{1}{\rho^2}\frac{\partial^2}{\partial\phi^2}$ disappears. The unperturbed Hamiltonian reads

$$\hat{H}_0 = \hat{T}_R + \hat{T}_z + \hat{T}_\rho + V_C \tag{2.3.13}$$

$$\hat{V}_C = \frac{1}{R} - \frac{1}{\sqrt{\rho^2 + (z+R/2)^2}} - \frac{1}{\sqrt{\rho^2 + (z-R/2)^2}} \tag{2.3.14}$$

$$\hat{T}_R = -\frac{1}{M_X}\frac{\partial^2}{\partial R^2} \tag{2.3.15}$$

$$\hat{T}_z = -\frac{1}{2\mu_e}\frac{\partial^2}{\partial z^2} \tag{2.3.16}$$

$$\hat{T}_\rho = -\frac{1}{2\mu_e}\left(\frac{\partial^2}{\partial\rho^2} + \frac{1}{\rho}\frac{\partial}{\partial\rho}\right) \tag{2.3.17}$$

where $\hat{T}_R$, $\hat{T}_z$ and $\hat{T}_\rho$ are the kinetic energy operators associated to the $R$, $z$ and $\rho$ coordinates, respectively, and $\hat{V}_C$ is the Coulomb interaction term between the particles and the normalization condition is chosen so that

$$\iiint dR\rho d\rho dz \left|\Psi(R,\rho,z)\right|^2 = 1. \tag{2.3.18}$$

With the appropriate unperturbed Hamiltonian (Eq. (2.3.13)) and with the interaction potential in length gauge (Eq. (2.3.11)) we are ready to face the practical challenges that the resolution of the TDSE of this system presents. We will refer to this model of the one-electron diatomic homonuclear molecule as the 3D model.

#### 2.3.2.1 *Selection rules*

The Hamiltonian of the $H_2^+$ molecule commutes with the parity operator $\hat{P}$. We can label the eigenstates of the field-free Hamiltonian as being $g$ (*gerade*) or $u$ (*ungerade*) states. Starting from a $g$ state and after the absorption of an odd number of photons, the molecule will change it symmetry to $u$ symmetry. On the other hand, the absorption of an even number of photons will not change the symmetry of the initial wavefunction. If we perform an expansion on spherical harmonics $Y_l^{m=0}$, absorption of an odd number of photons will lead to a combination of spherical harmonics with odd $l$ and, in the same way, absorption of an even number of photons will lead to a combination of spherical harmonics with even $l$.

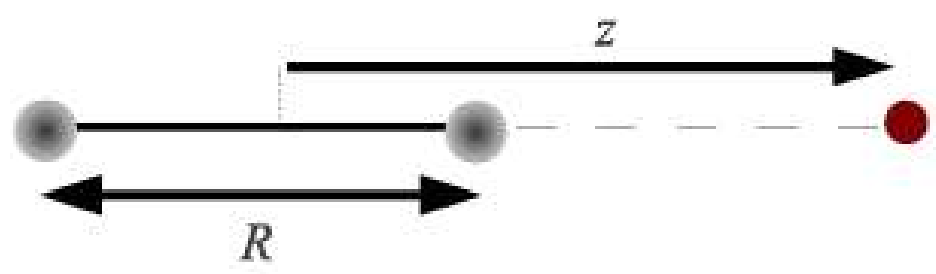

Figure 2.3.2.: Coordinate system used to describe the $H_2^+$ system in a reduced dimensionality model.

### 2.3.3 *Reduced dimensionality model*

We will also work with a reduced dimensional model for the $H_2^+$ molecule, that is based on [17]. In previous sections, the theoretical treatment has been given for the 3D model $(R, \rho, z)$. In the reduced model, we ignore the $\rho$ coordinate and use the soft-core potential described in [17]. We will refer to this model as the 1+1D model (see Fig. 2.3.2), in opposition to the 3D model. The soft-core potential is given by

$$V_{SC}(z,R) = \frac{-1}{1/a(R) - a(R)/b + \sqrt{z_-^2 + [a(R)/b]^2}} \quad (2.3.19)$$

$$+ \frac{-1}{1/a(R) - a(R)/b + \sqrt{z_+^2 + [a(R)/b]^2}} \quad (2.3.20)$$

$$+ 1/R \quad (2.3.21)$$

$$z_\pm = z \pm R/2 \quad (2.3.22)$$

where $b = 5$ and the function $a(R)$ is adjusted to reproduce exactly the 3D potential energy curve of $1s\sigma_g$.

This reduced dimensionality model of the $H_2^+$ molecule cannot give us accurate quantitative results. However, since we are working with very strong laser fields that are oriented along the internuclear axis, the electron dynamics will mainly occur in the $z$ coordinate. In this case, this reduced dimensionality model can be a good approximation to obtain the qualitative picture, reducing significantly the computational effort.

### 2.3.4 *Dipole operator and its derivatives*

As we will see in Chapter 5, the dipole mean value and its temporal derivatives are the observables that are used in the computation of the HHG spectrum. In the following, we assume that the interaction term of the Hamiltonian is in length gauge. The dipole operator of the $H_2^+$ molecule in the laboratory frame is given by

$$\boldsymbol{d} = \boldsymbol{R_a} + \boldsymbol{R_b} - \boldsymbol{r}_{1'} \tag{2.3.23}$$

and making the change of coordinates

$$\begin{pmatrix} \boldsymbol{R} \\ \boldsymbol{r}_1 \\ \boldsymbol{R}_{CM} \end{pmatrix} = \begin{pmatrix} -1 & 1 & 0 \\ -1/2 & -1/2 & 1 \\ \frac{M_X}{2M_X+m} & \frac{M_X}{2M_X+m} & \frac{m}{2M_X+m} \end{pmatrix} \begin{pmatrix} \boldsymbol{R_a} \\ \boldsymbol{R_b} \\ \boldsymbol{r}_{1'} \end{pmatrix} \tag{2.3.24}$$

$$\begin{pmatrix} \boldsymbol{R_a} \\ \boldsymbol{R_b} \\ \boldsymbol{r}_{1'} \end{pmatrix} = \begin{pmatrix} -1/2 & \frac{-m}{2M_X+m} & 1 \\ 1/2 & \frac{-m}{2M_X+m} & 1 \\ 0 & \frac{2M_X}{2M_X+m} & 1 \end{pmatrix} \begin{pmatrix} \boldsymbol{R} \\ \boldsymbol{r}_1 \\ \boldsymbol{R}_{CM} \end{pmatrix} \tag{2.3.25}$$

the dipole operator is written as

$$\boldsymbol{d} = \frac{-2m - 2M_X}{2M_X + m}\boldsymbol{r}_1 + \boldsymbol{R}_{CM}. \tag{2.3.26}$$

To evaluate the dipole in velocity form, $\boldsymbol{v} \equiv \frac{d}{dt}\boldsymbol{d}$, and the dipole in acceleration form, $\boldsymbol{a} \equiv \frac{d^2}{dt^2}\boldsymbol{d}$, we use the Ehrenfest theorem. The time derivative of an expectation value is given by

$$\frac{d}{dt}\langle \boldsymbol{d} \rangle = -i \left\langle \left[ \boldsymbol{d}, \hat{H}(t) \right] \right\rangle + \left\langle \frac{\partial}{\partial t}\boldsymbol{d} \right\rangle. \tag{2.3.27}$$

This way we can evaluate both velocity and acceleration form of the dipole as

$$\langle \boldsymbol{v} \rangle = \frac{-2m - 2M_X}{2M_X}\langle \boldsymbol{p}_1 \rangle + \langle \boldsymbol{p}_{CM} \rangle \frac{1}{2M_X + m} \tag{2.3.28}$$

$$\langle \boldsymbol{a} \rangle = \frac{-m - M_X}{M_X m}\langle \nabla_1 V_C \rangle + \frac{M_X + 2m}{M_X m}\boldsymbol{E}(t) \tag{2.3.29}$$

where $\boldsymbol{p}_1 = -i\nabla_1$ and $\boldsymbol{p}_{CM} = -i\nabla_{CM}$. The three forms of the dipole provide us three alternative ways to obtain the HHG spectra [39]. In Chapter 5, we will see that the acceleration form of the HHG spectra is the more appropriate.

### 2.3.5 *Kinetic and potential energy operators in a non-uniform grid*

The treatment given in this subsection is based on [40]. The time-independent Schrödinger equation can be formulated in a variational way. We know that the eigenstates of an Hamiltonian can be obtained by minimizing the energy functional, i.e.

$$\frac{\delta E\left[\Psi\right]}{\delta\Psi} = 0 \tag{2.3.30}$$

$$E\left[\Psi\right] = \frac{\langle\Psi|\hat{H}|\Psi\rangle}{\langle\Psi|\Psi\rangle}, \tag{2.3.31}$$

and for a one-dimensional case we have

$$E\left[\Psi\right] = \frac{\int\left(\frac{1}{2\mu}\frac{d\Psi^*}{dx}\frac{d\Psi}{dx} + V\right)\xi\left(x\right)dx}{\int\Psi^*\Psi\xi\left(x\right)dx} \tag{2.3.32}$$

where $\mu$ is the mass of the particle and $\xi\left(x\right)dx$ is the volume element. It must be noticed that Eq. (2.3.32) can be simply generalized to the multi-dimensional case. To keep it in a simpler and clear way, the following analysis will be done for the one-dimensional case.

To discretize this equation we will replace any integral by the midpoint rule

$$\int f\left(x\right)dx \to \sum_{i=1}^{N} f\left(x_i\right)\left(x_{i+(1/2)} - x_{i-(1/2)}\right) \tag{2.3.33}$$

and the derivatives by the central differencing scheme

$$\left.\frac{df}{dx}\right|_{x_i} \to \frac{f_{i+(1/2)} - f_{i-(1/2)}}{x_{i+(1/2)} - x_{i-(1/2)}}. \tag{2.3.34}$$

In this treatment we will work with an arbitrary grid. The variational principle can also be written in the following form

$$\frac{\partial E\left[\Psi\right]}{\partial\Psi_j^*} = 0. \tag{2.3.35}$$

Substituting Eqs. (2.3.33) and (2.3.34) in Eq. (2.3.32), we obtain

$$\begin{aligned} E\Psi_j\xi_j\left(x_{j+(1/2)} - x_{j-(1/2)}\right) &= \frac{1}{2\mu}\left[\frac{\Psi_j - \Psi_{j-1}}{x_j - x_{j-1}}\xi_{j-(1/2)} - \frac{\Psi_{j+1} - \Psi_j}{x_{j+1} - x_j}\xi_{j+(1/2)}\right] \\ &+ V_j\Psi_j\xi_j\left(x_{j+(1/2)} - x_{j-(1/2)}\right) \end{aligned} \quad (2.3.36)$$

where $j$ is the spatial index. Rewriting as a matrix equation we get

$$\boldsymbol{H}\vec{\Psi} = E\boldsymbol{S}\vec{\Psi} \quad (2.3.37)$$

where $\boldsymbol{H}$ is the Hamiltonian matrix and $\boldsymbol{S}$ is the overlap matrix. The wavefunction is written as $\vec{\Psi}$ to remark the fact that after discretization we can express the wavefunction as a column vector. The Hamiltonian matrix is a tridiagonal, non-Hermitian matrix. The overlap matrix is a diagonal matrix.

We must impose boundary conditions. In this case our boundary conditions can be the continuity or the differentiability of the wavefunction at the boundaries. If we impose that the wavefunction must be zero at the boundaries, we then must set $\Psi_0$ (and $\Psi_{N+1}$) as zero. No other changes in the matrix are needed. On the other hand, if we impose that the derivative of the wavefunction at the boundaries is zero we must impose that $\Psi_0 = \Psi_1$ (or $\Psi_{N+1} = \Psi_N$). This requires that a $\frac{1}{2\mu}\frac{\xi_{1/2}}{x_1 - x_o}$ term should be subtracted from the first diagonal element of the Hamiltonian matrix. A similar term must be subtracted of the last diagonal element of the Hamiltonian. We will choose this last option for the boundary conditions.

We have arrived to a generalized eigenvalue problem. We can rewrite the overlap matrix as

$$\boldsymbol{S} = \boldsymbol{L}\boldsymbol{L}^T \quad (2.3.38)$$

and since $\boldsymbol{S}$ is a diagonal matrix we obtain

$$L_i = \sqrt{S_i}. \quad (2.3.39)$$

We can transform the Hamiltonian matrix and the wavefunction accordingly and transform the generalized eigenvalue problem into a standard eigenvalue problem,

$$\begin{aligned} \boldsymbol{H}' &= \boldsymbol{L}^{-1}\boldsymbol{H}\boldsymbol{L}^{-T} && (2.3.40) \\ \vec{\Psi}' &= \boldsymbol{L}^T\vec{\Psi} && (2.3.41) \\ \boldsymbol{H}'\vec{\Psi}' &= E\vec{\Psi}' && (2.3.42) \end{aligned}$$

This transformation will produce a Hermitian Hamiltonian. We can write explicitly the value of the transformed Hamiltonian. It will be a tridiagonal matrix, just like before the transformation. The diagonal elements of the new Hamiltonian will be

$$H'_{ii} = \frac{1}{2\mu}\left[\frac{1}{x_i - x_{i-1}}\xi_{i-(1/2)} - \frac{1}{x_{i+1} - x_i}\xi_{i+(1/2)}\right]\frac{1}{\xi_i\left(x_{i+(1/2)} - x_{i-(1/2)}\right)} + V_i, \tag{2.3.43}$$

and the non-diagonal elements of the matrix will be

$$H'_{i,i+1} = -\frac{1}{2\mu}\frac{\xi_{i+(1/2)}}{\left(x_{i+1} - x_i\right)\sqrt{\xi_i\left(x_{i+(1/2)} - x_{i-(1/2)}\right)\xi_{i+1}\left(x_{i+(3/2)} - x_{i+(1/2)}\right)}} + V_i \tag{2.3.44}$$

where the other terms can be obtained by the symmetry of the $\boldsymbol{H}'$ matrix ($H'_{i,i+1} = H'_{i+1,i}$). The wavefunction will also be transformed. The normalization condition of our wavefunction becomes

$$1 = \sum_{j=1}^{N}\left|\Psi_j\right|^2\xi_j\left(x_{j+(1/2)} - x_{j-(1/2)}\right) = \sum_{j=1}^{N}\left|\Psi'_j\right|^2. \tag{2.3.45}$$

For the sake of clarity, in our case for the $z$ and $R$ coordinate $\xi = 1$. For the $\rho$ coordinate we have that $\xi = \rho$. Our wavefunction will be also transformed

$$\Psi'\left(R,\rho,z\right) = \Psi\left(R,\rho,z\right)\sqrt{\rho dRdzd\rho}. \tag{2.3.46}$$

This method allows the treatment of quantum systems in non-linear and non-cartesian grids avoiding the non-Hermitian discretized Hamiltonian. In the 1+1D model, however, we will use a homogeneous grid.

#### 2.3.5.1 *Cubic grid in the 3D model*

For each coordinate we use a cubic grid and the position of each grid point $x$ is given by the following formula

$$x\left(n\right) = an^3 + cn \tag{2.3.47}$$

$$x'\left(n\right) = 3an^2 + c \tag{2.3.48}$$

The values of $a$ and $c$ are fixed by defining the grid separation at the middle of the grid (at $x = 0$) $\Delta x$ and the place where the grid separation is $2\Delta x$ labelled as $\Delta\Delta x$. The relation between $a$,$c$ and $\Delta x$ and $\Delta\Delta x$ is obtained by

$$x\left(n_1\right) = 0 \tag{2.3.49}$$
$$x'\left(n_1\right) = \Delta x \tag{2.3.50}$$
$$x\left(n_2\right) = \Delta\Delta x \tag{2.3.51}$$
$$x'\left(n_2\right) = 2\Delta x \tag{2.3.52}$$

and if we solve these equations we get that

$$c = \Delta x \tag{2.3.53}$$
$$a = \frac{16\left(\Delta x\right)^3}{27\left(\Delta\Delta x\right)^2}. \tag{2.3.54}$$

In the 3D calculations one has to choose values of $\Delta x$ and $\Delta\Delta x$ to define the non-linear grid.

## 2.4 SOLVING THE TDSE

### 2.4.1 *Time-Dependent Schrödinger Equation*

We start with the Schrödinger equation.

$$i\frac{\partial}{\partial t}\Psi\left(\boldsymbol{r},t\right) = \hat{H}\left(t\right)\Psi\left(\boldsymbol{r},t\right). \tag{2.4.1}$$

A formal integration of (2.4.1) gives [41]

$$\Psi\left(\boldsymbol{r},t\right) = \hat{P}\exp\left[-i\int_{t_0}^{t}dt'\hat{H}\left(t'\right)\right]\Psi\left(\boldsymbol{r},t_0\right) \tag{2.4.2}$$

where $\hat{P}$ is the Dyson time-ordering operator. Eq. (2.4.2) is another way to write the TDSE, but the time-ordering operator is very difficult to evaluate. In some cases, the Dyson operator becomes a more feasible object, such as in a time-independent Hamiltonian where Eq. (2.4.2) becomes

$$\Psi\left(\boldsymbol{r},t\right) = \exp\left[-i\left(t-t_0\right)\hat{H}\right]\Psi\left(\boldsymbol{r},t_0\right). \tag{2.4.3}$$

For an interval $[t, t+\Delta t]$ where $\Delta t$ is a very small time step, the integral in Eq. (2.4.2) can be approximated by $\hat{H}(t+\Delta t/2)\Delta t$. In this case

$$\Psi(\boldsymbol{r}, t+\Delta t) = \exp\left(-i\Delta t \hat{H}(t+\Delta t/2)\right)\Psi(\boldsymbol{r}, t) \tag{2.4.4}$$

But how can we evaluate the exponential of the Hamiltonian? We can, in principle, diagonalize the Hamiltonian, and write the exponential of the Hamiltonian in terms of its eigenvalues in a straightforward way[4]. However, since we want to work on a numerical grid, where we explicitly avoid the diagonalization of our Hamiltonian, we must use another method. We will apply the Crank-Nicholson method.

### 2.4.2 *Crank-Nicolson Method*

Our first approach to the question that arose in the previous subsection would be to expand the exponential in a Taylor series, truncating the series at a certain order. However, we will not expand in a Taylor series but into a Padé Approximant. The problem of the expansion in a Taylor series is the fact that the approximated exponential operator would not be unitary. In fact, it can be easily verified that in first order, $\left(\hat{1} - i\Delta t\hat{H}\right)\left(\hat{1} - i\Delta t\hat{H}\right)^{\dagger} \neq \hat{1}$.

The Padé Approximant is a rational approximation to a function. In general, a function $f(z)$ can be written as

$$f(z) = \frac{P_n(z)}{Q_m(z)} + \mathcal{O}\left(z^{n+m+1}\right) \tag{2.4.5}$$

where $P_n$ and $Q_m$ are two polynomials of order $n$ and $m$. Evaluating the exponential with $n = 1$ and $m = 1$ we get

$$\exp(z) = \frac{2+z}{2-z} + \mathcal{O}\left(z^3\right). \tag{2.4.6}$$

Applying this to Eq. (2.4.4), we obtain

$$\Psi(\boldsymbol{r}, t+\Delta t) = \left(\frac{1 - i\hat{H}\Delta t/2}{1 + i\hat{H}\Delta t/2}\right)\Psi(\boldsymbol{r}, t) \tag{2.4.7}$$

4 The exponential of a diagonal matrix is a diagonal matrix whose terms are the exponentials of the diagonal terms of the original matrix.

where $\hat{H}$ is evaluated at $t+\Delta t/2$. With this expansion we introduce an error $\mathcal{O}\left(\Delta t^3\right)$. It is important to notice that this propagator will now be unitary, since $\left(\frac{1-i\hat{H}\Delta t/2}{1+i\hat{H}\Delta t/2}\right)\left(\frac{1-i\hat{H}\Delta t/2}{1+i\hat{H}\Delta t/2}\right)^{\dagger}=\hat{1}$. To calculate $\Psi\left(\boldsymbol{r},t+\Delta t\right)$ we can write Eq. (2.4.7) in the following form

$$\left(1+i\hat{H}\Delta t/2\right)\Psi\left(\boldsymbol{r},t+\Delta t\right)=\left(1-i\hat{H}\Delta t/2\right)\Psi\left(\boldsymbol{r},t\right) \tag{2.4.8}$$

and in a numerical grid this will become a set of linear equations. Since the complexity of this problem increases with the dimensionality of the problem we can separate the propagation of each time step into pieces, in the so-called Split-Operator Method.

### 2.4.3 *Split-Operator Method*

Making use of Eq. (2.4.7) without further approximations requires the solution of a linear set of $N$ equations, that in general does not scale linearly with $N$. We know that for the one-dimensional case, the kinetic energy operator, in a numerical grid and applying a three-point difference scheme, can be expressed as a tridiagonal matrix. The potential operator is diagonal and the resulting Hamiltonian is a tridiagonal matrix. This can lead to a substantial reduction of computational effort, since there is a very efficient algorithm to solve tridiagonal systems of equations that scales with $N$. The algorithm is known as the tridiagonal matrix algorithm or Thomas algorithm [42].

In order to decrease the computational effort, it would be enough to split the propagator into one-dimensional propagators. We start by looking at (2.4.4) and expressing the Hamiltonian as the sum of a kinetic energy operator $\hat{T}$ and a potential energy operator $\hat{V}$. The Zassenhaus formula [43] provides a way to calculate the exponential of the sum of two or three operators

$$e^{t\left(\hat{X}+\hat{Y}\right)} = e^{t\hat{X}}e^{t\hat{Y}}e^{-\frac{t^2}{2}\left[\hat{X},\hat{Y}\right]}e^{\mathcal{O}\left(t^3\right)} \tag{2.4.9}$$

$$e^{t\left(\hat{X}+\hat{Y}+\hat{Z}\right)} = e^{t\hat{X}}e^{t\hat{Y}}e^{t\hat{Z}}e^{-\frac{t^2}{2}\left(\left[\hat{X},\hat{Y}\right]+\left[\hat{Y},\hat{Z}\right]+\left[\hat{X},\hat{Z}\right]\right)}e^{\mathcal{O}\left(t^3\right)} \tag{2.4.10}$$

If we replace $e^{-i\Delta t\left(\hat{T}+\hat{V}\right)}$ by $e^{-i\Delta t\hat{T}}e^{-i\Delta t\hat{V}}$, we are making an error of $\mathcal{O}\left(\Delta t^2\right)$. Instead if we use

$$e^{-i\Delta t\left(\hat{V}/2+\hat{T}+\hat{V}/2\right)} \to e^{-i\frac{\Delta t}{2}\hat{V}}e^{-i\Delta t\hat{T}}e^{-i\frac{\Delta t}{2}\hat{V}} \tag{2.4.11}$$

since $\left[\hat{V}/2, \hat{T}\right] + \left[\hat{T}, \hat{V}/2\right] + \left[\hat{V}/2, \hat{V}/2\right] = 0$, we will only make an error of $\mathcal{O}\left(\Delta t^3\right)$. Since the kinetic energy operator, in our case, is $\hat{T} = \hat{T}_R + \hat{T}_\rho + \hat{T}_z$ and since each one of the kinetic energy operators commutes with the others, we can factorize the exponential without making an additional approximation. In this way, our propagator will be written as

$$e^{-i\frac{\Delta t}{2}\hat{V}} e^{-i\Delta t \hat{T}_R} e^{-i\Delta t \hat{T}_\rho} e^{-i\Delta t \hat{T}_z} e^{-i\frac{\Delta t}{2}\hat{V}} \tag{2.4.12}$$

We know that the time scale for the electronic motion is shorter than the time scale for the motion of the nuclei, so we can take two time steps for the propagation: $\Delta t_{elec}$ for propagating the potential and the electronic kinetic energy operator and $\Delta t_{nuc} = k\Delta t_{elec}$ for propagating the nuclear kinetic energy operator (in particular, we use $k = 10$). Every exponential is evaluated using the Crank-Nicolson method. The explicit form of each kinetic energy operator is explained in Subsection 2.4.4.

### 2.4.4 *Propagation in Imaginary Time*

To solve the TDSE we must impose an initial condition, which usually implies imposing an initial wavefunction at the beginning of the pulse. The initial wavefunction is typically the molecular groundstate or a linear combination of states. The obtention of the molecular groundstate is the subject treated in this subsection.

There are two standard procedures to obtain the eigenstates of a Hamiltonian: the direct diagonalization of the unperturbed Hamiltonian (that is described in Subsection 2.3.5) using the routines that are available at SLEPc [44], and the propagation of a wavefunction in imaginary time, which we will discuss here. A general wavefunction can be expanded in a basis of eigenstates of the Hamiltonian. For a time independent Hamiltonian,

$$\Psi\left(\boldsymbol{r}, t\right) = \sum_{n=0}^{\infty} c_n \psi_n\left(\boldsymbol{r}\right) e^{-iE_n t}. \tag{2.4.13}$$

If we propagate the TDSE with a negative imaginary time, $t \to -i\tau$, Eq. (2.4.13) becomes

$$\Psi\left(\boldsymbol{r}, \tau\right) = \sum_{n=0}^{\infty} c_n \psi_n\left(\boldsymbol{r}\right) e^{-E_n \tau}, \tag{2.4.14}$$

and for a very long $\tau$, the expression will be dominated by the term $n = 0$,

$$\lim_{\tau \to +\infty} \Psi(\boldsymbol{r}, \tau) = c_0 \psi_0(\boldsymbol{r}) e^{-E_0 \tau}, \tag{2.4.15}$$

since all the terms $e^{-E_n \tau}$ decay faster than $e^{-E_0 \tau}$. To obtain the groundstate, we insert at the beginning a trial wavefunction, that should share some general features with the groundstate to ensure a faster convergence. For example, to find the groundstate of the hydrogen atom (an atomic 1*s* orbital) we can place a Gaussian function centred at the nucleus. The propagation is carried out with the previous scheme. For practical reasons, the wavefunction is renormalized every $n$ steps. To ensure convergence we can check the mean value of the energy and verify if it is converged, i.e.

$$|E[\Psi(\boldsymbol{r}, \tau)] - E[\Psi(\boldsymbol{r}, \tau + n\Delta t)]| < \varepsilon \tag{2.4.16}$$

for a small $\varepsilon$, where $\varepsilon$ is an energy that must be small compared to the energy difference between two eigenstates. We can also compare the scalar product between the wavefunctions at successive time steps

$$1 - \frac{|\langle \Psi(\boldsymbol{r}, \tau) | \Psi(\boldsymbol{r}, \tau + n\Delta t) \rangle|^2}{||\Psi(\boldsymbol{r}, \tau)||^2 \, ||\Psi(\boldsymbol{r}, \tau + n\Delta t)||^2} < \delta, \tag{2.4.17}$$

where $\delta << 1$ .

The excited states can be found iteratively. Having the groundstate wavefunction $\psi_0$, we can do the following transformation, to remove the contribution of the groundstate

$$\Psi(\boldsymbol{r}, \tau = 0) = \Phi_{trial}(\boldsymbol{r}) - \langle \psi_0 | \Phi_{trial} \rangle \psi_0(\boldsymbol{r}) \tag{2.4.18}$$

so $c_0 = 0$, and the propagation will lead to the first excited state. To obtain the $i^{th}$ excited state we must calculate all the previously eigenstates before and apply the transformation

$$\Psi(\boldsymbol{r}, \tau = 0) = \Phi_{trial}(\boldsymbol{r}) - \sum_{n=0}^{i-1} \langle \psi_n | \Phi_{trial} \rangle \psi_n(\boldsymbol{r}) \tag{2.4.19}$$

and then propagate in imaginary time. In principle, one could calculate by this method as many eigenstates as needed, but in practice numerical noise limits that number, since lower eigenstates are reintroduced and they must be removed during the propagation. For states that are very close in energy or even degenerate, this method does not provide trustable results. However, we can obtain very good results for the groundstate and first excited state of the $H_2^+$ molecule.

### 2.4.5 *Absorbers*

Until now, we have assumed that our wavefunction is fully contained in the numerical grid, assuming that it is sufficiently large to contain all the probability density of the wavefunction. This assumption requires a huge box of hundreds or even thousands of atomic units, since with a laser pulse we can populate continuum states and create an unbound wavepacket that will travel and eventually reach the boundaries, where it will be reflected, producing an artificial effect in the calculation. To avoid this we should calculate in a very large box, but this is unfeasible in most of the situations.

To solve this problem we can absorb the wavefunction near the boundaries of our numerical grid. We have two ways of doing that. The first one is placing an imaginary potential, which will produce a non-Hermitian Hamiltonian, causing a non-conservation of the norm of our wavefunction.

Let us examine in which way the addition of an imaginary potential (also called complex absorbing potential CAP) can affect the propagation. If we have our hermitian Hamiltonian $\hat{H}$, and our imaginary potential $-iW(\boldsymbol{r})$, and following a split-operator scheme we will have that

$$\Psi(\boldsymbol{r}, t + \Delta t) = e^{-W(\boldsymbol{r})\frac{\Delta t}{2}} e^{-i\hat{H}\Delta t} e^{-W(\boldsymbol{r})\frac{\Delta t}{2}} \Psi(\boldsymbol{r}, t) \tag{2.4.20}$$

and here we can see that the probability density is exponentially damped where $W(\boldsymbol{r}) > 0$. In this way we can define an optical potential that is only different from zero near the boundaries, and we call that region the absorbing region. We will use this kind of absorbing potentials for the calculations in the 1+1D model for the $H_2^+$ molecule.

We can also use mask functions. In this method we only multiply our wavefunction by a exponential decaying function, $f(z)$, in the absorbing region each time step propagation. This functions will damp the wavefunction in that region.

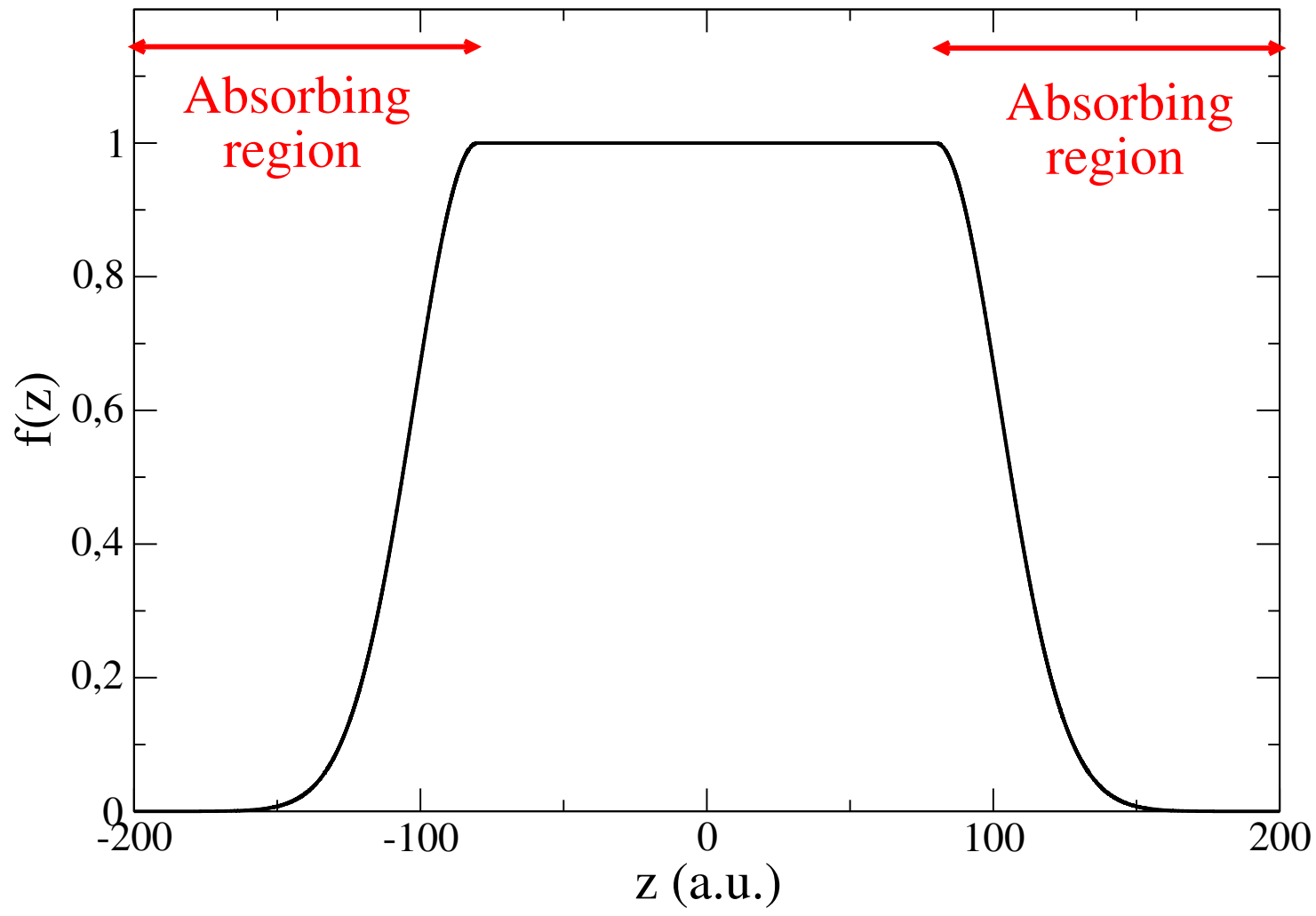

Figure 2.4.1.: Example of a mask function. In this case $z_0 = 80$ a.u. and $\alpha = 0.001$ a.u.

We will use this kind of approach in the 3D calculations for the $H_2^+$ molecule. In particular, we will use a mask function of the form

$$f(z) = \begin{cases} 1 & |z| \leq z_0 \\ \exp\left(-\alpha \Delta t \left|z - z_0\right|^2\right) & z > z_0 \\ \exp\left(-\alpha \Delta t \left|z + z_0\right|^2\right) & z < -z_0 \end{cases} . \quad (2.4.21)$$

in each coordinate. The CAP can be seen as the differential version of the mask function.

If we are interested in studying excitation of bound states or even oscillations of the electronic wavefunction around the nuclear centers, the presence of absorbers should not affect the calculation. However, if we are interested in ionization, it is clear that the absorbers will kill the ionizing part of the wavefunction and this information will be lost. To avoid this problem, we should apply the Virtual Detector method, which is introduced in Chapter 3, to obtain the spectra.

The parameters of these absorbers must be chosen such that they are strong enough to absorb even the fastest parts of the wavefunction, but not too much so that the slowest parts are not reflected. Besides, their spatial width should be long enough to enclose the particle wavelength of the slowest parts of the wavefunction.

# 3

# COMPUTATION OF SPECTRUM OF ELECTRONIC AND IONIC FRAGMENTS

In Chapter 2, we have established the dynamical equations in order to get the time evolution of the wavefunction. But this effort can be pointless if we can not extract observables from our computations. The analysis of molecular dynamics in the strong-field regime is typically done by looking at three different observables: high-harmonic generation (HHG) [45, 46], the kinetic energy release (KER) of the molecular target [47] and the photoelectron-kinetic energy spectrum (EKE) [48].

The description of how the energetic spectrum of fragments can be obtained within a grid method is presented in this section. We must mention other methods that can provide us the energetic spectra of our fragments such as the t-SURFF [23, 49, 50]. In this work, we will provide two different ways to extract this information and several correlated differential quantities that are now measured at the labs with the COLTRIMS technique [16].

## 3.1 MOLECULAR PROCESSES INDUCED BY A LASER FIELD IN $H_2^+$ RESULTING IN FRAGMENTS

### 3.1.1 *Dissociation*

$$H_2^+ + n\hbar\omega \to H^* + H^+ \tag{3.1.1}$$

The molecule absorbs energy that leads to the breakup of the molecule in one proton and one H atom, either in the groundstate or in an excited state. The total kinetic energy of the two fragments can be measured in the center-of-mass frame for different channels of the excited H atom. We will refer to this spectrum as the nuclear kinetic energy spectrum (NKE).

### 3.1.2 *Dissociative ionization*

$$H_2^+ + n\hbar\omega \rightarrow H^+ + H^+ + e^- \tag{3.1.2}$$

The molecule absorbs energy that leads to the ionization of the molecule. After the electron leaves, the parent ion is going to breakup into two protons by Coulomb explosion. In this case we can fully characterize this process by knowing the electronic kinetic energy, $E_e$, the nuclear kinetic energy, $E_N$, and the direction of the emission of the electron, $\Omega$.

## 3.2 VIRTUAL DETECTOR METHOD

The Virtual Detector Method was first introduced by Feuerstein *et al* [51]. The idea is closely related to the hydrodynamical formulation of Quantum Mechanics [52]. How can the momentum distribution of the fragments be determined in an experiment? We can place a detector very far away from the system and measure the flux through a certain area and at the same time determine its momentum. We can then make a histogram and construct the spectrum. The virtual detector method is based on the same idea, but now we do not need a physical machine but instead a surface is chosen where the calculation of the flux is done. The absorber is placed beyond the surface.

We can write the wavefunction as

$$\Psi(\boldsymbol{r},t) = A(\boldsymbol{r},t)\exp(i\phi(\boldsymbol{r},t)) \tag{3.2.1}$$

where $A(\boldsymbol{r},t)$ and $\phi(\boldsymbol{r},t)$ are real-valued functions. We know that the flux in quantum mechanics can be derived from the continuity equation. The probability flux is calculated as

$$\boldsymbol{j}(\boldsymbol{r},t) = \mathrm{Re}\left(\Psi^* \frac{1}{im}\nabla\Psi\right) \tag{3.2.2}$$

$$= \frac{|A(\boldsymbol{r},t)|^2}{m}\nabla\phi(\boldsymbol{r},t). \tag{3.2.3}$$

where $m$ is the mass of the particle. Here we will make a physical assumption. We can think on the probability flux as a local velocity $\boldsymbol{v}(\boldsymbol{r},t)$ times a probability density $|A(\boldsymbol{r},t)|^2$. The local momentum can be defined as

$$\boldsymbol{k}(\boldsymbol{r},t) = m\boldsymbol{v}(\boldsymbol{r},t) = \nabla\phi(\boldsymbol{r},t). \tag{3.2.4}$$

so $\boldsymbol{v}(\boldsymbol{r},t) = \boldsymbol{j}(\boldsymbol{r},t) / |A(\boldsymbol{r},t)|^2$, where $\boldsymbol{j}(\boldsymbol{r},t)$ is obtained with Eq. (3.2.2).

To obtain the momentum distribution, we define a surface on which we calculate the flux and momentum for each point and time step, and then apply a binning procedure. We are assuming that the particle will behave just like a free particle after crossing the virtual detector. Corrections must be added if the pulse is not over when the particle reaches the detector and if the interaction potential is not negligible.

For the sake of clarity, we will explicitly write the equations for our case. The electron momentum is calculated (in the virtual detector, denoted as $\boldsymbol{r}_d$) as

$$p_z(\boldsymbol{r}_d,t) = \nabla_z\phi(\boldsymbol{r}_d,t) - A(t) \tag{3.2.5}$$

$$p_\rho(\boldsymbol{r}_d,t) = \nabla_\rho\phi(\boldsymbol{r}_d,t) \tag{3.2.6}$$

where $A(t)$ is the vector potential. This correction is made if the pulse is not over at the time of measurement, and it is simply the classical change in momentum when an electron is interacting with the field after passing through the virtual detector, see Appendix B. In the $\rho$ direction there is no need of this correction, since the field is polarized in the $z$ direction. The electronic energy is then

$$E_{elec} = \frac{p_z^2}{2} + \frac{p_\rho^2}{2} + V_{eN}. \tag{3.2.7}$$

In the limit, $V_{eN} \to 0$ for $z, \rho \to \infty$.

After ionization the nuclei will evolve along the Coulomb explosion curve. Since this curve is always dissociative, the nuclear momentum can be computed. We will have then a nuclear momentum distribution. When the electron is detected in the virtual detector the nuclei have not had enough time to dissociate, so the interaction potential between the nuclei cannot be neglected. We can calculate the nuclear momentum at the virtual detector, compute the kinetic energy associated and then sum the interaction potential to account for this extra kinetic energy. The nuclear energy will be computed as

$$E_N = \frac{\nabla_R\phi(\boldsymbol{r}_d,t)^2}{2\mu} + \frac{1}{R} \tag{3.2.8}$$

and the nuclear momentum in the asymptotic region as

$$P_N = \sqrt{2\mu E_N} \tag{3.2.9}$$

where $\mu$ is the reduced mass of the nuclei. This method allows one to calculate the photoelectron spectra and nuclear kinetic energy spectra.

## 3.3 RESOLVENT OPERATOR METHOD

In this section, we present an alternative method to the Virtual Detector Method, for extracting observables from a grid calculation. We will present an extension for molecules of a technique used to obtain the photoelectron spectra in atoms, the Resolvent Operator Method.

### 3.3.1 *Total energy distribution*

The Resolvent Operator Method is useful to extract the energy spectrum at the end of the laser pulse, and it requires that all the wavefunction is contained in the grid. It was first introduced by Kulander *et al* [26]. The first calculations based on this technique were compared with experiments in arbitrary units [53–55]. Twenty years later the proportionality constant was described, making this technique able to calculate the probability density [27]. We will show the theory that already existed, which can be applied to study atomic photoionization.

The idea beyond this method is to use an operator to select only states that lie in a particular energy range. We can start by assuming that the wavefunction at the end of the pulse is a sum over bound states and an integral over continuum states, all of them eigenstates of the field-free Hamiltonian. The bound states are normalized to unity, and the continuum states normalized to the Dirac's delta function. In the following, $|b\rangle$ represents bound states and $|\varepsilon\rangle$ represents continuum states. For the sake of simplicity, we will drop the indexes for other quantum numbers. Therefore

$$|\Psi\rangle = \sum_b c_b |b\rangle + \int d\varepsilon c(\varepsilon) |\varepsilon\rangle . \tag{3.3.1}$$

Denoting the field-free Hamiltonian as $\hat{H}_0$, we will have the following relationships for the $|b\rangle$ and $|\varepsilon\rangle$ states:

$$\begin{aligned} \hat{H}_0 |b\rangle &= E_b |b\rangle && (3.3.2) \\ \hat{H}_0 |\varepsilon\rangle &= \varepsilon |\varepsilon\rangle && (3.3.3) \\ \langle\varepsilon|\varepsilon'\rangle &= \delta\left(\varepsilon - \varepsilon'\right) && (3.3.4) \\ \langle b|b'\rangle &= \delta_{b,b'} && (3.3.5) \\ \langle\varepsilon|b\rangle &= 0. && (3.3.6) \end{aligned}$$

The normalization condition of the wavefunction will be

$$1 = \sum_b |c_b|^2 + \int dE \rho\left(E\right), \quad (3.3.7)$$

where we define a probability density $\rho\left(E\right) = \left|c\left(E\right)\right|^2$ for the continuum states. In both cases, the eigenvalue of the Hamiltonian represents the total energy of the system. In the case of a one-electron atom we can directly link this energy to the photoelectron energy.

Since we are in a grid method, we do not have a explicit knowledge of the eigenstates. We can start by defining the resolvent operator as

$$\hat{R}^n_\delta\left(E\right) = \frac{\delta^n}{\left(E - \hat{H}_0\right)^n - i\delta^n} \quad (3.3.8)$$

and its adjoint

$$\hat{R}^n_\delta\left(E\right)^\dagger = \frac{\delta^n}{\left(E - \hat{H}_0\right)^n + i\delta^n}, \quad (3.3.9)$$

where $n$ is the order of the resolvent operator, $\delta$ its resolution and $E$ is the selected energy.

We can write the resolvent operator as a product of Green operators,

$$\hat{R}^n_\delta\left(E\right) = \frac{\delta^n}{\left(E - \hat{H}_0\right)^n - i\delta^n} \equiv \prod_{j=1}^{n} \delta G\left(E - q_j\delta\right), \quad (3.3.10)$$

where

$$G\left(E - q_j\delta\right) = \frac{1}{E - q_j\delta - \hat{H}_0} \quad (3.3.11)$$

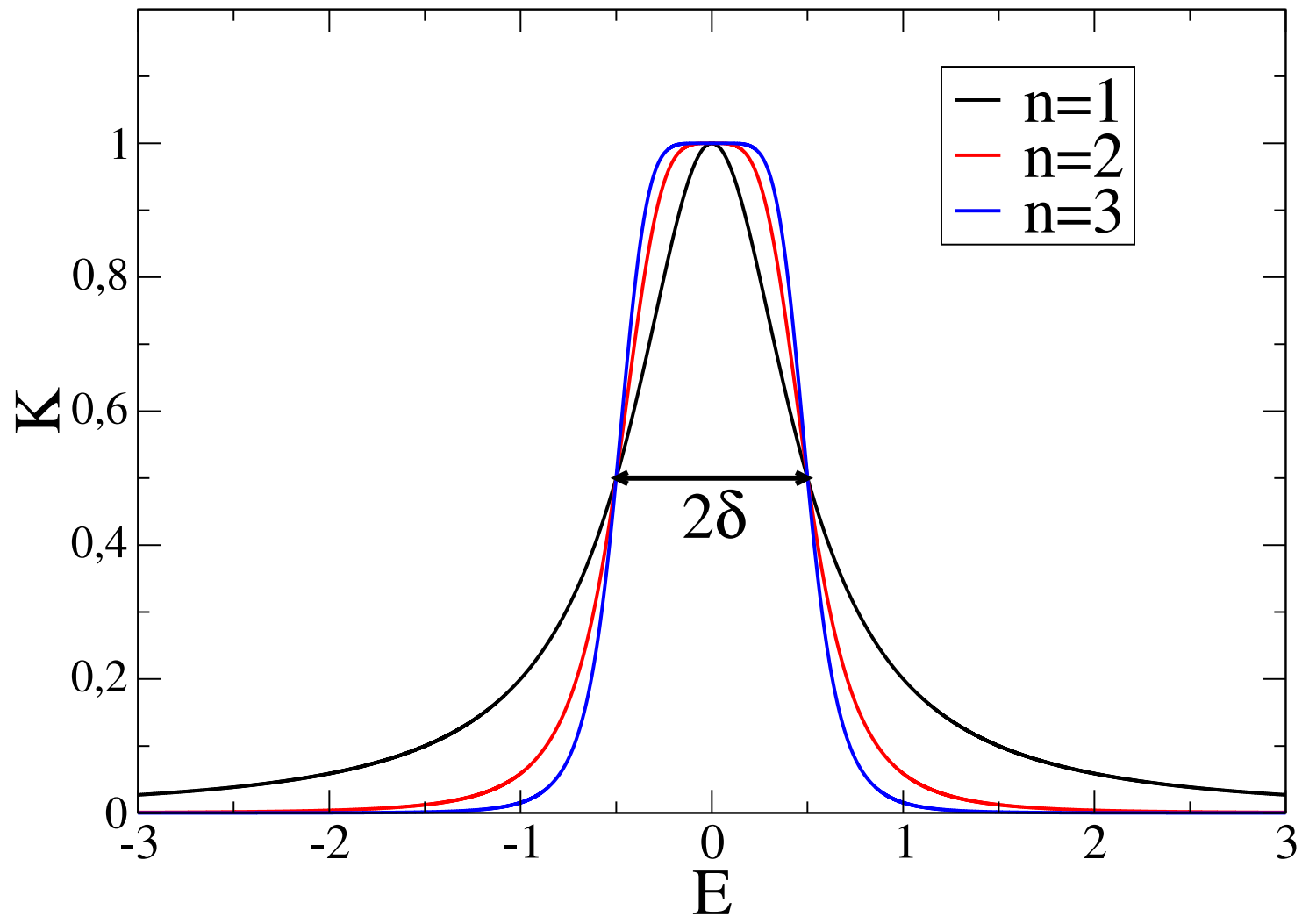

Figure 3.3.1.: Plot of $K(E, 0.5, n)$ for different values of $n$.

and $q_j = i^{1/n} z_j$, where $z_j = e^{i2\pi\frac{j-1}{n}}$ is the $j^{th}$ complex root of the unity. This is an important simplification in the computational implementation of the method, since it allows to treat it as an inversion problem. We can define $|\Psi'\rangle \equiv \hat{R}^n_\delta(E)|\Psi\rangle$, which is the result of applying the resolvent operator to a general state. The modulus square of $|\Psi'\rangle$ is proportional to the probability density. We will now prove the previous statement and find the proportionality constant:

$$\begin{aligned}
\langle\Psi'|\Psi'\rangle &= \langle\Psi|\hat{R}^n_\delta(E)^\dagger \hat{R}^n_\delta(E)|\Psi\rangle = \left\langle \Psi \left| \frac{\delta^{2n}}{\left(E-\hat{H}_0\right)^{2n}+\delta^{2n}} \right| \Psi \right\rangle && (3.3.12)\\
&= \sum_b |c_b|^2 K(E-E_b, \delta, n) + \int d\varepsilon\, |c(\varepsilon)|^2 K(E-\varepsilon, \delta, n) && (3.3.13)\\
&\equiv P(E, \delta, n), && (3.3.14)
\end{aligned}$$

where we have defined the quantity $K(E, \delta, n) = \frac{\delta^{2n}}{E^{2n}+\delta^{2n}}$. This function has always a maximum at $E = 0$, where $K(0, \delta, n) = 1$, see Fig. 3.3.1. At $E = \pm\delta$, $K(\pm\delta, \delta, n) = 0.5$. At $\pm\infty$ it goes to zero. Its most interesting feature is that this function is very similar to the boxcar function $\Pi_{a,b}(E)$, a function that is 1 in the interval $[a, b]$ and zero elsewhere, being $a = -\delta$ and $b = \delta$. In fact, $\lim_{n\to\infty} K(E, \delta, n) = \Pi_{-\delta,\delta}(E)$.

Taking into account the properties of $K(E, \delta, n)$ we can conclude that the quantity $\langle\Psi'|\Psi'\rangle$ is dominated by the states that are close to the energy that appears in the resolvent operator. Only energies in the interval $[E-\delta, E+\delta]$ have a signifi-cant weight in the sum. For any energy close to a bound energy $\langle \Psi' | \Psi' \rangle \propto |c_b|^2$, and for any energy in the continuum, $\langle \Psi' | \Psi' \rangle \propto |c(E)|^2 = \rho(E)$. The problem now relies on the obtention of the constant that relates this quantity to the real probability density [27]. We can suppose without loss of generality that the continuum states are all states with $E > 0$. If we take a positive energy in the resolvent operator, the contribution of the sum over bound states will be negligible. In this way Eq. (3.3.13) can be approximated by

$$P(E, \delta, n) \approx \int_{\varepsilon > 0} d\varepsilon \, |c(\varepsilon)|^2 K(E - \varepsilon, \delta, n) \tag{3.3.15}$$

$$\approx |c(E)|^2 \int_{\varepsilon > 0} d\varepsilon K(E - \varepsilon, \delta, n) \tag{3.3.16}$$

$$\approx |c(E)|^2 \int_{-\infty}^{\infty} d\varepsilon K(E - \varepsilon, \delta, n) \tag{3.3.17}$$

The first approximation neglects the sum over the bound states for a positive energy. The second approximation is reasonable in the sense that if $\delta$ is small enough, $|c(E)|^2$ has the same value in the interval $[E - \delta, E + \delta]$ and it can be extracted from the integral. The last approximation is based on the fact that the integral in a region far from $E$ is negligible. The proportionality constant that relates $P(E, \delta, n)$ to the probability $|c(E)|^2$ is thus

$$\int_{-\infty}^{\infty} d\varepsilon K(E - \varepsilon, \delta, n) = \frac{\pi}{n} \delta \csc\left(\frac{\pi}{2n}\right) \tag{3.3.18}$$

The limit of this when $n \to \infty$ is $\lim_{n \to \infty} \int_{-\infty}^{\infty} d\varepsilon K(E - \varepsilon, \delta, n) = \int_{-\infty}^{\infty} \Pi_{-\delta,\delta}(E - \varepsilon) d\varepsilon = 2\delta$, as expected. Therefore,

$$\rho(E) = \frac{P(E, \delta, n)}{\frac{\pi}{n} \delta \csc\left(\frac{\pi}{2n}\right)} \tag{3.3.19}$$

A similar treatment can be applied to bound states. In this case only the integration $\int_{E_b - \delta}^{E_b + \delta} d\varepsilon \rho(\varepsilon) \approx |c_b|^2$ is meaningful, since the integral provides an approximation to the probability of that bound state.

In the calculations we will not have a real continuum, since all calculations are performed in a finite box and all states are discretized. In this way, the choice of $\delta$ must obey to some considerations. If it is too small, it will show the peaks that correspond to the discretization of the box. If it is too large, we will lose resolution in the spectrum and features like ATI peaks will be lost. So a criterion can be that the interval $\delta$ of energy must contain at least one discretized state. This means that the density of states, $\rho_{st}$, multiplied by $\delta$, should be approximately one ($\rho_{st}\delta \approx 1$).

This technique can be a very useful tool to study and calculate the photoelectron spectrum. On the other hand, one of the drawbacks of this method is that it is only applicable when we have only one fragment to analyze. We will show in this work that we can also study molecular systems (ionization and dissociation) by using several extensions in the method.

### 3.3.2 *Differential energy distribution of molecular fragments*

The Resolvent Operator Method described before for atoms can be extended to molecules [28, 31], within the Born-Oppenheimer approximation. The BO approximation is based on the fact that the electronic dynamics is much faster than the nuclear dynamics. In this approximation a stationary wavefunction of the molecular Hamiltonian can be written as

$$\Psi_i^\nu (\mathbf{r}, \mathbf{R}) = \Phi_i (\mathbf{r}; \mathbf{R}) \chi_i^\nu (\mathbf{R}), \tag{3.3.20}$$

where the electronic wavefunction depends parametrically on the nuclear coordinate,

$$\hat{H}_{el} \Phi_i (\mathbf{r}; \mathbf{R}) = E_i (\mathbf{R}) \Phi_i (\mathbf{r}; \mathbf{R}), \tag{3.3.21}$$

and the nuclear wavefunction satisfies

$$\left[\hat{T}_N + E_i (\mathbf{R})\right] \chi_i^\nu (\mathbf{R}) = W_i^\nu \chi_i^\nu (\mathbf{R}). \tag{3.3.22}$$

The latter two equations are respectively the electronic and nuclear Schrödinger equations. The molecular Hamiltonian is the sum over the nuclear kinetic energy operator and the electronic Hamiltonian, $\hat{H} = \hat{H}_{el} + \hat{T}_N$. Using (3.3.20)

$$\begin{aligned} \hat{H}\Psi_i^\nu (\mathbf{r}, \mathbf{R}) &= \left[\hat{H}_{el} + \hat{T}_N\right] \Psi_i^\nu (\mathbf{r}, \mathbf{R}) & (3.3.23) \\ &= \left[\hat{H}_{el}\Phi_i (\mathbf{r}; \mathbf{R}) \chi_i^\nu (\mathbf{R}) + \hat{T}_N \Phi_i (\mathbf{r}; \mathbf{R}) \chi_i^\nu (\mathbf{R})\right] & (3.3.24) \\ &= \left[E_i (\mathbf{R}) \Phi_i (\mathbf{r}; \mathbf{R}) \chi_i^\nu (\mathbf{R}) + \hat{T}_N \Phi_i (\mathbf{r}; \mathbf{R}) \chi_i^\nu (\mathbf{R})\right] & (3.3.25) \end{aligned}$$

The ansatz (3.3.20) is only a molecular eigenstate if $\hat{T}_N \Phi_i (\mathbf{r}; \mathbf{R}) \chi_i^\nu (\mathbf{R}) \approx \Phi (\mathbf{r}; \mathbf{R}) \hat{T}_N \chi_i^\nu (\mathbf{R})$. This means that the BO approximation is valid when the electronic wavefunction varies slowly with the nuclear coordinates.

We can define the resolvent operator in the BO approximation,

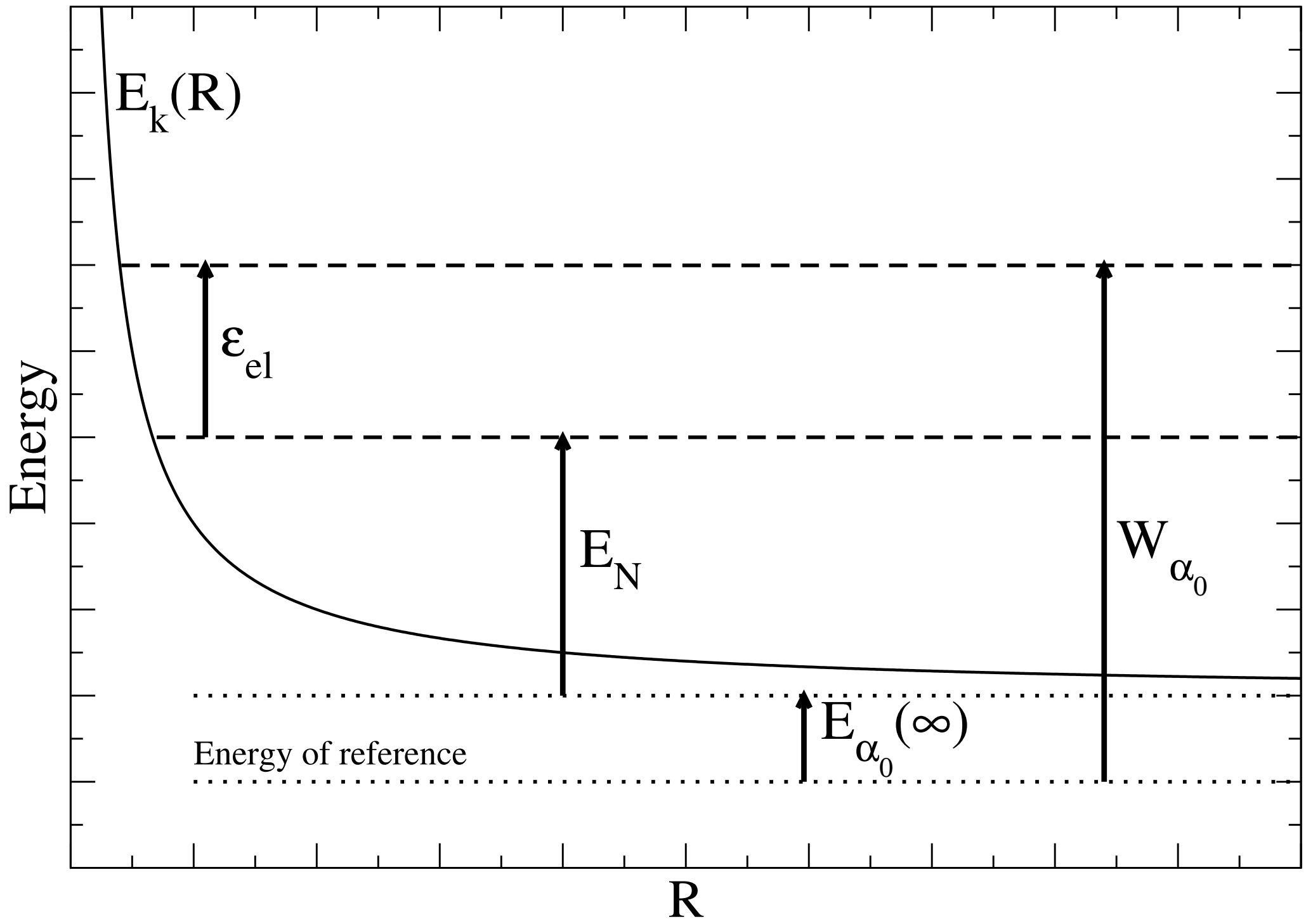

Figure 3.3.2.: Schematics of the different quantities used in the ROM analysis. $\varepsilon_{ele}$ is the electron energy in the continuum associated to the ionization potential energy curve $E_{\alpha_0}(R)$. $E_N$ is the nuclear energy in the asymptotic region. $W_{\alpha_0}$ is the vibronic energy referred to a given energy (in this example $E_{\alpha_0}(\infty)$), so that $W_{\alpha_0} = \varepsilon_{ele} + E_N + E_{\alpha_0}(\infty)$. In the case of Coulomb explosion (ionization) of $H_2^+$, $W_{\alpha_0} = \varepsilon_{ele} + E_N$.

$$\begin{aligned}\hat{R}\left(E_N, \varepsilon_{ele}, \alpha_0, \delta_N, \delta_e, n_N, n_e\right) &= \frac{\delta_N^{n_N}}{\left[\left(\hat{T}_N + E_{\alpha_0}\left(R\right) + \varepsilon_{ele}\right) - E_N - E_{\alpha_0}(\infty) - \varepsilon_{ele}\right]^{n_N} - i\delta_N^{n_N}} \\ &\times \frac{\delta_e^{n_e}}{\left[\hat{H}_{el} - E_{\alpha_0}\left(R\right) - \varepsilon_{ele}\right]^{n_e} - i\delta_e^{n_e}} \qquad (3.3.26)\end{aligned}$$

where $E_N$ and $\varepsilon_{ele}$ are the selected nuclear and electronic energies, respectively, $\alpha_0$ is the potential energy curve of a given electronic state of $H_2^+$ or $H_2^{2+}$, $\delta_N$ and $\delta_e$ are the resolution , and $n_N$ and $n_e$ are the orders of the resolvent operator for nuclear and electronic energies, respectively. The resolvent operator is then the product of a nuclear resolvent ($\hat{R}_N$) and an electronic resolvent ($\hat{R}_{ele}$). The meanings of the different energies used in these equations are depicted in the schematic given in Fig. 3.3.2.

The idea is the following: first, an electronic curve $\alpha_0$ is selected with a given electronic energy, and then the proper vibrational energy, $E_N$, is chosen. As shown below, when applied to the wavefunction, the resolvent operator will provide the necessary information to evaluate the probability of producing an electron of kinetic energy $\varepsilon_{ele}$ and protons with center-of-mass energy $E_N$. If $\alpha_0$ is an electronic bound curve, one must take $\varepsilon_{ele} = 0$, and if it is an electronic continuum state (like the $1/R$ Coulomb explosion curve) one must take $\varepsilon_{ele} \geq 0$. For $\varepsilon_{ele} \geq 0$ all electronic curves have identical shape since they are obtained by summing to $1/R$ the energy of the electron that is in the continuum.

The operator (3.3.26) selects first one particular electronic state, and then it resolves the vibronic energy. In the expansion of a general wavefunction as a sum of Born-Oppenheimer states:

$$\begin{aligned} |\Psi\rangle &= \sum_{\alpha=bound} |\,\Phi_\alpha(R)\rangle \left\{ \sum_\nu c_\alpha^\nu \,|\, \chi_\alpha^\nu\rangle + \int_{W_\alpha > E_\alpha(\infty)} dW_\alpha c_\alpha(W_\alpha) \,|\, \chi_\alpha^{W_\alpha}\rangle \right\} \\ &+ \sum_{\alpha=continuum} \int_{\varepsilon>0} d\varepsilon \,|\, \Phi_{\alpha,\varepsilon}(R)\rangle \\ &\times \left\{ \sum_\nu c_\alpha^\nu(\varepsilon) \,|\, \chi_{\alpha,\varepsilon}^\nu\rangle + \int_{W_{\alpha,\varepsilon} > E_{\alpha,\varepsilon}(\infty)} dW_{\alpha,\varepsilon} c_\alpha(\varepsilon, W_{\alpha,\varepsilon}) \,|\, \chi_{\alpha,\varepsilon}^{W_{\alpha,\varepsilon}}\rangle \right\} \qquad (3.3.27) \\ &\equiv |b,b\rangle + |b,c\rangle + |c,b\rangle + |c,c\rangle \qquad (3.3.28) \end{aligned}$$

Four different terms are present: states that are electronically unbound and vibrationally bound, which will be denoted by $|c,b\rangle$, and states that are both electronically and vibrationally bound, $|b,b\rangle$ and $|c,c\rangle$ and $|b,c\rangle$ states, defined analogously. Note that, for example, non-dissociative ionization is included in $|c,b\rangle$ states.

In $H_2^+$ there are no bound electronic curves in the continuum, only the dissociative Coulomb explosion curve. For this reason, here we will ignore the $|c,b\rangle$ states.

From now on we will apply the BO approximation. As in the previous section, the physically interesting quantity is $\langle\Psi|\hat{R}^{\dagger}\hat{R}|\Psi\rangle$. We will calculate $\hat{R}\,|\Psi\rangle$ by parts. We will start with $\hat{R}\,|b,b\rangle$, where $\varepsilon_{ele}=0$:

$$\begin{aligned}\hat{R}\,|b,b\rangle &= \sum_{\nu}\frac{\delta_N^{n_N}}{\left[\left(\hat{T}_N+E_{\alpha_0}(R)\right)-E_N-E_{\alpha_0}(\infty)\right]^{n_N}-i\delta_N^{n_N}} \\ &\times \frac{\delta_e^{n_e}c_{\alpha_0}^{\nu}\,|\Phi_{\alpha_0}(R)\rangle\,|\chi_{\alpha_0}^{\nu}(R)\rangle}{\left[E_{\alpha_0}(R)-E_{\alpha_0}(R)\right]^{n_e}-i\delta_e^{n_e}} \\ &+ \sum_{\alpha\neq\alpha_0,\nu}\hat{R}_N\frac{\delta_e^{n_e}c_{\alpha_0}^{\nu}\,|\Phi_{\alpha_0}(R)\rangle\,|\chi_{\alpha}^{\nu}(R)\rangle}{\left[E_{\alpha}(R)-E_{\alpha_0}(R)\right]^{n_e}-i\delta_e^{n_e}}\end{aligned} \tag{3.3.29}$$

One reasonable simplification is to neglect the second term, since for $\alpha\neq\alpha_0$ and if $\delta_e\ll E_\alpha(R)-E_{\alpha_0}(R)$ we know from the previous section that we can neglect that contribution to the resolvent operator. We obtain

$$\hat{R}\,|b,b\rangle=\sum_{\nu}\frac{\delta_N^{n_N}}{\left[\left(W_{\alpha_0}^{\nu}\right)-E_N-E_{\alpha_0}(\infty)\right]^{n_N}-i\delta_N^{n_N}}\,\frac{c_{\alpha_0}^{\nu}\,|\Phi_{\alpha_0}(R)\rangle\,|\chi_{\alpha_0}^{\nu}(R)\rangle}{-i} \tag{3.3.30}$$

where $W_{\alpha_0}^{\nu}$ is the vibronic (nuclear plus electronic) energy. Proceeding in the same way for $\hat{R}\,|b,c\rangle$ we obtain

$$\begin{aligned}\hat{R}\,|b,c\rangle &= \int_{W_{\alpha_0}>E_{\alpha_0}(\infty)}dW_{\alpha_0}\frac{\delta_N^{n_N}}{\left[\left(W_{\alpha_0}\right)-E_N-E_{\alpha_0}(\infty)\right]^{n_N}-i\delta_N^{n_N}} \\ &\times \frac{c_{\alpha_0}(W_{\alpha_0})\,|\Phi_{\alpha_0}(R)\rangle\,|\chi_{\alpha_0}^{\nu}(R)\rangle}{-i}\end{aligned} \tag{3.3.31}$$

For calculating $\hat{R}\,|c,c\rangle$, the sum over electronic curves that are not included in the resolvent operator can be neglected. For that reason, we will drop the index for the electronic curve:

$$\begin{aligned}
\hat{R}\left|c,c\right\rangle &= \int d\varepsilon \int_{W_\varepsilon > \varepsilon + E(\infty)} dW_\varepsilon \hat{R}_N \hat{R}_{ele} c\left(\varepsilon, W_\varepsilon\right) \left|\Phi_\varepsilon\left(R\right)\right\rangle \left|\chi_\varepsilon^{W_\varepsilon}\right\rangle \\
&= \int d\varepsilon \int_{W_\varepsilon > \varepsilon + E(\infty)} dW_\varepsilon \frac{\delta_N^{n_N}}{\left[\left(\hat{T}_N + E\left(R\right) + \varepsilon_{ele}\right) - E_N - E\left(\infty\right) - \varepsilon_{ele}\right]^{n_N} - i\delta_N^{n_N}} \\
&\times \frac{\delta_e^{n_e}}{\left[\hat{H}_{el} - E\left(R\right) - \varepsilon_{ele}\right]^{n_e} - i\delta_e^{n_e}} c\left(\varepsilon, W_\varepsilon\right) \left|\Phi_\varepsilon\left(R\right)\right\rangle \left|\chi_\varepsilon^{W_\varepsilon}\right\rangle \\
&= \int d\varepsilon \int_{W_\varepsilon > \varepsilon + E(\infty)} dW_\varepsilon \frac{\delta_N^{n_N}}{\left[W_\varepsilon - E_N - E\left(\infty\right) - \varepsilon_{ele}\right]^{n_N} - i\delta_N^{n_N}} \\
&\times \frac{\delta_e^{n_e}}{\left[\varepsilon - \varepsilon_{ele}\right]^{n_e} - i\delta_e^{n_e}} c\left(\varepsilon, W_\varepsilon\right) \left|\Phi_\varepsilon\left(R\right)\right\rangle \left|\chi_\varepsilon^{W_\varepsilon}\right\rangle
\end{aligned} \tag{3.3.32}$$

Note that the vibronic energy is the sum of the nuclear energy and the electronic energy,

$$W_{\alpha_0,\varepsilon_{ele}} = \varepsilon_{ele} + E_N + E_{\alpha_0}\left(\infty\right). \tag{3.3.33}$$

The next step is the calculation of $\langle\Psi|\hat{R}^\dagger\hat{R}|\Psi\rangle$. Since all Born-Oppenheimer states are orthogonal to each other we will have that

$$\langle\Psi|\hat{R}^\dagger\hat{R}|\Psi\rangle = \langle b,b|\hat{R}^\dagger\hat{R}|b,b\rangle + \langle b,c|\hat{R}^\dagger\hat{R}|b,c\rangle + \langle c,c|\hat{R}^\dagger\hat{R}|c,c\rangle \tag{3.3.34}$$

The first two terms in the sum are associated with non-ionizing channels and can be treated exactly as in the Resolvent Technique for atoms. Thus the differential probability in the nuclear energy for a bound electronic curve, $\frac{dP^{\alpha_0}}{dE_N}$, is only affected by a constant,

$$\frac{dP^{\alpha_0}}{dE_N} = \langle\Psi|\hat{R}^\dagger\hat{R}|\Psi\rangle \frac{1}{\frac{\pi}{n_N}\delta_N \csc\left(\frac{\pi}{2n_N}\right)}. \tag{3.3.35}$$

In this way, we will be able to calculate the differential probability of the nuclear energy for each electronic curve and this quantity is the NKE and is the useful observable for the non-ionizing dissociative channels.

Now if we focus on the dissociative ionization channel, in the third term, Eq. (3.3.34), we must find the relationship between the doubly differential probability (in electron and nuclear kinetic energies) and $\langle c,c|\hat{R}^\dagger\hat{R}|c,c\rangle$,

$$\begin{aligned}\langle c,c|\hat{R}^{\dagger}\hat{R}|c,c\rangle &= \int d\varepsilon \int_{W_\varepsilon>\varepsilon+E(\infty)} dW_\varepsilon K\left(W_\varepsilon - W_{\varepsilon_{ele}},\delta_N,n_N\right) &(3.3.36)\\ &\times\ K\left(\varepsilon-\varepsilon_{ele},\delta_e,n_e\right)\left|c\left(\varepsilon,W_\varepsilon\right)\right|^2\\ &\approx \left|c\left(\varepsilon_{ele},W_{\varepsilon_{ele}}\right)\right|^2 \int d\varepsilon \int_{W_\varepsilon>\varepsilon+E(\infty)} dW_\varepsilon &(3.3.37)\\ &\times\ K\left(W_\varepsilon - W_{\varepsilon_{ele}},\delta_N,n_N\right) K\left(\varepsilon-\varepsilon_{ele},\delta_e,n_e\right)\\ &\approx \left|c\left(\varepsilon_{ele},W_{\varepsilon_{ele}}\right)\right|^2 \int_{-\infty}^{+\infty} d\varepsilon \int_{-\infty}^{+\infty} dW_\varepsilon &(3.3.38)\\ &\times\ K\left(W_\varepsilon - W_{\varepsilon_{ele}},\delta_N,n_N\right) K\left(\varepsilon-\varepsilon_{ele},\delta_e,n_e\right)\\ &= \left|c\left(\varepsilon_{ele},W_{\varepsilon_{ele}}\right)\right|^2 \frac{1}{\frac{\pi}{n_N}\delta_N \csc\left(\frac{\pi}{2n_N}\right)} \frac{1}{\frac{\pi}{n_e}\delta_e \csc\left(\frac{\pi}{2n_e}\right)}. &(3.3.39)\end{aligned}$$

Here $\left|c\left(\varepsilon_{ele},W_{\varepsilon_{ele}}\right)\right|^2$ is the doubly differential probability, in the electronic and vibronic energy, for a particular electronic curve. Using Eq. (3.3.33), we conclude that

$$\frac{d^2P^{\alpha_0}}{dW_{\varepsilon_{ele}}d\varepsilon_{ele}} = \frac{d^2P^{\alpha_0}}{dE_N d\varepsilon_{ele}}, \tag{3.3.40}$$

therefore

$$\frac{d^2P^{\alpha_0}}{dE_N d\varepsilon_{ele}} = \langle\Psi|\hat{R}^{\dagger}\hat{R}|\Psi\rangle \frac{1}{\frac{\pi}{n_N}\delta_N \csc\left(\frac{\pi}{2n_N}\right)} \frac{1}{\frac{\pi}{n_e}\delta_e \csc\left(\frac{\pi}{2n_e}\right)}. \tag{3.3.41}$$

The previous formula will allow us to calculate the probability differential in the nuclear energy and the electronic energy, by selecting a particular electronic curve. In conclusion, we have derived a formal way to obtain the doubly differential probability. This quantity is called the correlated energy spectrum (CKE). By integrating this quantity over the electronic (nuclear) energy we obtain the EKE (NKE) spectrum.

#### 3.3.2.1 *Angular distributions*

After application of $\hat{R}$ to $|\Psi\rangle$ we obtain a molecular state in the $\alpha_0^{th}$ electronic state with an electronic energy ($\varepsilon_{ele}$) and nuclear energy ($E_N$). To obtain an angular spectra [26, 27, 29] we can apply to $\hat{R}\,|\Psi\rangle$, a projection operator that selects only a small interval of electron emission angles around the angle $\theta$. We define $\hat{P}_{[\theta-\Delta\theta/2,\theta+\Delta\theta/2]}$ as

$$\hat{P}_{[\theta-\Delta\theta/2,\theta+\Delta\theta/2]}\Psi\left(z,\rho,R\right)=\begin{cases}\Psi\left(z,\rho,R\right) & ,\theta-\Delta\theta/2\leq\arctan\left(\rho/z\right)\leq\theta-\Delta\theta/2\\ 0 & otherwise\end{cases}. \tag{3.3.42}$$

We can then obtain a full differential spectra both in energies of the fragments as well as in the emission angles of the electron:

$$\frac{d^3P}{dE_N d\varepsilon_e d\theta}=\lim_{\Delta\theta\to0}\frac{\langle\Psi|\hat{R}^{\dagger}\hat{P}_{[\theta-\Delta\theta/2,\theta+\Delta\theta/2]}\hat{R}|\Psi\rangle}{\Delta\theta\frac{\delta_e}{n_e}\pi\csc\left(\frac{\pi}{2n_e}\right)\frac{\delta_N}{n_N}\pi\csc\left(\frac{\pi}{2n_N}\right)}. \tag{3.3.43}$$

Having this observable, we can integrate over one variable to obtain the CKE, the correlated angular and nuclear kinetic energy, $\mathrm{CAK_N}$, and the correlated angular and nuclear kinetic energy, $\mathrm{CAK_e}$, spectra:

$$\frac{d^2P}{dE_N d\varepsilon_e} = \int d\theta\frac{d^3P}{dE_N d\varepsilon_e d\theta}, \tag{3.3.44}$$

$$\frac{d^2P}{dE_N d\theta} = \int d\varepsilon_e\frac{d^3P}{dE_N d\varepsilon_e d\theta}, \tag{3.3.45}$$

$$\frac{d^2P}{d\varepsilon_e d\theta} = \int dE_N\frac{d^3P}{dE_N d\varepsilon_e d\theta}, \tag{3.3.46}$$

being the CKE spectra calculated by Eq. (3.3.44), the $\mathrm{CAK_N}$ spectra by Eq. (3.3.45) and the $\mathrm{CAK_e}$ spectra by (3.3.46).

# 4

# STRONG FIELD IONIZATION

When a molecule or an atom is irradiated by a strong laser field it may ionize. In this Chapter we will introduce the qualitative aspects of the Keldysh theory [30,56,57] that allows one to qualitatively understand the physics of strong field ionization. Photoionization is a well known phenomenon that has being studied since the $19^{th}$ century and it was crucial for the understanding and acceptance of the quantum theory with the explanation of the photoelectric effect by Einstein.

When the energy required to ionize the system is larger than the energy of the photon, $I_p > \hbar\omega$, the ionization is a nonlinear process where more than one photon is absorbed. Multiphoton transitions can be studied within perturbation theory and are not restricted to the ionization process since we can also induce transitions from a bound state to another bound state by absorbing several photons. When perturbation theory fails to converge the multiphoton picture is replaced by a static picture of ionization by a laser field, i.e., tunneling ionization.

In the following, we are going to review the different regimes of strong field ionization.

## 4.1 TUNNELING IONIZATION

When a laser field, of frequency $\omega_0$ and field amplitude $E_0$, that is linearly polarized, is applied to an atom, it changes the potential that is felt by the electron. The electron is initially bound by the Coulomb potential but as the laser field is turned on, the electron can escape through the potential barrier to the continuum, see Fig. 4.1.1.

We can estimate the tunneling time as the ratio between the width of the potential barrier and the velocity of the electron at the barrier. Neglecting the Coulomb potential and assuming that the electron is initially located at $z = 0$, the width of the barrier is just $I_p/|q|E_0$, where $I_p$ is the ionization potential, $q$ is the electric charge and $E_0$ is the field amplitude. The electron velocity under the barrier is more complicated to evaluate. Under the barrier, the electron has a negative value of the kinetic energy and this implies an imaginary velocity. Let just take the absolute value of this imaginary velocity, $\sqrt{2I_p/m}$, as the velocity of the electron under the barrier. In this way, we can evaluate the tunneling time, $\tau_T$ as

$$\tau_T = \frac{I_p/|q|E_0}{\sqrt{2I_p/m}}. \tag{4.1.1}$$

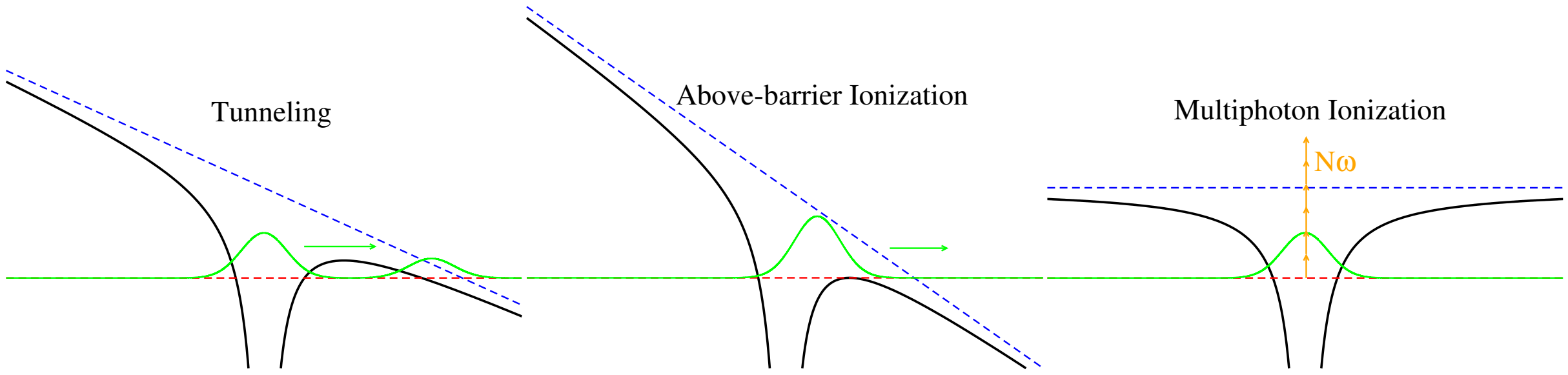

Figure 4.1.1.: Scheme for tunneling ionization, above barrier ionization and multiphoton ionization.

We can define a new parameter called the *Keldysh parameter*,

$$\gamma = \sqrt{\frac{I_p}{2U_P}} \tag{4.1.2}$$

where $U_p = q^2E_0^2/4m\omega_0^2$ is the ponderomotive energy [58,59]. Using Eqs. (4.1.1) and (4.1.2), the Keldysh parameter can be written as a ratio between the tunneling time, $\tau_T$, and the laser period, $T$:

$$\gamma = \frac{4\pi\tau_T}{T}. \tag{4.1.3}$$

We must expect an efficient tunneling ionization when the tunneling time is much smaller than the period of the laser field. So when $\gamma \ll 1$, we expect that tunneling ionization will occur. Under these circumstances, the laser field is sufficiently slow that can be approximated as a static field.

If the laser field amplitude is high enough, the barrier potential is below the bound electron energy. In this case we have above-barrier ionization, see Fig. 4.1.1.

## 4.2 MULTIPHOTON IONIZATION

If we are in the regime where $\gamma \gg 1$ and $I_p > \hbar\omega_0$, which happens for relatively weak fields and short wavelengths, ionization can only occur by absorbing several photons from the laser field. This process is known as multiphoton ionization, see Fig. 4.1.1. In this process, the electron absorbs a minimum number of photons, $N$, to reach the continuum. The final electron kinetic energy will be equal to $N\hbar\omega_0 - I_p$. For very weak fields, perturbation theory can be applied and the ionization rate, $P$, is going to depend on the intensity of the field, $I$, to the power of $N$, so $P \propto I^N$, being $N$ the minimum number of photons required to ionize the atom. In the case that $N = 1$, we have the single-photon ionization process.

If the intensity of the field is high enough, the perturbative picture is no longer valid and multiphoton ionization can give rise to several peaks in the photoelectron spectrum that are separated by the photon energy. This phenomena is known as above threshold ionization (ATI).

The physics of strong field ionization is then determined by three typical energies: the ponderomotive energy, $U_p$; the ionization potential, $I_p$; and the photon energy, $\hbar\omega_0$. Let's enumerate the main cases:

- $\gamma \gg 1 \vee U_p < \hbar\omega_o$: we are in the perturbutive regime. In this case we can apply lowest order perturbation theory (LOPT).

- $\gamma > 1 \vee U_p > \hbar\omega_o$: we are in the multiphoton ionization regime. We can no longer apply perturbation theory since we have a high ponderomotive energy.

- $\gamma \sim 1 \vee U_p > \hbar\omega_o$, we have the tunneling ionization regime.

To illustrate, let's take typical examples for the $H_2^+$ molecule, $I_p = 1.1$:

| | $I = 10^{13}\mathrm{W/cm}^2$ | $I = 3 \times 10^{14}\mathrm{W/cm}^2$ |
|---|---|---|
| $\hbar\omega_0 = 0.0569$ | Multiphoton ionization | Tunneling ionization |
| $\hbar\omega_0 = 1$ | Perturbative regime | Perturbative regime |

As we can see, for long wavelengths of the pulse with moderate/strong intensities we can reach the quasi-static ionization regime. For very short wavelengths, even with strong laser pulses we can still apply perturbation theory.

# 5

# HIGH HARMONIC GENERATION

Until now, we were analysing the strong-field phenomena by looking at the resulting fragments (electrons and ions). But it is widely known that we can extract structural information of atoms and molecules from the light that is either emitted or absorbed by matter. Until the mid-1900s this was achieved only at the linear regime, e.g., the response of the system was in the same frequency as the frequency of our light source. This was due to the low intensity of the light sources available at that time. This is the field of "usual" spectroscopy [60].

With the invention of the laser, new light sources were available. One of the first findings with lasers was the discovery of the second harmonic generation [4]. This was the first time that a non-linear response to light was observed. This non-linear response can be explained by using a perturbative expansion on the polarization response to the field. If we choose our medium to be an extreme non-linear medium and then irradiates it with a very strong infrared laser pulse, we can also observe the generation of very high harmonics of the infrared frequency. This was brilliantly explained by Corkum with the so-called three step model [7]. Lewenstein et al. developed an analytical quantum model for the HHG, called *Lewenstein model* [9]. The discovery of the High Harmonic generation was the birth of Attophysics.

In this Chapter, we will start by discussing the three step model. After that, we will derive the formula for the HHG spectra in the single atom response regime considering the EMF as a classical or as a quantum field. At the end, we will provide the numerical way to calculate the HHG spectra and we will introduce an important analysis tool in the HHG study, the Gabor profile, that allow us to see the harmonic generation process in the time domain.

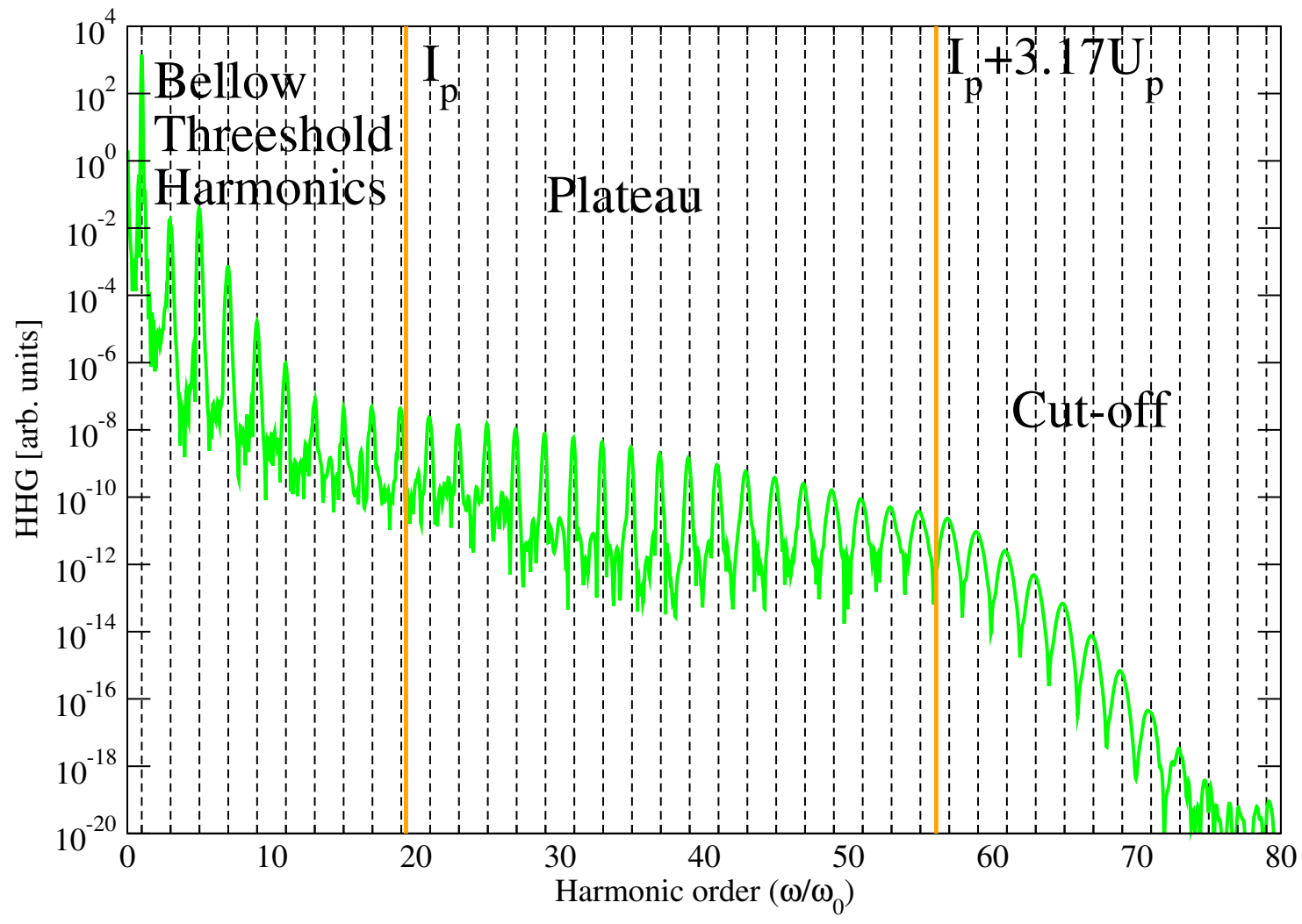

Figure 5.1.1.: HHG spectrum from a $D_2^+$ molecule for a laser pulse with $\lambda = 800\,\mathrm{nm}$, $I = 3 \times 10^{14}\,\mathrm{W/cm^2}$ and with a total duration of 5 optical cycles.

## 5.1 THREE STEP MODEL

The typical HHG spectrum (see Fig. 5.1.1) shows three different regions of the spectrum. The below threshold harmonics that are perturbative harmonics, the plateau harmonics that are intermediate harmonics that came from recombination of electrons that were ionized by tunneling and the cut-off harmonics that are located a $3.17U_p + I_p$. The three step model is a model that helps us to understand with a semiclassical picture the process of HHG and predicts very well the cut-off energy.

The three step model or the Simpleman's model [7] is a semiclassical model that explains the HHG process and can be very useful in the understanding of the physics behind it.

According to this model, see Fig. 5.1.2, in the first step, an electron is ionized to the continuum in a quasi static laser field (tunneling ionization) at the ion position with zero velocity. In the second step, the electron is accelerated by the laser field within the assumption that the core potential does not influence the motion of the electron in the continuum (propagation/acceleration in the continuum). In the final step, since the laser field changes direction, the electron comes back to the parent ion and can recombine with the parent ion and return to the groundstate (recombination). Upon recombination, a photon is liberated with an energy that is equal to the sum of the kinetic energy at the moment of recombination and the ionization potential.

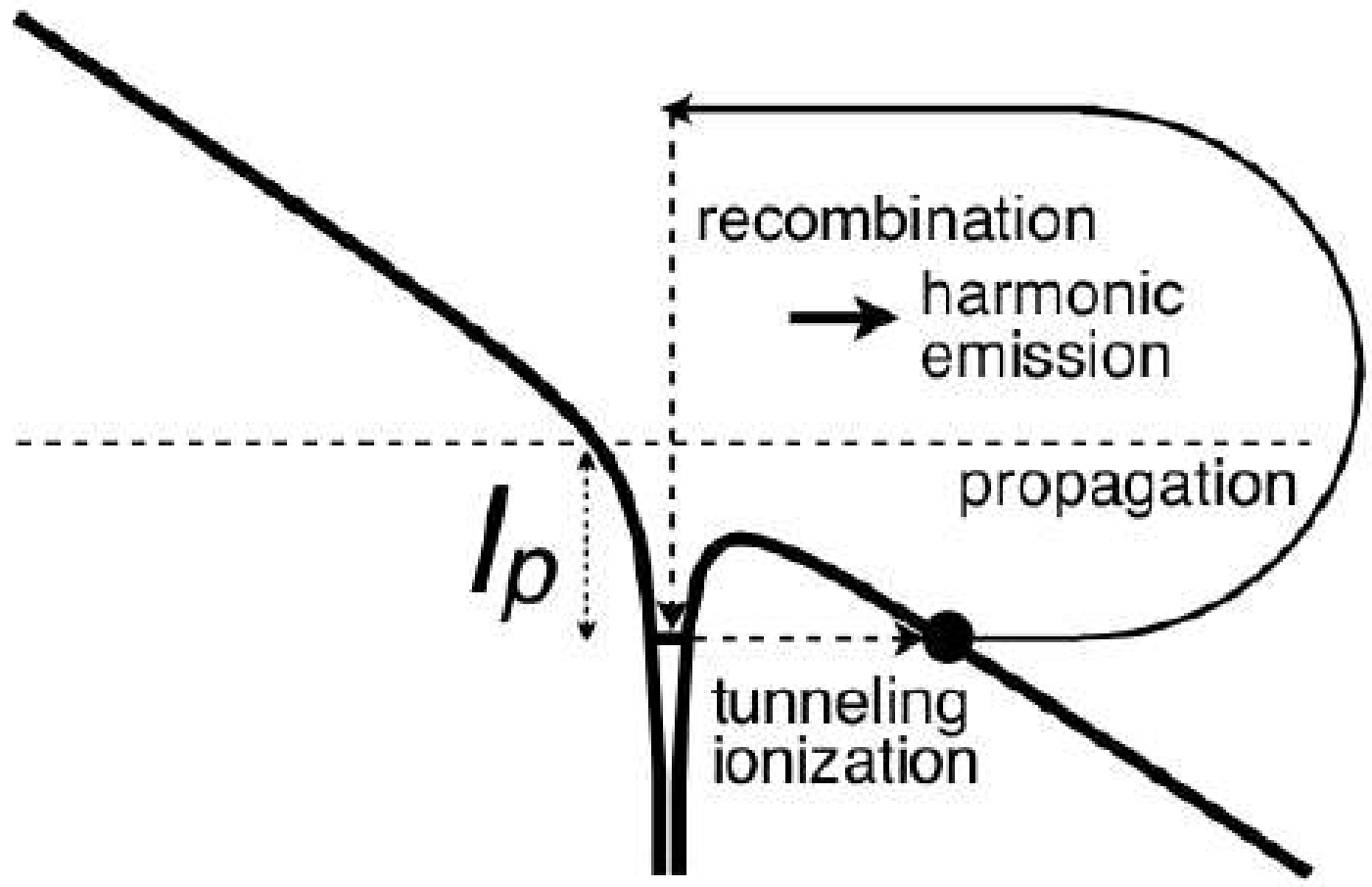

Figure 5.1.2.: Schematics of the three step model. Figure reproduced from [1].

This model is a semiclassical model because, although the ionization and recombination processes requires quantum mechanics, the propagation of the electron in the continuum is treated classically.

Let us consider the motion of a free electron in a time-dependent electric field. We consider that the electric field, $E(t)$, is linearly polarized along the $z$ direction and is given by

$$E(t) = E_0 \cos(\omega_0 t), \tag{5.1.1}$$

where $E_0$ denotes the field amplitude and $\omega_0$ the frequency. The electron is supposed to be ejected at $t = t_i$ with zero velocity. In this way, the initial conditions for the equation of motion are

$$z(t_i) = 0, \tag{5.1.2}$$

$$\dot{z}(t_i) = 0, \tag{5.1.3}$$

and integrating the equation of motion we get

$$z(t) = \frac{E_0}{\omega_0^2}\left[(\cos(\omega_0 t) - \cos(\omega_0 t_i)) + (\omega_0 t - \omega_0 t_i)\sin(\omega_0 t_i)\right]. \tag{5.1.4}$$

If we introduce the new phase variable $\theta \equiv \omega_0 t$, we can rewrite Eq. (5.1.4) as

$$z(\theta) = \frac{E_0}{\omega_0^2}\left[(\cos(\theta) - \cos(\theta_i)) + (\theta - \theta_i)\sin(\theta_i)\right]. \tag{5.1.5}$$

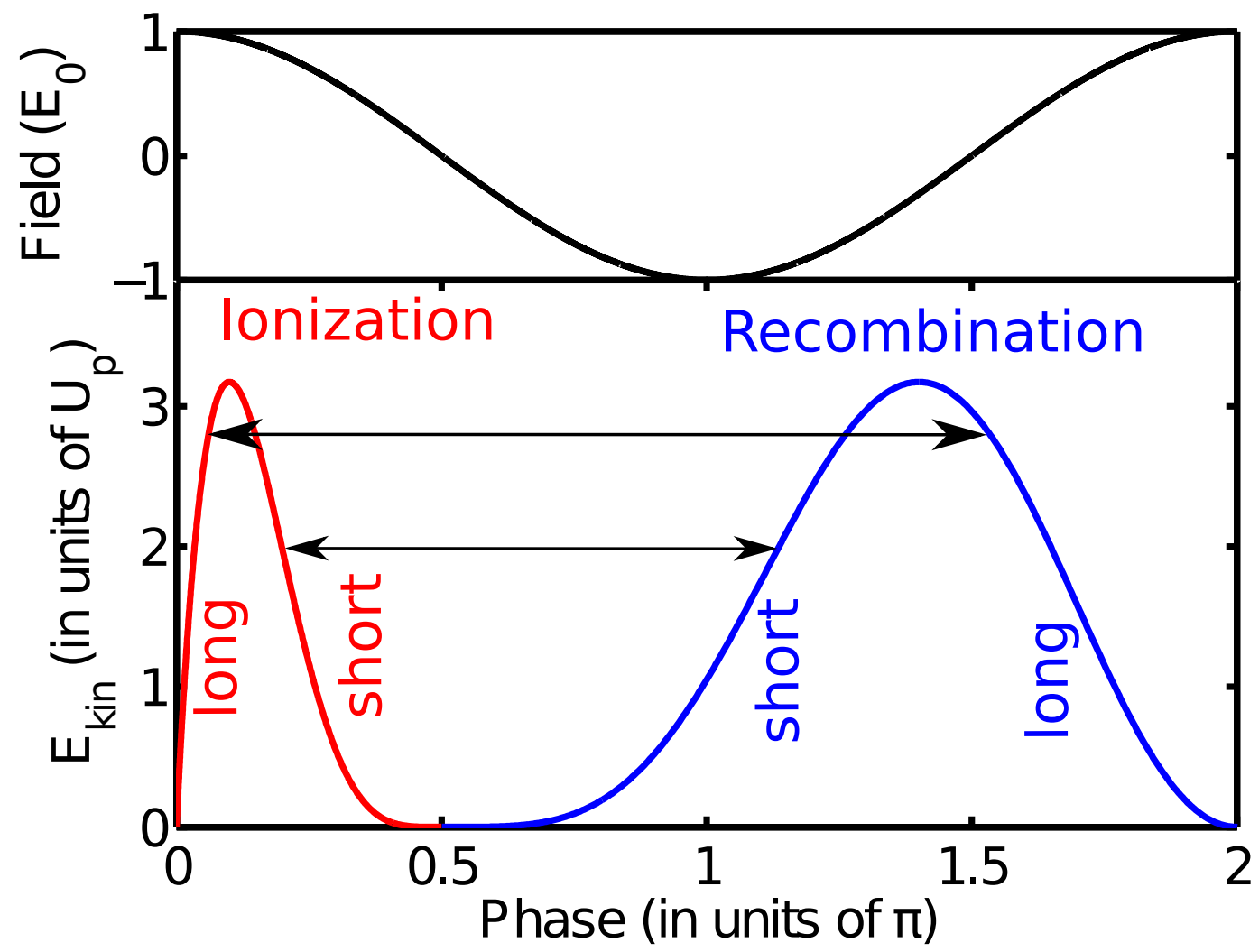

Figure 5.1.3.: Kinetic energy of the electron at recombination in function of ionization and recombination phase, $\theta_i$ and $\theta_r$. Two pairs of solutions, corresponding to a short and long trajectory, are indicated by the horizontal arrows.

The kinetic energy can be written as

$$E_{kin}(\theta) = 2U_p (\sin\theta - \sin\theta_i)^2 . \tag{5.1.6}$$

We want to find the times of recombination, $\theta_r$, and for that we must find the roots of Eq. (5.1.4) for $z(\theta_r) = 0$. We notice that the electron can recombine only when $0 < \theta_i < \pi/2$, otherwise, $\pi/2 < \theta_i < \pi$ the electron flies away from the nucleus and will never recombine.

For a given value of $E_{kin}$, we can view $\theta_i$ and $\theta_r$ as the solutions of the following coupled equations:

$$\cos\theta_r - \cos\theta_i = (\theta_i - \theta_r)\sin\theta_i, \tag{5.1.7}$$

$$(\sin\theta_r - \sin\theta_i)^2 = E_{kin}/2U_p \tag{5.1.8}$$

The path $z(\theta)$ that the electron follows from $\theta = \theta_i$ to $\theta_r$ is called trajectory. We must notice that for a given kinetic energy, we have two pairs of solutions $(\theta_i, \theta_r)$ that contribute to the same energy in the harmonic spectrum. We will call *short trajectories* to those where $17° < \theta_i < 90°$ and *long trajectories* for those where $0° < \theta_i < 17°$ (see Fig. 5.1.4).

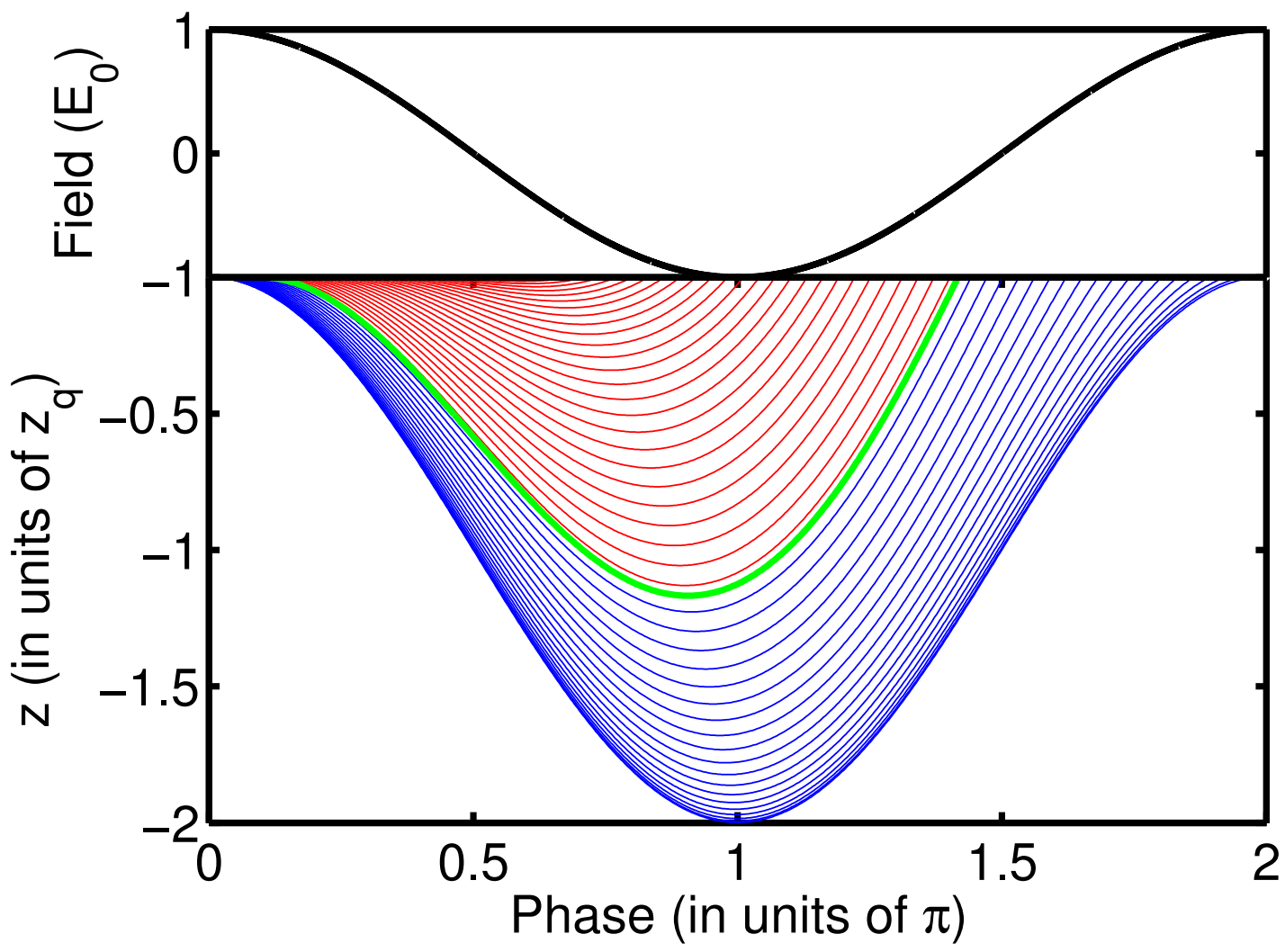

Figure 5.1.4.: Classical electron trajectories as a function of the phase of the field, for trajectories with ionization phase $0 < \theta_i < \pi/2$. In red the short trajectories, in blue the long trajectories and in green the trajectory corresponding to maximum energy gain ($3.17U_p$).

In Fig. 5.1.3 we show the $E_{kin}(\theta_r)$ as a function of the ionization phase, $\theta_i$, and the recombination phase, $\theta_r$. The maximum kinetic energy at recombination is $3.17U_p$. This occurs for $\theta_i = 17°$ and $\theta_r = 255°$. The energy of the photon emitted is then given by the sum of the kinetic energy at $\theta_r$ and the ionization potential, $I_p$. This explains why the cutoff energy is given at $I_p + 3.17U_p$.

If $(\theta_i, \theta_r)$ are solutions of Eqs. (5.1.7) and (5.1.8), $(\theta_i + m\pi, \theta_r + m\pi)$ are also solutions, being $m$ an integer. The trajectory associated to $(\theta_i + m\pi, \theta_r + m\pi)$, $z_m(\theta)$, is related to the trajectory of $(\theta_i, \theta_r)$, $z(\theta)$, by $z_m(\theta) = (-1)^m z(\theta - m\pi)$. We thus conclude that the harmonics emitted at each half cycle of the pulse have an alternating phase. This implies that the generated field obeys to

$$E(t) = -E(t + \pi/\omega_0). \tag{5.1.9}$$

We can show that the Fourier transform of such function takes nonzero values only at odd multiples of the fundamental frequency, $\omega_0$. In principle, this is true if we are in centrosymmetric medium.

## 5.2 HARMONIC SPECTRUM

### 5.2.1 *Classical radiation emitted by an oscillating electric dipole*[1]

The electric and magnetic radiation field produced by an electric dipole (confined in a region around the center of reference) at a position $\boldsymbol{r}$, that is very far away from the sources, assuming that the motion of the particles is non-relativistic and disregarding terms of the fields that go with $1/r^2$, can be written as [61]

$$\boldsymbol{E}(\boldsymbol{r},t) = \frac{\mu_0}{4\pi r}\left[\hat{\boldsymbol{r}}\times\left(\hat{\boldsymbol{r}}\times\ddot{\boldsymbol{d}}(t_0)\right)\right] \tag{5.2.1}$$

$$\boldsymbol{B}(\boldsymbol{r},t) = -\frac{\mu_0}{4\pi rc}\left[\hat{\boldsymbol{r}}\times\ddot{\boldsymbol{d}}(t_0)\right] \tag{5.2.2}$$

where $\ddot{\boldsymbol{d}}(t) = \frac{d^2}{dt^2}[\boldsymbol{d}(t)]$, $\boldsymbol{d}(t)$ is the electric dipole moment and $t_0 = t - r/c$ is the retarded time. We know that the radiated power in the differential solid angle, $d\Omega$, is just the scalar product of the Poynting vector, $\boldsymbol{S} = 1/\mu_0(\boldsymbol{E}\times\boldsymbol{B})$ with the differential area, $r^2 d\Omega$

$$\frac{dE}{dtd\Omega} = \frac{1}{\mu_0}(\boldsymbol{E}\times\boldsymbol{B}).\hat{\boldsymbol{r}}r^2 \tag{5.2.3}$$

$$= \frac{\mu_0}{(4\pi)^2 c}\left[\hat{\boldsymbol{r}}\times\ddot{\boldsymbol{d}}(t_0)\right]^2. \tag{5.2.4}$$

The above expression is similar to the Larmor formula. If we take two versors, $\hat{\boldsymbol{\varepsilon}}_1$ and $\hat{\boldsymbol{\varepsilon}}_2$, that are orthogonal to $\hat{\boldsymbol{r}}$ and $\hat{\boldsymbol{\varepsilon}}_1.\hat{\boldsymbol{\varepsilon}}_2 = 0$, we can express

$$\left[\hat{\boldsymbol{r}}\times\ddot{\boldsymbol{d}}(t_0)\right]^2 = \sum_{\lambda=1,2}\left[\hat{\boldsymbol{\varepsilon}}_\lambda.\ddot{\boldsymbol{d}}(t_0)\right]^2 \tag{5.2.5}$$

$$\frac{dE}{dtd\Omega} = \frac{\mu_0}{(4\pi)^2 c}\sum_{\lambda=1,2}\left[\hat{\boldsymbol{\varepsilon}}_\lambda.\ddot{\boldsymbol{d}}(t_0)\right]^2. \tag{5.2.6}$$

By using the Parseval-Plancherel identity [32], we can express the Larmor formula in the frequency domain just by taking its Fourier transform. We must divide then by $2\pi$ and multiply by a factor of 2. The $2\pi$ factor comes from the normalization of our Fourier transform and the factor of 2 comes from sum-

1 This subsection is in SI units

ming the $\omega$ and $-\omega$ contributions. Then, in the frequency domain, Eq. (5.2.6), becomes

$$\frac{dE}{d\omega d\Omega} = \frac{2\mu_0}{2\pi(4\pi)^2 c}\sum_{\lambda=1,2}\left[\hat{\varepsilon}_\lambda.\tilde{\ddot{\boldsymbol{d}}}(\omega)\right]^2 \tag{5.2.7}$$

$$\tilde{\ddot{\boldsymbol{d}}}(\omega) = \int_{-\infty}^{+\infty} dt\ddot{\boldsymbol{d}}(t)e^{-i\omega t}. \tag{5.2.8}$$

Let's consider the case where the electric dipole is oriented along the $z$ axis, $\ddot{\boldsymbol{d}}(t) = (0,0,\ddot{d}_z(t))$. In spherical coordinates [62], Eq. (5.2.7), takes the form

$$\frac{dE}{d\omega d\Omega} = \frac{2\mu_0}{2\pi(4\pi)^2 c}\left(\tilde{\ddot{d}}_z(\omega)\right)^2\sin^2(\theta) \tag{5.2.9}$$

and integrating over all solid angles we obtain the usual Larmor Formula

$$\begin{aligned}\frac{dE}{d\omega} &= \int_0^{2\pi}\int_0^{\pi}\frac{2\mu_0}{2\pi(4\pi)^2 c}\left(\tilde{\ddot{\boldsymbol{d}}}_z(\omega)\right)^2\sin^3(\theta)\,d\theta d\varphi\\ &= \frac{\mu_0}{6\pi^2 c}\left(\tilde{\ddot{\boldsymbol{d}}}_z(\omega)\right)^2\\ &= \frac{1}{6\pi^2\varepsilon_0 c^3}\left(\tilde{\ddot{\boldsymbol{d}}}_z(\omega)\right)^2.\end{aligned} \tag{5.2.10}$$

In our case, the dipole is always oriented along the $z$ axis so we will use this formula to evaluate the High Harmonic generation spectrum. In the following subsection the quantum description of the field will be taken into account and we will see that we get the same result as in the classical case.

### 5.2.2 *Quantum radiation emitted by an oscillating electric dipole*[2]

In this subsection, we will derive a formula for the HHG spectra in the single atom response regime [35, 63]. We will take into account the fact that the EMF is a quantum field. The Hamiltonian of a system of non-relativistic charged particles interacting with a quantum EMF in the Coulomb gauge is (Eq. A.16 of [32])

2 This subsection is in SI units. The notation adapted in this subsection is the same as the one used in [32, 63].

$$H = \sum_{\alpha} \frac{1}{2m_{\alpha}} \left[ \boldsymbol{p}_{\alpha} - q_{\alpha} \boldsymbol{A} \left( \boldsymbol{r}_{\alpha} \right) \right]^2 + \tag{5.2.11}$$

$$+ \quad V_C + \sum_i \hbar \omega_i \left( a_i^{\dagger} a_i + 1/2 \right) \tag{5.2.12}$$

where the first term is the kinetic energy term of the particles, the second term the Coulomb interaction and the third term is the energy of the transverse part of the EMF. The Hamiltonian determines the dynamics of the system. To treat this problem we will work on the Heisenberg picture where the state vector, $|Psi\rangle$, remains the same and the operators evolve in time obeying to the Heisenberg equation of motion

$$\frac{d}{dt} G(t) = \frac{1}{i\hbar} \left[ G(t), H(t) \right]. \tag{5.2.13}$$

From complement $B_{III}$ of [32], we have that the equation of motion of the destruction operator of a certain normal mode of the quantum electromagnetic field ($i = \left( k_x, k_y, k_z, \lambda \right)$, where $\vec{k}_i$ stands for the propagation vector and obeys to periodic boundary conditions in a 3D box of size $L$, i.e. $\vec{k}_i = \left( \frac{n_{x,i} 2\pi}{L}, \frac{n_{y,i} 2\pi}{L}, \frac{n_{z,i} 2\pi}{L} \right)$, and $\lambda$ is the polarization of the normal mode) $i$) is

$$\dot{a}_i + i\omega_i a_i = s_i \tag{5.2.14}$$

$$s_i = \frac{i}{\sqrt{2\varepsilon_0 \hbar \omega_i L^3}} \int d^3 r e^{-i\vec{k}_i . \vec{r}} \vec{\varepsilon}_i . \vec{j} \left( \vec{r}, t \right) \tag{5.2.15}$$

where $\omega_i = c \left| \vec{k}_i \right|$. This expression is similar to the equation of motion of the destruction operator for a quantum harmonic oscillator, where $s_i$ is the driven term. In the absence of sources, $s_i = 0$, and the solution is trivial $a_i(t) = a_i(0) e^{-i\omega_i t}$. In all the studies of HHG, it is enough to calculate the polarization of the medium by solving the TDSE and, after that, propagate the Maxwell's equations as if the medium was a classical one. In this way, the $\vec{j}(\vec{r}, t)$ operator becomes a classical function $\vec{j}_{cl}(\vec{r}, t)$. The current density of a system of particles is given by

$$\vec{j}_{cl}(\vec{r}, t) = \sum_{\alpha} q_{\alpha} \vec{v}_{\alpha}(t) \, \delta\left( \vec{r} - \vec{r}_{\alpha}(t) \right) \tag{5.2.16}$$

where $\alpha$ is the index for the particles, $q_\alpha$ is the electric charge of the particle and $\vec{r}_\alpha(t)$ the position of particle $\alpha$ at time $t$. The electric dipole operator and its time derivative and second time derivative, with respect to the origin, are defined as

$$\vec{d}(t) = \sum_\alpha q_\alpha \vec{r}_\alpha(t) \tag{5.2.17}$$

$$\frac{d\left(\vec{d}(t)\right)}{dt} \equiv \vec{d}_v(t) = \sum_\alpha q_\alpha \vec{v}_\alpha(t) \tag{5.2.18}$$

$$\frac{d^2\left(\vec{d}(t)\right)}{dt^2} \equiv \vec{d}_a(t) = \sum_\alpha q_\alpha \vec{a}_\alpha(t) \tag{5.2.19}$$

In the dipole approximation, $e^{-i\vec{k}_i.\vec{r}} \approx 1$ and replacing the current operator, $\vec{j}(\vec{r},t)$, by its classical function, $\vec{j}_{cl}(\vec{r},t)$, Eq. (5.2.15) becomes

$$s_i^{cl} = \frac{i}{\sqrt{2\varepsilon_0 \hbar \omega_i L^3}} \int d^3r \sum_\alpha q_\alpha \delta\left(\vec{r} - \vec{r}_\alpha(t)\right) \vec{\varepsilon}_i.\vec{v}_\alpha(t) \tag{5.2.20}$$

$$= \frac{i}{\sqrt{2\varepsilon_0 \hbar \omega_i L^3}} \vec{\varepsilon}_i. \left(\sum_\alpha q_\alpha \vec{v}_\alpha(t)\right) \tag{5.2.21}$$

$$= \frac{i}{\sqrt{2\varepsilon_0 \hbar \omega_i L^3}} \vec{\varepsilon}_i.\vec{d}_v(t) \tag{5.2.22}$$

and replacing this in the equation of motion for the destruction operator and integrating this equation of motion we have that

$$a_i(t) = a_i(0) e^{-i\omega_i t} + \int_0^t dt' \frac{i}{\sqrt{2\varepsilon_0 \hbar \omega_i L^3}} \vec{\varepsilon}_i.\vec{d}_v(t) e^{-i\omega_i(t-t')}. \tag{5.2.23}$$

The associated number operator is $N_i(t) = a_i^\dagger(t) a_i(t)$, where the mean value of $N_i(t)$ is the number of photons in the $i$ mode of the EMF, and the number operator becomes

$$
\begin{aligned}
N_i(t) &= a_i^{\dagger}(0)\, a_i(0) && (5.2.24)\\
&+ \left(\int_0^t dt' \frac{-i}{\sqrt{2\varepsilon_0 \hbar\omega_i L^3}} \vec{\varepsilon}_i.\vec{d}_v(t)\, e^{i\omega_i(t-t')}\right) a_i(0)\, e^{-i\omega_i t} && (5.2.25)\\
&+ a_i^{\dagger}(0)\, e^{i\omega_i t} \left(\int_0^t dt' \frac{i}{\sqrt{2\varepsilon_0 \hbar\omega_i L^3}} \vec{\varepsilon}_i.\vec{d}_v(t)\, e^{-i\omega_i(t-t')}\right) && (5.2.26)\\
&+ \left(\int_0^t dt' \frac{-i}{\sqrt{2\varepsilon_0 \hbar\omega_i L^3}} \vec{\varepsilon}_i.\vec{d}_v(t)\, e^{i\omega_i(t-t')}\right) \times && (5.2.27)\\
&\times \left(\int_0^t dt' \frac{i}{\sqrt{2\varepsilon_0 \hbar\omega_i L^3}} \vec{\varepsilon}_i.\vec{d}_v(t)\, e^{-i\omega_i(t-t')}\right) && (5.2.28)
\end{aligned}
$$

We want to calculate the mean value of the number operator and we assume that the radiation state at $t = 0$ is the vacuum state and since we are working on the Heisenberg picture $|\Psi(t)\rangle = |\Psi(0)\rangle = |0,0,0...0,0,0\rangle$, so we suppose that the quantum EMF is unpopulated at the beginning. To calculate the number of photons in the mode $i$ at time $t$, we just need to calculate $< N_i(t) >= \langle 0,0,0...0,0,0| N_i(t) |0,0,0...0,0,0\rangle$, and only the last term survives, since $a_i(0)\,|\{\rangle\, \Psi(t)\} = \langle \Psi(t)|\, a_i^{\dagger}(0) = 0$, so

$$
\begin{aligned}
\langle N_i(t)\rangle &= \left(\int_0^t dt' \frac{-i}{\sqrt{2\varepsilon_0 \hbar\omega_i L^3}} \vec{\varepsilon}_i.\vec{d}_v(t)\, e^{i\omega_i(t-t')}\right) \times && (5.2.29)\\
&\times \left(\int_0^t dt' \frac{i}{\sqrt{2\varepsilon_0 \hbar\omega_i L^3}} \vec{\varepsilon}_i.\vec{d}_v(t)\, e^{-i\omega_i(t-t')}\right) && (5.2.30)\\
&= \frac{1}{2\varepsilon_0 \hbar\omega_i L^3} \left| \int_0^t dt' \vec{\varepsilon}_i.\vec{d}_v(t)\, e^{-i\omega_i t'} \right|^2 && (5.2.31)
\end{aligned}
$$

and the mean value of the energy of each mode $i$, $E_i$, is obtained by just multiplying $\langle N_i(t)\rangle$ by the energy of the photon, $\hbar\omega_i$,

$$\langle E_i(t) \rangle = \hbar\omega_i \langle N_i(t) \rangle = \frac{1}{2\varepsilon_0 L^3} \left| \int_0^t dt' \vec{\varepsilon}_i . \vec{d}_v(t) \, e^{-i\omega_i t'} \right|^2 . \tag{5.2.32}$$

To take the continuum limit, $L \to \infty$, we must take Eq. C.43 I.C.6 of [32]

$$a_i = a_{(\vec{k},\lambda)} \left( \frac{2\pi}{L} \right)^{3/2} \tag{5.2.33}$$

and in this way

$$\left\langle E_{(\vec{k},\lambda)}(t) \right\rangle = \frac{1}{(2\pi)^3 \, 2\varepsilon_0} \left| \int_0^t dt' \vec{\varepsilon}_{(\vec{k},\lambda)} . \vec{d}_v(t') \, e^{-i\omega_i t'} \right|^2 \tag{5.2.34}$$

so the density of energy of the radiation, expressing the $\vec{k}$ vector in spherical coordinates, is given by

$$\frac{dE_{(\vec{k},\lambda)}}{k^2 dk \sin(\theta) \, d\theta d\varphi} = \frac{c^3 dE_{(\omega,\theta,\varphi,\lambda)}}{\omega^2 d\omega \sin(\theta) \, d\theta d\varphi} \tag{5.2.35}$$

so the density of energy emitted for a given polarization $\lambda$, differential in $\omega$ and in the solid angle $d\Omega = \sin(\theta) \, d\theta d\varphi$ is

$$\frac{dE}{d\omega d\Omega} = \frac{\omega^2}{(2\pi c)^3 \, 2\varepsilon_0} \left| \int_0^t dt' \vec{\varepsilon}_{(\vec{k},\lambda)} . \vec{d}_v(t') \, e^{-i\omega_i t'} \right|^2 \tag{5.2.36}$$

Now take the assumption that $\vec{d}_v(t)$ is only along the $z$ axis $(0, 0, d_v^z(t))$. Expressing $\vec{k}$ in spherical coordinates and choosing two orthogonal polarization vectors

$$\vec{k} = \left| \vec{k} \right| (\sin\theta \cos\varphi, \sin\theta \sin\varphi, \cos\theta) \tag{5.2.37}$$

$$\vec{\varepsilon}_{(\vec{k},1)} = (\cos\theta \cos\varphi, \cos\theta \sin\varphi, -\sin\theta) \tag{5.2.38}$$

$$\vec{\varepsilon}_{(\vec{k},2)} = (-\sin\varphi, \cos\varphi, 0) \tag{5.2.39}$$

only the first polarization is not vanishing and integrating over the solid angle we get that

$$\frac{dE}{d\omega} = \frac{\omega^2}{(2\pi c)^3 2\varepsilon_0} \int_0^{2\pi}\int_0^{\pi} \left| \int_0^t d_v^z(t') e^{i\omega t'} dt' \right| \sin^3\theta d\theta d\varphi \tag{5.2.40}$$

$$= \frac{2\pi\omega^2 4}{3(2\pi c)^3 2\varepsilon_0} \left| \tilde{d}_v^z(\omega) \right|^2 \tag{5.2.41}$$

$$= \frac{\omega^2}{6\pi^2 c^3 \varepsilon_0} \left| \tilde{d}_v^z(\omega) \right|^2. \tag{5.2.42}$$

If the dipole velocity vanishes at the extremes of the pulse, $d_v^z(t) = d_v^z(0) = 0$, we have that the square of the Fourier transform of the dipole velocity and the dipole acceleration are related by $\left|\tilde{d}_v^z(\omega)\right|^2 \omega^2 = \left|\tilde{d}_a^z(\omega)\right|^2$. In this case Eqs. (5.2.10) and (5.2.42) are equivalent.

### 5.2.3 *Acceleration form of the HHG spectrum*

In the previous section, we have shown that the acceleration form is the correct form to evaluate the HHG spectrum [64]. The way in which we calculate the dipole acceleration is described in SubSec. 2.3.4. If we perform directly the Fourier transform of the dipole acceleration, numerical noise will appear due to the fact that the dipole acceleration does not vanish smoothly at the end of the pulse. To avoid this problem, the dipole acceleration is multiplied by a window function in order to remove this numerical noise from the spectrum. The HHG spectrum, $S(\omega)$, is then calculated as (in atomic units)

$$S(\omega) = \frac{2}{3\pi c^3} \left| \widetilde{\ddot{d}}(\omega) \right|^2 \tag{5.2.43}$$

$$\widetilde{\ddot{d}}(\omega) = \int_0^T \ddot{d}(t) f_{\sigma,t_i,t_f}(t) e^{-i\omega t} dt. \tag{5.2.44}$$

$$f_{\sigma,t_i,t_f}(t) = \begin{cases} 0 & t < t_i \lor t > t_f \\ \sin^2\left(\frac{\pi(t-t_i)}{2\sigma}\right) & t > t_i \land t < t_i + \sigma \\ 1 & t > t_i + \sigma \land t < t_f - \sigma \\ \sin^2\left(\frac{\pi(t_f - t)}{2\sigma}\right) & t > t_f - \sigma \land t < t_f \end{cases} \tag{5.2.45}$$

where $t_i = 0$, $t_f = T$ and $\sigma = 1$ fs, being $T$ the total duration of the pulse. The value of $\sigma$ is always 1 fs otherwise stated. This value of $\sigma$ was chosen such that it does not introduce any artifacts in the HHG spectra. The envelope function is shown in Fig. 5.2.1. This envelope should be used to avoid numerical noise due to the discontinuity of the dipole acceleration at the extremes of the pulse.

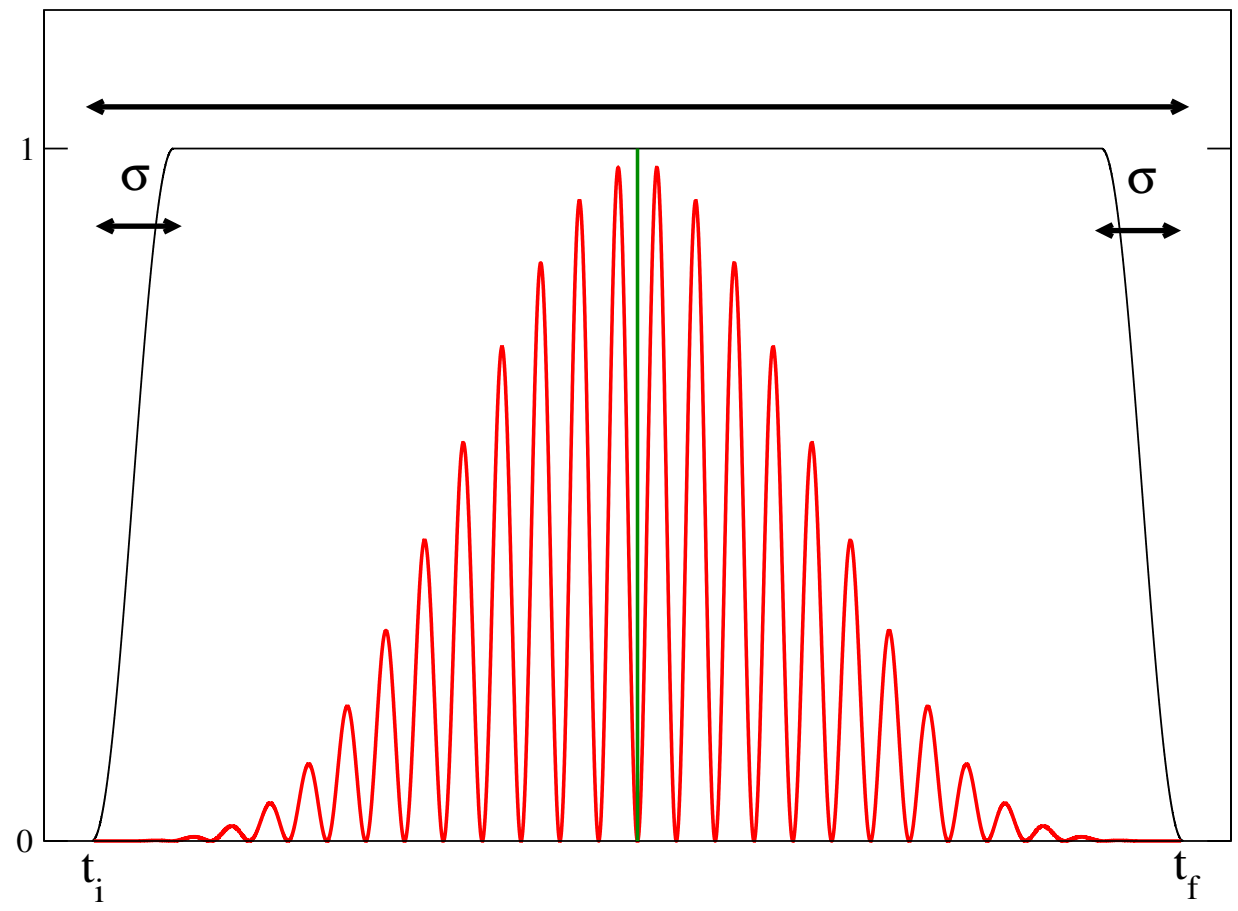

Figure 5.2.1.: Envelope function. The red line represents the square of the electric field, $E(t)^2$, for a 14 cycles pulse.

### 5.2.4 *Gabor profile*

To have an additional insight on the HHG process we can perform a Gabor analysis to know which are the frequencies that are being emitted by the molecule at a given time. The Gabor profile is just a short-time Fourier transform. We just multiply our function in the temporal domain by a gaussian centered at $t_0$. The Gabor profile, $G(\omega, t_0)$, is calculated as

$$G(\omega, t_0) = \frac{2}{3\pi c^3} \left| \int_0^T \ddot{d}(t)\, e^{-i\omega t} e^{-(t-t_0)^2/2\alpha^2} dt \right|^2 \tag{5.2.46}$$

$$\alpha = \frac{1}{3\omega_0} \tag{5.2.47}$$

where $\omega_0$ is the frequency of the infrared laser field. By choosing this particular value of $\alpha$, the Gabor transform provides satisfactory resolution for the whole harmonic range in all cases studied in this work, and it is not necessary to consider more sophisticated techniques such as wavelet theory [65]. A typical Gabor profile is shown in Fig. 5.2.2.

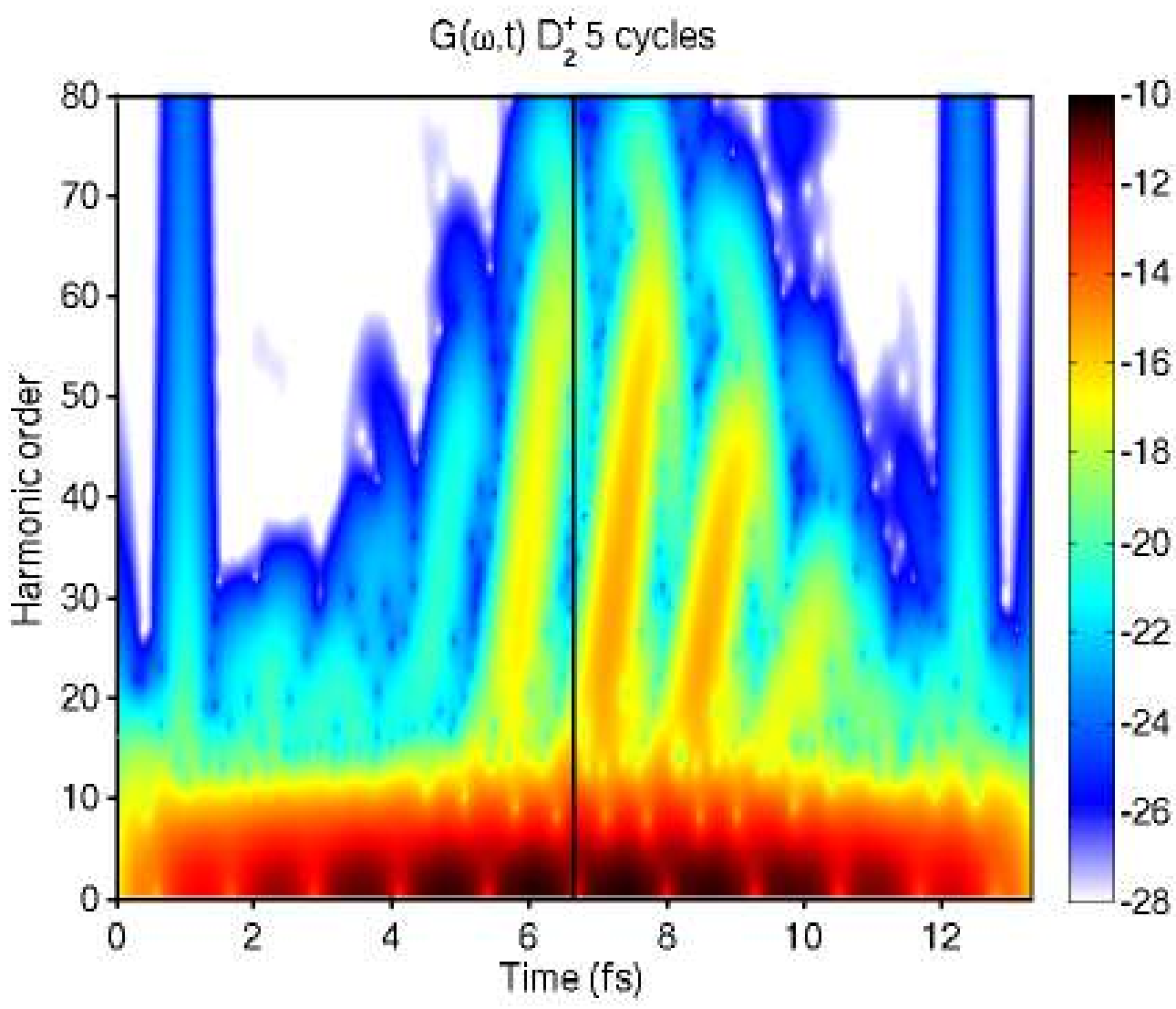

Figure 5.2.2.: Gabor profile from a $D_2^+$ molecule for a laser pulse with $\lambda = 800\,\text{nm}$, $I = 3 \times 10^{14}\,\text{W/cm}^2$ and with a total duration of 5 optical cycles. The result is shown in logarithmic scale.

# Part III

# RESULTS

# 6

# RESOLVENT OPERATOR METHOD ON A 1+1D CALCULATION ON $H_2^+$

In this Chapter, we present the results obtained with the ROM in a 1+1D calculation of the $H_2^+$ molecule. We present results for electron- (EKE), nuclear- (NKE), and correlated electron- and nuclear-kinetic energy (CKE) spectra for $H_2^+$ ionization and dissociation by ultrashort laser pulses. We study the validity of the differential ROM, as well as its comparison with the total ROM. In order to do that, we consider strong ($> 10^{13}$ W/cm$^2$) ultrashort (few femtoseconds) laser pulses with frequencies ranging from XUV to IR. It must be noted that the ROM can be applied in a wide range of intensities (in principle, up to the validity of the nonrelativistic TDSE equation) and frequencies, as well as pulse durations, from attosecond laser pulses to longer pulses. The upper limitation to the pulse duration is that the wave function must be fully contained inside the box where the ROM analysis is performed. Therefore, for longer pulses, bigger boxes are needed.

We have used a box with $|z| < 1500$ a.u. and $R < 30$ a.u., with uniform grid spacings of $\Delta z = 0.1$ a.u. and $\Delta R = 0.05$ a.u. The propagation was performed by using the Crank-Nicolson split-operator method with $\Delta t = 0.02$ a.u. We have checked that by increasing the electronic box size by 13%, the nuclear box size by 17%, and the density of the spatial and temporal grids by 50%, the results do not change. The laser pulses have a photon energy $\Omega$, a total pulse duration $T$, and a peak intensity $I$. A $\sin^2$ envelope was used in all cases. The propagation is performed until the end of the pulse or later, when the required observables are extracted with the resolvent-operator method (ROM). The initial state is the groundstate of the $H_2^+$ molecule.

In regard to the ROM, we have chosen the values of $n_e = n_N = 2$ and $\delta_e = \delta_N = 0.004$ [see Eq. (3.3.26)]. The ROM analysis is performed at the end of the pulse to a wavefunction for which the contribution to the groundstate was removed. We have checked that the results are converged.

The results that are presented in this Chapter were published in [28].

## 6.1 FEW-PHOTON ABSORPTION

The simplest case corresponds to the absorption of one or two photons with an ultrashort XUV laser pulse. Results for a central frequency $\Omega = 1.37$ a.u., a total pulse duration of $16$ fs, and peak intensity $10^{14}$W/cm$^2$ are presented in Fig. 6.1.1. These parameters correspond to a purely multiphotonic regime since the Keldysh parameter is $\gamma = 38 \gg 1$, where $\gamma^2 = I_p/2Upp_p$, with $I_p$ the ionization potential at the equilibrium distance $R_{eq} = 1.9$ a.u., and $U_p = I/4\Omega^2$ the ponderomotive energy.

The BO potential energy curves of the $H_2^+$ molecule are shown in Fig. 6.1.1(a), where the Franck-Condon region lies between the vertical dashed lines. The corresponding CKE spectrum is shown in Fig. 6.1.1(b), where the ionization probability is plotted (in log scale) as a function of the electron- and nuclear-kinetic energy. We expect to observe energy conservation lines, which obey $N\omega = E_e + E_N + D_{2H^+}$, where $D_{2H^+} = -E_0 = 0.597$ a.u. is the threshold energy required to produce two protons at infinite internuclear distance, and $E_0$ is the energy of the initial state. Therefore, in a 2D $(E_N, E_e)$ plot they correspond to lines with slope $-1$. Two of these lines are visible in the figure, corresponding to one- and two-photon absorption ($N = 1, 2$). The one-photon absorption line exhibits a minimum at a nuclear energy around $0.56$ a.u., which corresponds to an internuclear distance of $R \approx 1.78$ a.u. in the Coulomb explosion curve $1/R$ [green line in Fig. 6.1.1(a)]. This minimum is clearly visible in the integrated nuclear-kinetic energy (NKE) distribution shown in Fig. 6.1.1(b), top panel.

The origin of such a minimum in the CKE spectrum can be understood if we resort to one-dimensional calculations in which the internuclear distance is fixed. The corresponding ionization probability, which has been obtained by using the same soft-core potential as in the (1+1)D calculations, is shown in Fig. 6.1.1(c) as a function of the internuclear distance. This ionization probability exhibits a pronounced minimum at $R = 1.8$ a.u., which explains the depletion observed in the CKE spectrum. A similar depletion has been observed in $H_2^+$ ionization probabilities resulting from full dimensional calculations [18]. As in the present case, this is due to a minimum in the dipole-coupling matrix element connecting the ground and the final continuum state of $H_2^+$.

To get further insight on the origin of this minimum, we have performed model calculations in the framework of first-order perturbation theory. In these calculations, we have used the representation of the initial and final states proposed by Cohen and Fano [66]: The final state is described by a plane wave and the initial state by a linear combination of $1s$ orbitals corresponding to the actual nuclear charge $Z = 1$, but also to an $R$-dependent effective charge chosen to reproduce the exact ground-state potential energy curve. The ionization probability is obtained by weighting the square of the corresponding dipole matrix element with the ground-state nuclear probability density (reflection approximation [67]). A comparison with the results of the TDSE calculations (see Fig. 6.1.2) shows that the model with effective charges catches the essential features of the full calculation. Hence, the minimum in the ionization probability can unambiguously be attributed to the interference resulting from the coherent electron emission from the two molecular centers.

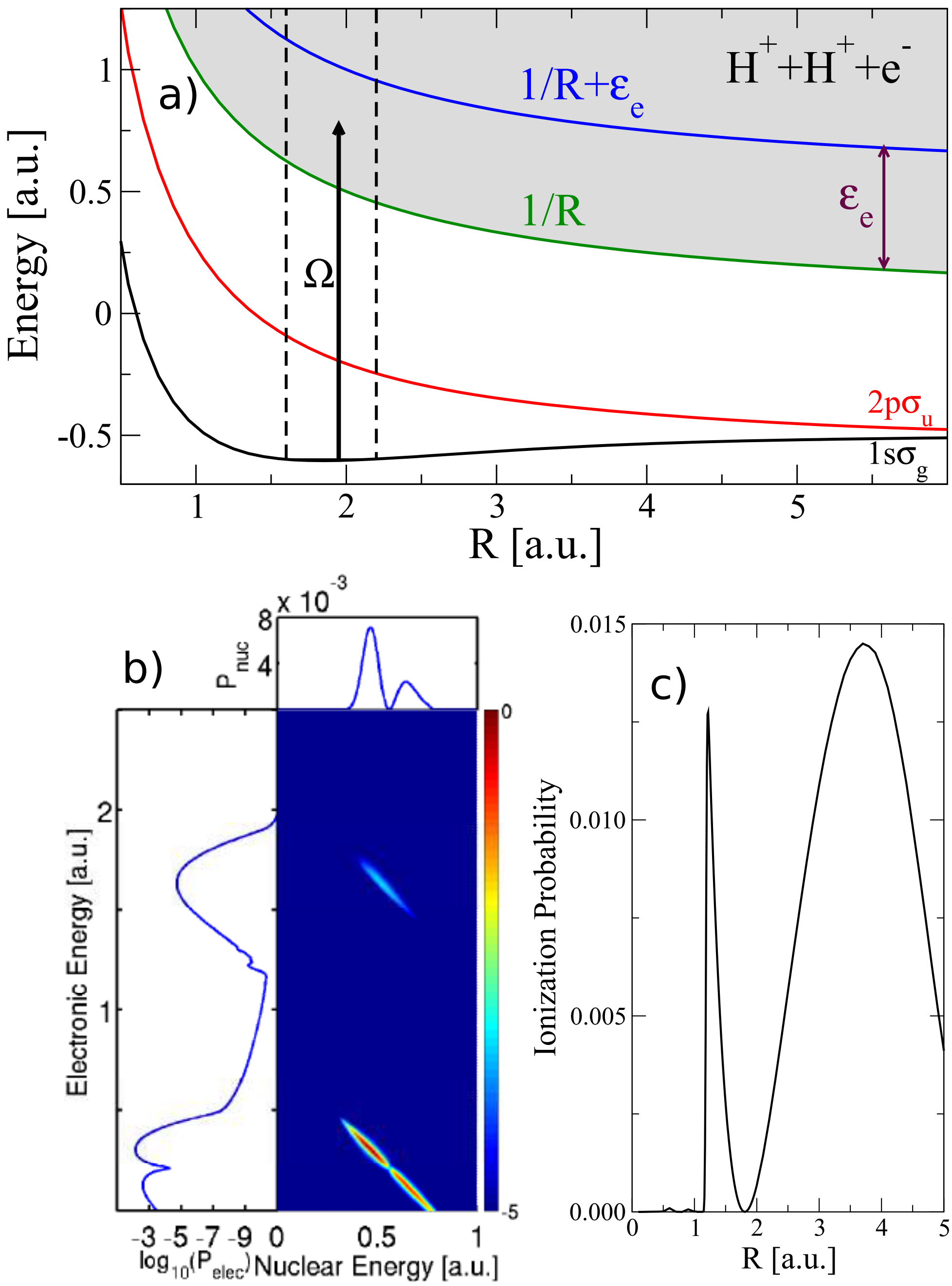

Figure 6.1.1.: (a) Born-Oppenheimer potential energy curves for the $H_2^+$ molecule. The black arrow represents a vertical transition from the $H_2^+$ ground state to the ionization channel with a photon energy $\Omega = 1.37$ a.u. The dashed lines represent the limits of the Franck-Condon region. (b) CKE for a pulse with $\Omega = 1.37$ a.u., total duration 16 fs and $I = 10^{14}$W/cm$^2$. The corresponding projections (singly differential probabilities) in electronic energy ($P_{elec}$) and nuclear energy ($P_{nuc}$) are shown on the left and on top of the figure. (c) Ionization probability as a function of the internuclear distance $R$ using the same soft-core potential as in the (1+1)D calculations.

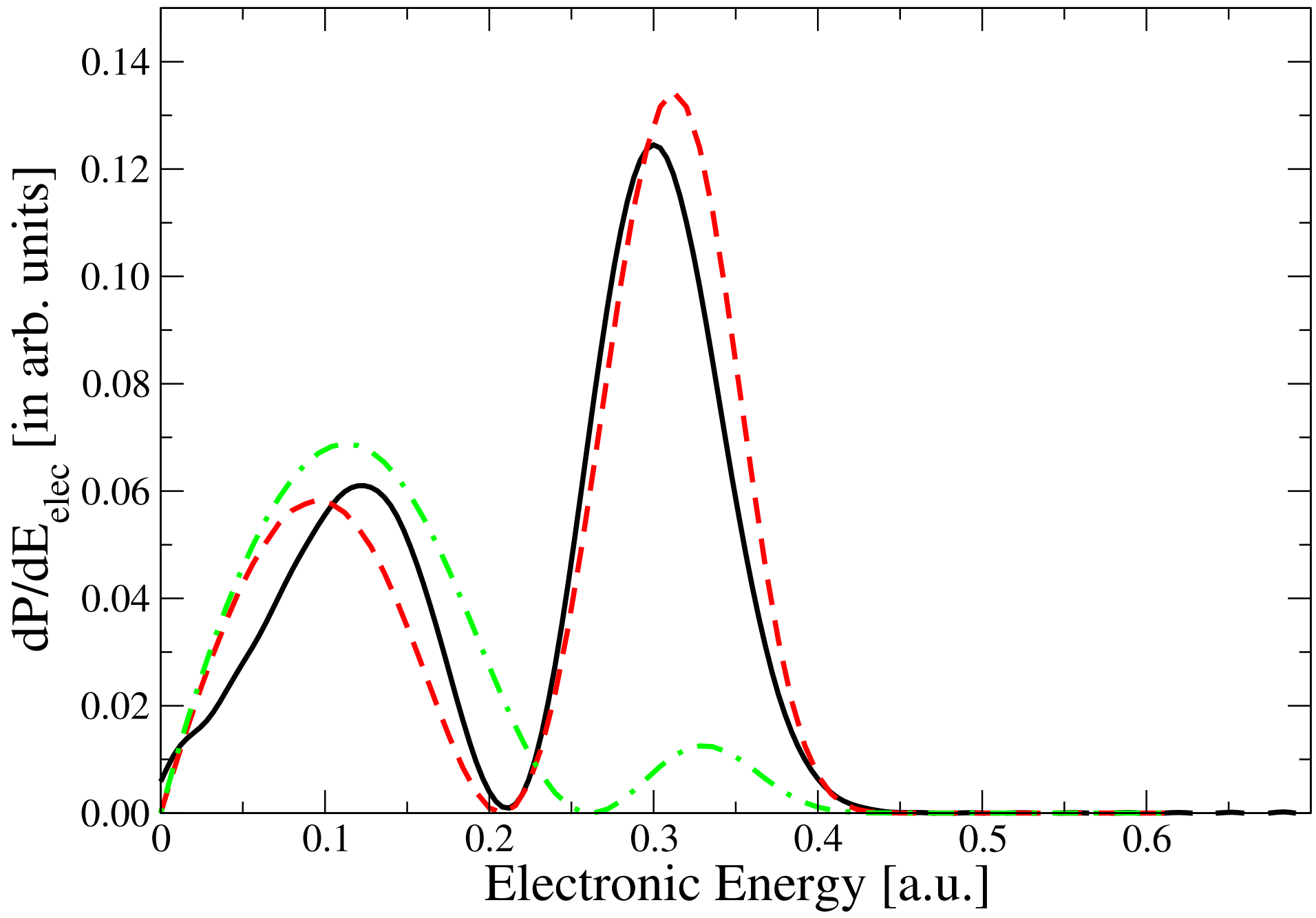

Figure 6.1.2.: Ionization probability as a function electron energy. Comparison between the results of the full calculations shown in Fig. 6.1.1 (black solid curve) and those of the first-order perturbative model described in text with $Z = 1$ (green dash-dotted curve) and effective charge (red dashed curve).

## 6.2 RESONANT TRANSITION

The differential ROM makes it possible to extract the contributions to the energy spectrum arising from different electronic states of the molecule. To illustrate this we choose a laser pulse with the same duration and intensity as before, but whose energy is resonant to a particular electronic transition, $1s\sigma_g \rightarrow 2p\sigma_u$, instead of leading directly to ionization. At the internuclear equilibrium distance, the energy difference between those states is $E_{2p\sigma_u} - E_{1s\sigma_g} \approx 0.4$ a.u. The seventh harmonic of a 800-nm pulse ( $\Omega = 0.398$ a.u.) matches well that energy difference. As shown in Fig. 6.2.1(a), the final kinetic energy of the nuclei in this situation is expected to be $T_N \sim 0.29$ a.u. The NKE spectrum for the $2p\sigma_u$ state calculated with the differential ROM, shown in Fig. 6.2.1(c), confirms this prediction. However, in this case we can also have ionization through absorption of three or more photons, via resonance-enhanced multiphoton ionization (REMPI). This can be seen in the CKE spectrum shown in Fig. 6.2.1(b), where several lines of energy conservation can be observed.

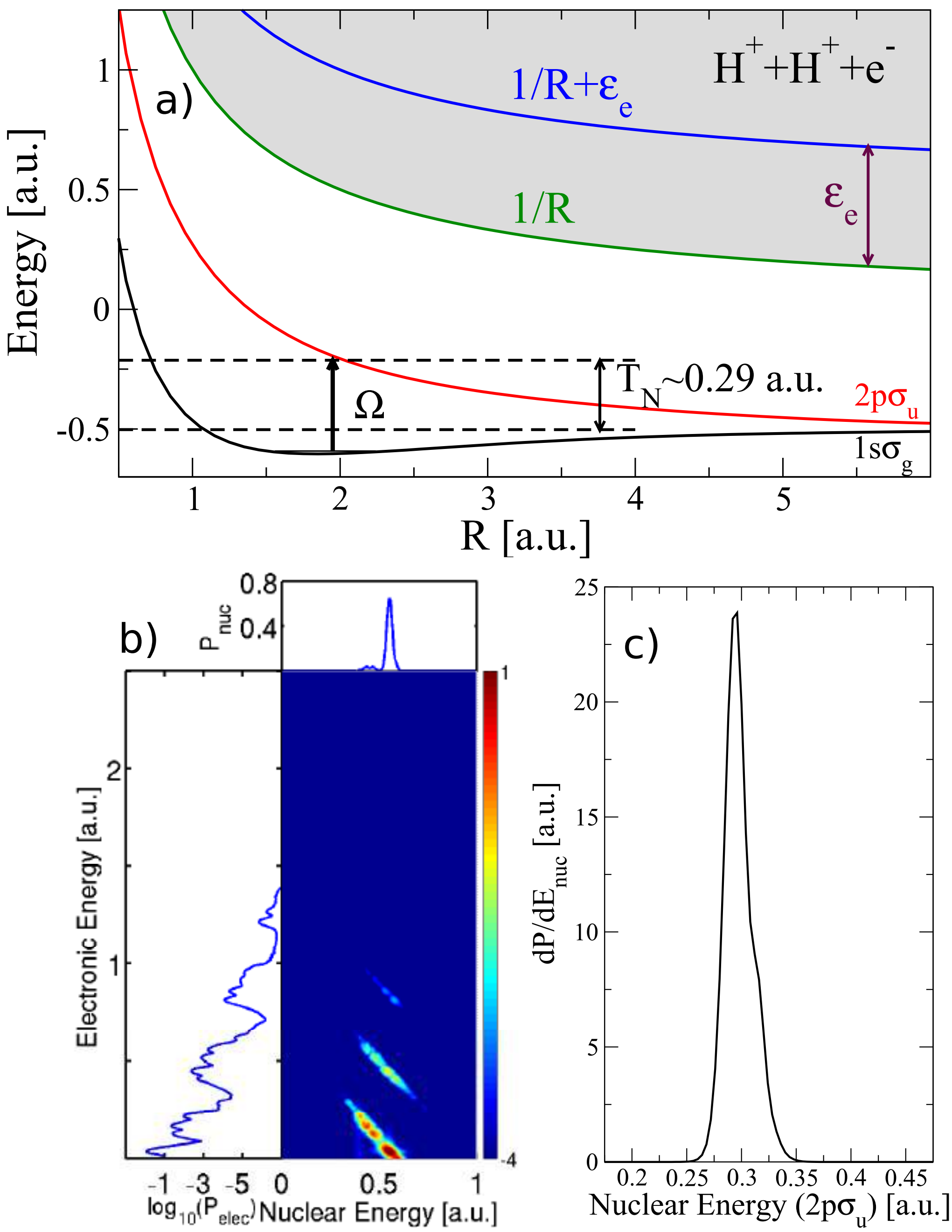

Figure 6.2.1.: (a) Born-Oppenheimer curves of the $H_2^+$ molecule. The black arrow represents a vertical transition from the $H_2^+$ ground state to the first excited state ($1s\sigma_g \to 2p\sigma_u$) with a photon energy $\Omega = 0.398$ a.u. (b) CKE for a pulse with $T = 16$ fs and $I = 10^{14}$W/cm$^2$. The corresponding projections (singly differential probabilities) in electronic energy ($P_{elec}$) and nuclear energy ($P_{nuc}$) are shown on the left and on top of each panel. (c) NKE spectrum of the $2p\sigma_u$ electronic state.

The integrated EKE spectrum [Fig. 6.2.1(b), left panel] is more complex than in the UV case shown in Fig. 6.1.1. The maxima corresponding to multiphoton absorption ($N = 3$) show a splitting of the peaks due to the Autler-Townes effect and the nuclear motion [22]. In the case of atomic resonant ionization, this splitting can be interpreted in terms of dressed states induced by the field [68]: Each state splits in two with an energy separation equal to the Rabi frequency $\Omega_R = \mu E$, where $\mu$ is the dipole matrix element between the resonantly coupled states and $E$ is the maximum amplitude of the electric field. The molecular case is not that simple. For the photon energy used here, $\Omega = 0.398$ a.u., the two lowest electronic states are resonantly coupled; see Fig. 6.2.2. The dressed $2p\sigma_u$ curve (shifted by $-\Omega$) splits into two different curves, so that the energy difference between them is $\Omega_R$. This energy difference depends on the internuclear distance because the dipole matrix element $\mu(R) =< 2p\sigma_u|z|1s\sigma_g > |_R$ does. An order of magnitude of $\Omega_R$ can be obtained by taking the value of $\mu$ at the equilibrium distance $R_0 = 1.9$ a.u., which is $\mu(R_0) = 1.084$ a.u., thus leading to $\Omega_R = 0.058$ a.u. for $10^{14}$W/cm$^2$. This value of $R$ is compatible with the observed splitting. Now the nuclear motion further complicates this picture. When the nuclear wave packet is centered around $R \sim R_1$, the electron is preferentially emitted from the lower curve and, therefore, has smaller kinetic energy than when the nuclear wave packet is centered around $R \sim R_2$, since in the latter case the electron is preferentially emitted from the upper curve [Fig. 6.2.2(a)]. The energy difference between both channels projected onto the ionization potential curve is $\Delta E \sim 1/R_1 - 1/R_2 = 0.059$ a.u., which is very similar to $\Omega_R$. As a consequence of this, additional structures due to the different kinetic energies that the electron can acquire as the nuclear wave packet moves are expected to appear in an energy interval similar to that dictated by the Autler-Townes splitting.

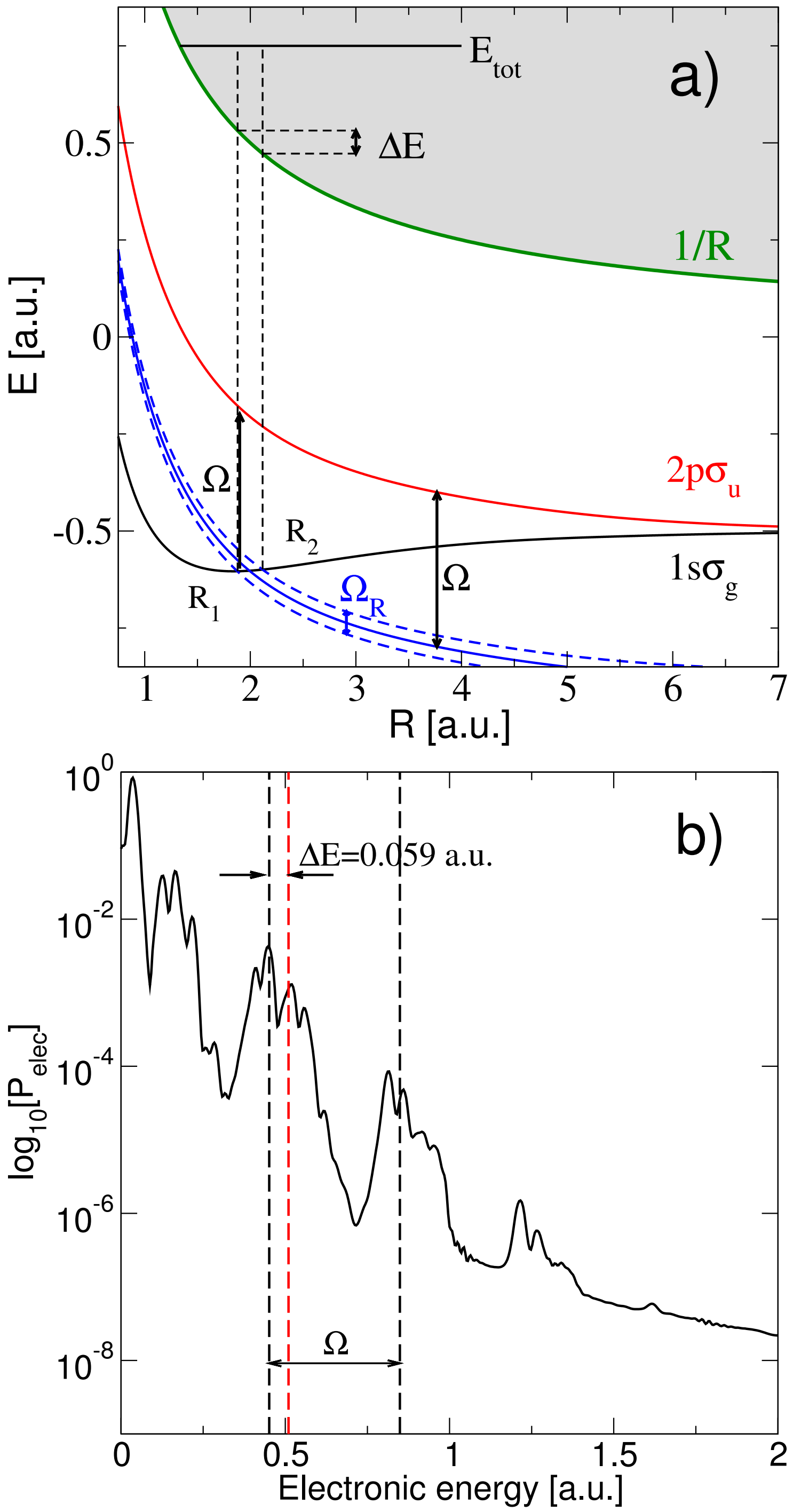

Figure 6.2.2.: (a) Potential energy curves corresponding to ionization ($1/R$), $2p\sigma_u$ and $1s\sigma_g$ as a function of the internuclear distance. The dressed states of the $2p\sigma_u$ curve are shifted by $\Omega_R/2$ (upward and downward). The diabatic coupling of these states is represented by the vertical dashed lines. (b) EKE shown in Fig. 6.1.1(b).

## 6.3 TOTAL ROM VS DIFFERENTIAL ROM

Finally, we discuss the results obtained with the total ROM in comparison to those obtained by integrating the differential ROM. In the latter case, the total probability density for each total energy $E_{tot}$ results from integration over all possible energy pairs $(E_e, E_N)$ such that $E_{tot} = E_e + E_N$. To compare the spectra, we have considered the first 15 bound curves of the $H_2^+$ molecule, as well as all positive electron energies up to 3 a.u. To distinguish the contribution to the total spectrum that comes from ionization, we also calculate the ROM spectrum due only to the ionization channel.

In Fig. 6.3.1 we consider the same configuration as in Fig. 6.1.1, $\Omega = 1.37$ a.u. and $10^{14}$W/cm$^2$, but with total pulse durations of 4 fs (a) and 1 fs (b), corresponding to 36 and 9 optical cycles, respectively. In both cases, the agreement between the results obtained by using the total ROM (black) and those obtained by integrating the differential ROM (red) is excellent for high positive energies. In Fig. 6.3.1(a) we observe four main peaks. The leftmost peak, just above the groundstate energy ($-0.597$ a.u.), corresponds to excited vibrational states of the ground electronic state (remember that the groundstate was removed from the final wave function before performing the ROM analysis). These states are populated via absorption and emission of a photon: $\Psi_0^{\nu=0} \xrightarrow{+\omega} \Psi' \xrightarrow{-\omega'} \Psi_0^{\nu>0}$, with $\omega \gtrsim \omega'$, corresponding to the population of vibrational states by Raman processes. Since the laser pulse is short, its spectral bandwidth is wide enough to host photons with energies different enough to allow such transitions. The second peak is located around the $2p\sigma_u$ energy at the equilibrium internuclear distance $R_0 = 1.9$ a.u. ($-0.183$ a.u.). The remaining two peaks, at positive energies, correspond to ionization via one- and two-photon absorption from the initial state. The Fourier transform of the laser field is shown in orange in Fig. 6.3.1. The small lobes in its spectrum, due to aliasing, are transferred to the molecular spectrum. We note that the population of the $2p\sigma_u$ state is due to one-photon absorption from the $1s\sigma_g$ state with low-energy photons, which correspond to the left energy tail of the pulse.

The first conclusion is that in the regions where the population is high, around these four peaks, the comparison between differential ROM (red) and total ROM (black) is very good. In the regions between peaks there are tails, whose widths are determined by the choice of $\delta$ in the ROM analysis. As shown here, a correct choice of $\delta$ in both the differential ROM and the total ROM provides very similar spectra. When the pulse is shorter [Fig. 6.3.1(b)], the energy bandwidth of the pulse is larger. In this situation, the $2p\sigma_u$ state and in general the whole spec-trum has a significant population, and the differential and total ROM spectra are indistinguishable.

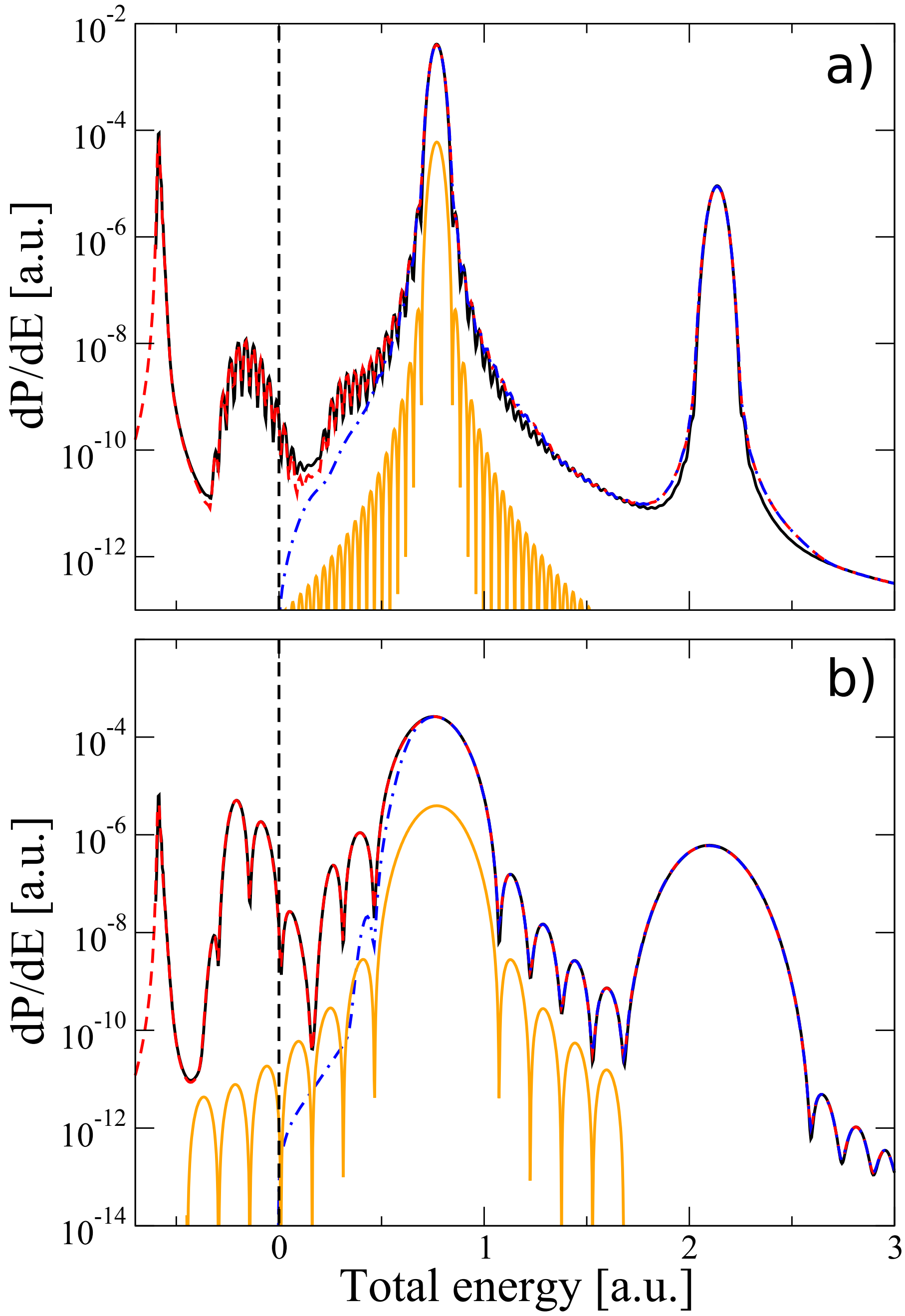

Figure 6.3.1.: Total density probability calculated by using the total ROM (black solid line) using $n = 2$ and $\delta = 0.004$ a.u. or by integrating the differential ROM, either including all channels (red dashed line) or only the ionization channel (blue dash-dotted line), for an XUV pulse with $\Omega = 1.37$ a.u. and $I = 10^{14}$ W/cm$^2$. The bottom orange line represents the Fourier transform of the pulse, $|E(\omega)|^2$, which has been shifted downward for clarity. Total durations: $T = 4$ fs (top) and 1 fs (bottom).

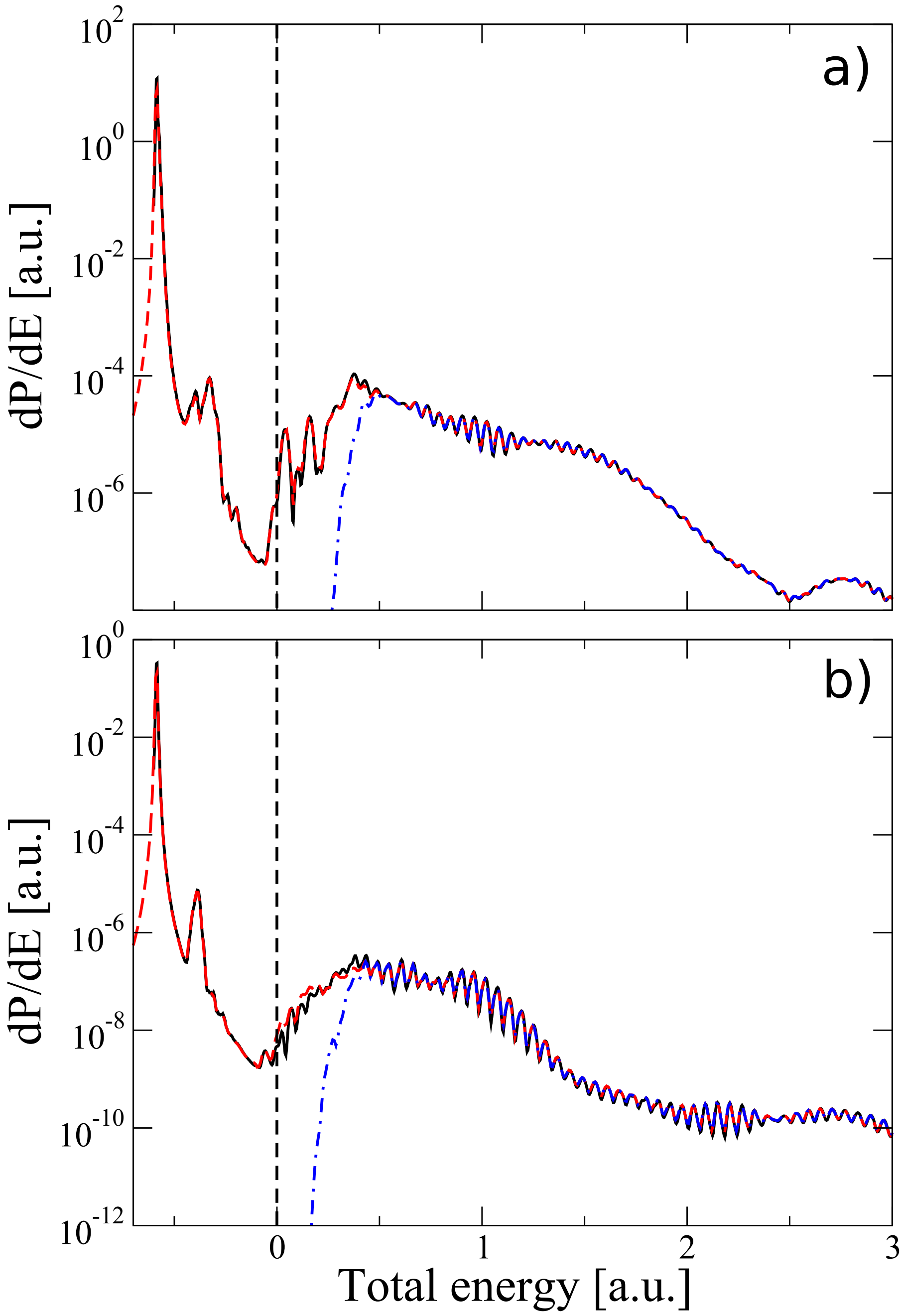

Figure 6.3.2.: Same as in Fig. 6.3.1 for a 800-nm pulse with $T = 16$ fs and $I = 2 \times 10^{14}\mathrm{W/cm^2}$ (top) and $T = 32$ fs and $I = 10^{14}\mathrm{W/cm^2}$ (bottom).

The contribution from ionizing states is shown in blue in Fig. 6.3.1. Ionization takes place whenever an electron is released, and this can happen even for $E_{tot} \approx 0$ for very high internuclear distances. However, most ionization takes place within the Franck-Condon region. The right limit to that region is located around the outer classical turning point at $R = 2.2$ a.u., which corresponds to $1/R = 0.45$ a.u. in the Coulomb explosion curve. For that total energy, ionization becomes the dominant channel, as can be observed in Figs. 6.3.1 and 6.3.2.

In Fig. 6.3.2(a) we show the multiphoton absorption spectrum of an 800-nm pulse with total pulse duration $T = 16$ fs and $I = 2 \times 10^{14}$W/cm$^2$, obtained with the differential ROM (red solid line) and total ROM (black dashed line). The ionization channel is shown again in a blue dash-dotted line. In the ionization region, the spectrum shows a typical ATI shape, with multiple peaks separated by the photon energy $\Omega = 0.057$ a.u. The differential ROM and total ROM calculations are indistinguishable in this region. In the nonionizing region ($E_{tot} < 0.45$ a.u.) the comparison is still excellent.

In Fig. 6.3.2(b) we show results for a longer pulse, with total duration 32 fs, and intensity $10^{14}$ W/cm$^2$. Here the peaks are narrower due to the smaller spectral width of the pulse. Both methods are still indistinguishable in the ionization region and at negative energies. Small differences can be seen in the region just above the ionization threshold, where ejected electrons are very slow and Rydberg states are expected to play some role. At first sight, one might be tempted to attribute these differences to nonadiabatic couplings between Rydberg states (whose potential energy curves lie just below that of the ionization threshold and run almost parallel to it) and the ionization continuum [69]. Such nonadiabatic effects might not be accounted for by the differential ROM when applied just at the end of the pulse. To check if this is the case, we have performed the ROM analysis at different times after the end of the pulse (see Fig. 6.3.3). One can see that the total energy spectrum resulting from the differential ROM is already converged at the end of the pulse. These results point out that the small differences observed near the ionization threshold (between 0 and 0.5 a.u.) are not due to nonadiabatic couplings. In this region, the main process is dissociation through Rydberg states. For these states, the interplay between the chosen value of $\delta$ in the ROM definition, the energy spacing resulting from discretization in the finite box, and the pulse duration determine the accuracy of the results. For instance, as can be seen in Fig. 6.3.2(a), there is no difference between total and differential ROM probabilities when a shorter pulse is used, because in this case the relevant Rydberg states do not need to be resolved individually due to the relatively large laser energy bandwidth. Since the present work focuses on ionization and the proper description of Rydberg states needs a specific treatment (for this work and any other method), it is not further discussed in the present work. In any case, it must be pointed out that, in the energy region where ionization is the dominant channel, the results obtained from the differential and the total ROM are indistinguishable. The same occurs in the region where the lower bound excited states, like the $2p\sigma_u$ one, are populated. These results prove that ROM spectra, in which the total energy is split into electronic and nuclear parts, are accurate and that the normalization of both differential and total ROM is correct.

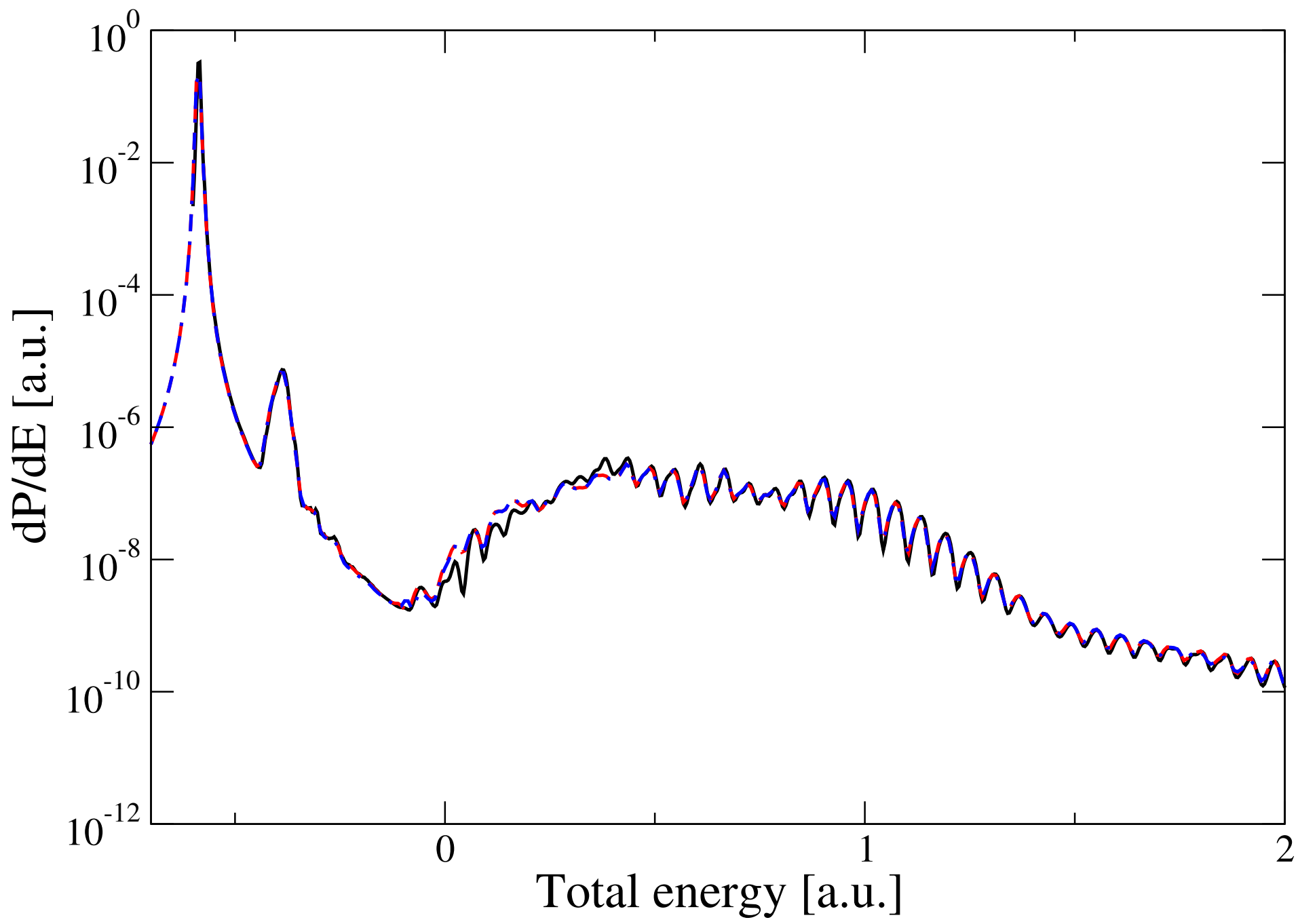

Figure 6.3.3.: Total energy spectra for a 800 nm pulse with $I = 10^{14}\text{W/cm}^2$ and $T = 32$ fs. Total ROM analysis is performed at the end of the pulse (black solid line) and differential ROM analysis is performed at the end of the pulse (red dashed line) and 5.14 fs after the end of the pulse (blue dash-dotted line).

## 6.4 isotopic effects

The nuclear motion has an important role in the shape of the correlated photoionization spectra. Therefore, we expect to find differences when the mass of the molecule changes. This can be tested by using different isotopes of the $H_2^+$ molecule. In Fig. 6.4.1 we show results for $H_2^+$ ($\mu = 918.0$ a.u., black lines) and $D_2^+$ ($\mu = 1835.2$ a.u., red lines) for two different pulse durations: $T = 16$ fs (a) and $T = 32$ fs (b). For both pulse durations, the ionization probability is higher for $H_2^+$ than for $D_2^+$. This is due to the fact that when the nuclei move faster, they access higher internuclear distances, for which the number of photons that is necessary to ionize the molecule is smaller.

The average value of the nuclear energy for each case, $\langle E_N \rangle$, is shown by vertical lines. We observe that the NKE distribution is shifted towards lower energy values in $H_2^+$, which corresponds to higher internuclear distances. This is also due to the faster nuclear motion in $H_2^+$. When the pulse is longer [Fig. 6.4.1(b)], the nuclei have time to reach even larger internuclear distances, which implies a shift towards lower NKE values for both isotopes.

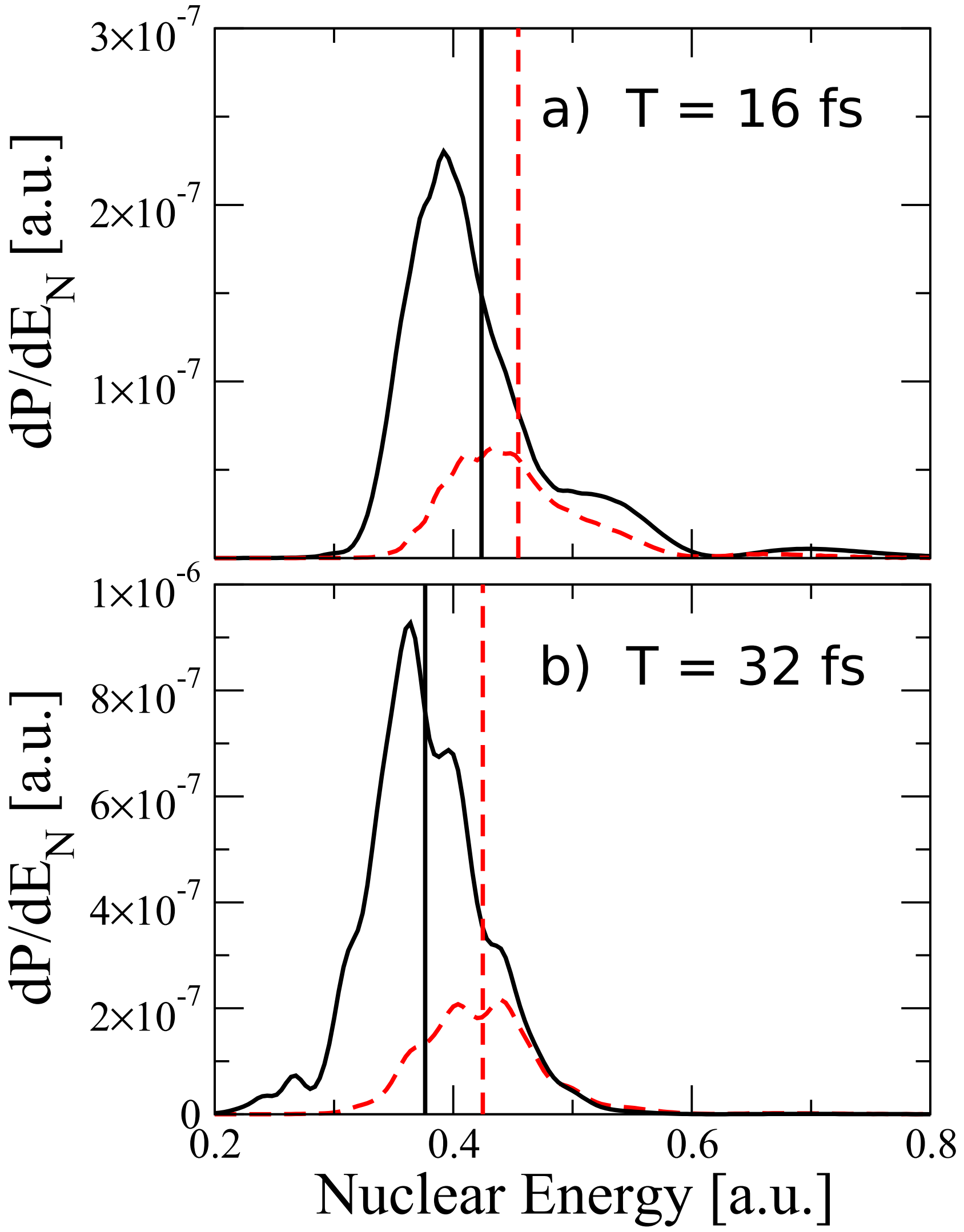

Figure 6.4.1.: NKE in the ionization channel for $H_2^+$ (black solid lines) and $D_2^+$ (red dashed lines) for a laser pulse of 800 nm and $10^{14}$W/cm$^2$, with total duration $T = 16$ fs (a) and $T = 32 fs$ (b). The vertical lines show the value of $E_N$ for each case.

# 7

# TRANSITION REGIME BETWEEN MULTIPHOTON IONIZATION AND TUNNELING IONIZATION

In this Chapter, we present a theoretical study of strong field ionization of $H_2^+$ by looking at the CKE spectra of ionization. The results show two different ionization mechanisms, tunnel and multiphoton ionization, and by looking at the correlated spectra we observe that for the two mechanisms the sharing of the energy between nuclei and electrons is not the same.

The results were obtained with the ROM in a 1+1D calculation of the $H_2^+$ molecule. We have used a box with $|z| < 3000$ a.u. and $R < 30$ a.u., with uniform grid spacings of $\Delta z = 0.1$ a.u. and $\Delta R = 0.05$ a.u. The propagation was performed by using the Crank-Nicolson split-operator method with $\Delta t = 0.02$ a.u. The initial state is the groundstate of the $H_2^+$ molecule.

In regard to the ROM, we have chosen the values of $n_e = n_N = 2$, $\delta_e = 0.004$ and $\delta_N = 0.02$. The ROM analysis is performed at the end of the pulse to a wavefunction for which the contribution to the groundstate was removed. We have checked that the results are converged.

The results that are presented in this Chapter were published in [31].

## 7.1 CORRELATED SPECTRA IN STRONG FIELD IONIZATION

Figure 7.1.1 shows the correlated photoelectron-nuclear kinetic energy spectra resulting from four different pulses with wavelength ($\lambda$), duration ($T$), and intensity ($I$). They correspond, respectively, to values of the Keldysh parameter $\gamma = \sqrt{I_p/2U_p} = 3.2,\ 1.6,\ 1.6,\ 1.1$ ($I_p$ is the ionization potential at the equilibrium internuclear distance), which cover different ionization regimes, from the multiphoton regime [Fig. 7.1.1(a)] to the tunneling regime [Fig. 7.1.1(d)]. In Fig. 7.1.1(a) one can see energy conservation lines satisfying the formula $N\omega = E_e + E_N + U_p + D_{2H^+}$, where $D_{2H^+}$ is the energy required to produce two protons at infinite internuclear distance, and $N$ indicates the number of ab-sorbed photons. The appearance of these lines indicates that the excess photon energy is shared between the ejected electron and protons. This energy sharing is only efficient within the Franck-Condon region, i.e., in the interval of nuclear-kinetic energies 0.25-0.55 a.u. This is the usual behaviour observed in weak-field ionization of diatomic molecules [70–72], and also, as pointed out very recently, in multiphoton above threshold ionization of $H_2^+$ [20]. As one increases the intensity without changing the photon energy [Fig. 7.1.1(b)], the calculated 2D spectrum exhibits a more complex structure. In this case, $\gamma$ is closer to 1, so the tunneling process is expected to occur. In addition to the diagonal lines observed in the previous case, one can observe horizontal lines approximately separated by $\omega$ indicating that the energy taken by the electrons from the field is not shared with the nuclei. Horizontal lines are the dominant pattern for NKE between 0.2 and 0.3 a.u., and they are visible up to photoelectron energies close to $10Up_p$. For NKE around 0.4 a.u., there is no clear dominant pattern. The spectrum shown in Fig. 7.1.1(c) for the same value of $\gamma$ but obtained with an IR pulse that is four times less intense and contains photons with half the energy is dominated by horizontal lines. The same occurs if one considers even higher intensities [Fig. 7.1.1(d)]. In the latter case, as the Keldysh parameter is close to 1 and $U_p$ is about eight times the photon energy at maximum intensity, the ionization mechanism is clearly dominated by nonperturbative effects.

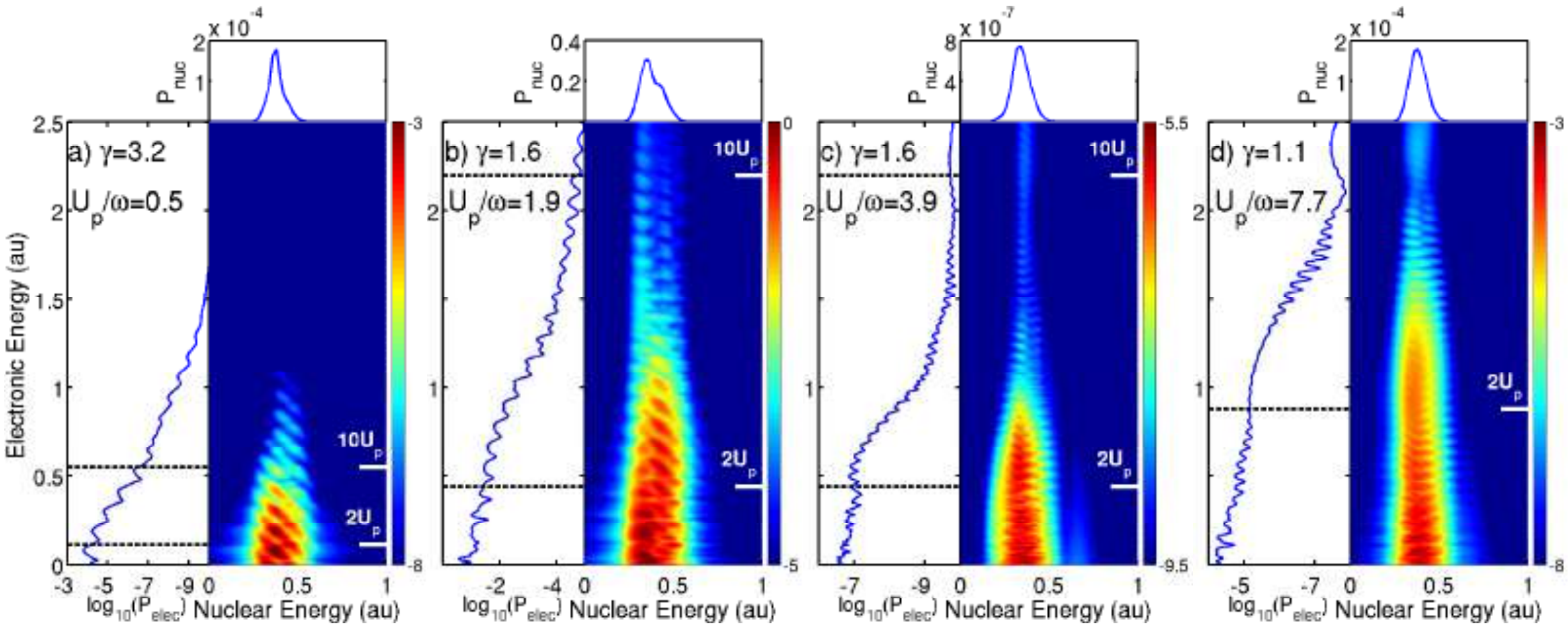

Figure 7.1.1.: Density plots for the correlated photoelectron and nuclear-kinetic energy spectra resulting from $H_2^+$ photo-ionization by using the following pulses: (a) $\lambda = 400$ nm, $T = 16$ fs, and $I = 10^{14}\text{W/cm}^2$, (b) $\lambda = 400$ nm, $T = 16$ fs, and $I = 4 \times 10^{14}\text{W/cm}^2$, (c) $\lambda = 800$ nm, $T = 32$ fs, and $I = 10^{14}\text{W/cm}^2$ and (d) $\lambda = 800$ nm, $T = 16$ fs, and $I = 2 \times 10^{14}\text{W/cm}^2$. The corresponding projections (singly differential probabilities) in electronic energy ($P_{elec}$) and nuclear energy ($P_{nuc}$) are shown on the left and on top of each panel. All panels include the values of the Keldysh parameter $\gamma$, the ratio between the ponderomotive energy and the photon energy $U_p/\omega$, and two and ten times $U_p$.

## 7.2 DYNAMICAL PICTURE

To understand how the different structures build up in the 2D spectra, we have evaluated in the velocity gauge the ionization probabilities at the zeros of the vector potential $A(t)$ for the case $\lambda = 400$ nm, $T = 16$ fs, and $I = 10^{14}$W/cm$^2$ [Fig. 7.1.1(b)]. In this gauge, the laser-molecule interaction vanishes at the zeros of $A$, and consequently, the Hamiltonian is identical to that of a free molecule. Also, the kinematic momentum of the electron coincides with its canonical momentum, which is convenient to compare CKE spectra obtained at different times with that obtained at the end of the simulation. Figure 7.2.1(a) shows the CKE spectrum after the first few cycles. During this time interval, the intensity does not reach a large enough value to induce tunneling, and consequently, the resulting spectra is very similar to that obtained in the pure multiphoton regime. When the peak intensity is reached, one can see the appearance of horizontal fringes below $2U_p$ in the region of NKE$\sim$ 0.25 a.u. [see Fig. 7.2.1(b)]. This is the signature of tunneling electrons directly escaping from the molecule (DE). A cycle later, the CKE spectrum shows the appearance of horizontal fringes up to $10U_p$ [see Fig. 7.2.1(c)], which is the signature of tunneling electrons driven back by the field and subsequently rescattered (RE). According to the three- step model [7], these electrons are expected to appear $\sim 0.65\,(2\pi/\omega)$ after the DE. The subsequent evolution of the system repeatedly generates similar patterns, which interfere with each other and thus lead to the complex 2D spectrum of Fig. 7.1.1(b).

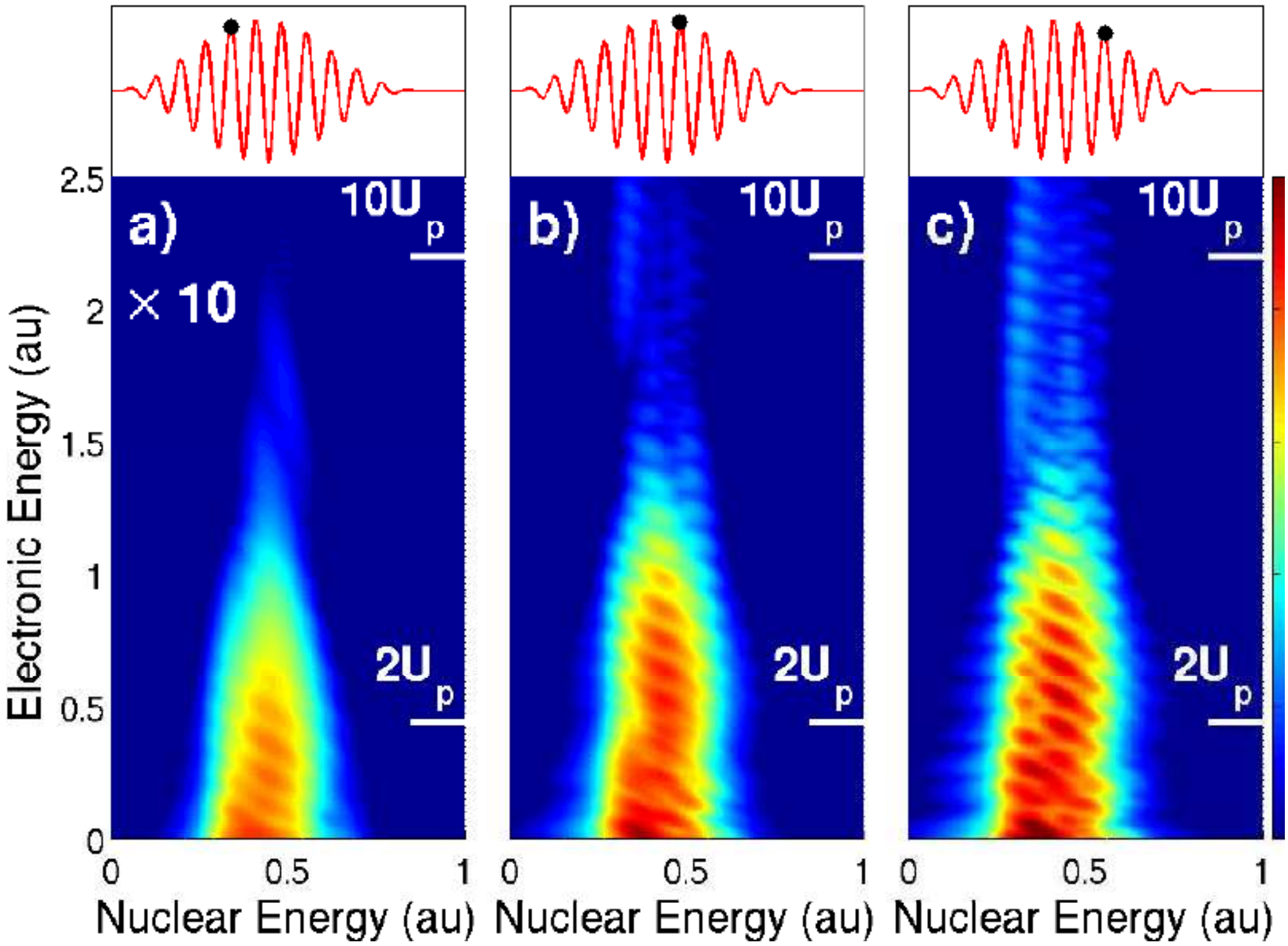

Figure 7.2.1.: Time evolution of the density plots for the correlated photoelectron and nuclear-kinetic energy spectra resulting from $H_2^+$ photoionization by using a pulse with $\lambda = 400$ nm, $T = 16$ fs, and $I = 4 \times 10^{14}\text{W}/\text{cm}^2$. The time values are indicated by black dots on the electric field displayed on top of each panel.

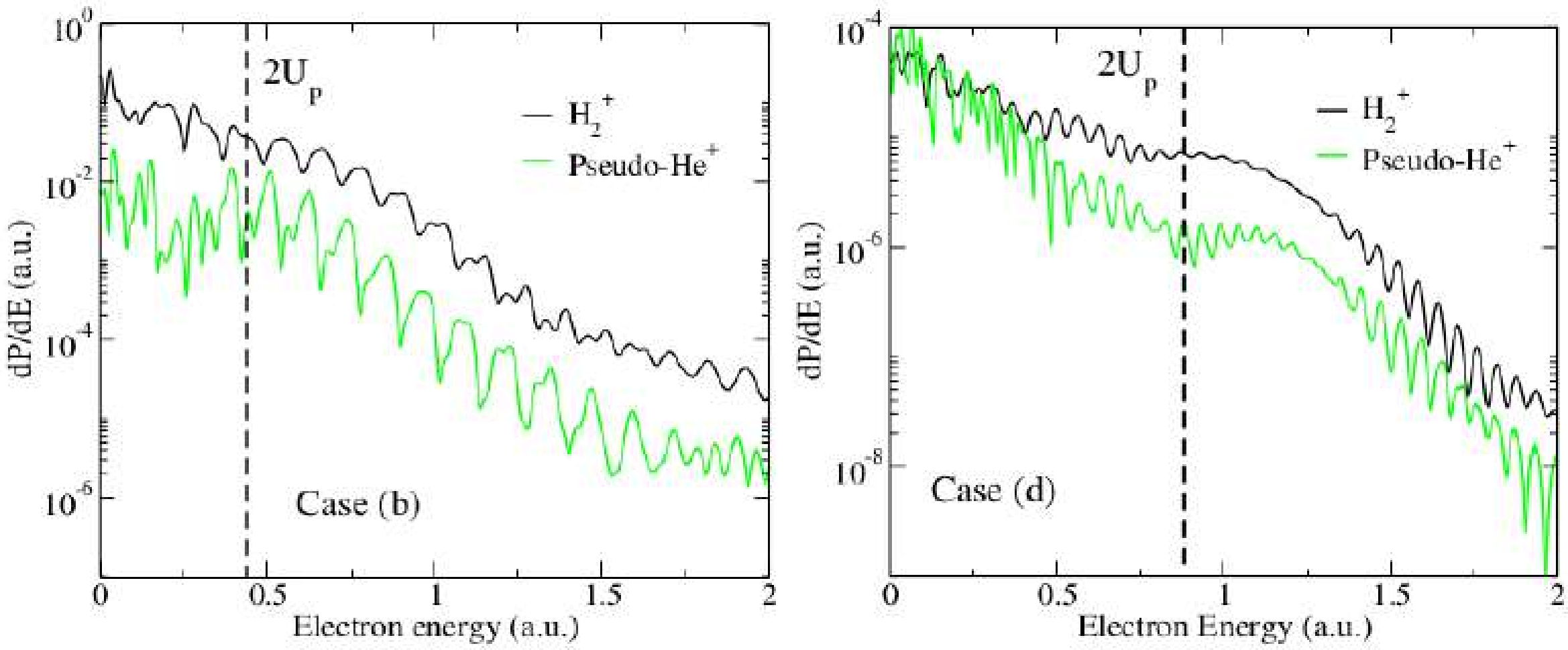

Figure 7.2.2.: Photoelectron kinetic energy spectra for $H_2^+$ (black) and a pseudo-$He^+$ atom (green) for laser parameters corresponding to Fig.7.1.1(b) (left) and Fig.7.1.1(d) (right) of the paper. The probabilities for pseudo-$He^+$ are scaled.

That the origin of the horizontal fringes is tunneling ionization is confirmed by calculations performed on a one-dimensional $He^+$ system represented by a soft Coulomb potential that provides the same ionization energy $I_p$ as for $H_2^+$. The resulting $He^+$ electron kinetic energy spectra, Fig. 7.2.2, resemble those of $H_2^+$ for electron energies well above $2U_p$, i.e., in the region where ionization comes almost exclusively from tunneling electrons. At low electron kinetic energies, interferences between multiphoton and tunneling ionization leads to complex patterns that are different in $He^+$ and $H_2^+$ due to the molecular character of multiphoton ionization.

An interesting feature of the CKE spectrum shown in Fig. 7.1.1(b) is that the NKE distribution exhibits some structure and is significantly wider than that expected from the Franck-Condon principle (which is close to the NKE distributions observed in the other three cases). This can be explained with the help of Fig. 7.2.3, which shows the variation with time of ionization and vibrational-excitation probabilities, as well as that of the average value of the internuclear distance, for the 400 nm, $4 \times 10^{14}$W/cm$^2$, and 800 nm, $10^{14}$W/cm$^2$ cases [Figs. 7.1.1(b) and 7.1.1(c)]. As can be seen, for 400 nm, there is a large degree of vibrational excitation, which is the consequence of Rabi oscillations between the $1s\sigma_g$ and $2p\sigma_u$ states following a stepladder mechanism similar to that previously described in $H_2^+$ [22]. The coherent population of the different vibrational states launches a nuclear wave packet in the $1s\sigma_g$ state that moves considerably during the pulse duration. As a consequence of this, the average internuclear distance increases from $\sim$ 1.9 a.u. up to $\sim$ 2.7 a.u. at the end of the pulse. Thus, when ionization occurs in the second half of the pulse, the available NKE is smaller because the internuclear distance is larger. This leads to the broadening of the NKE distribution when tunneling electrons appear. For 800 nm (and all the other cases), the effect is negligible and the NKE distribution follows a typical Franck-Condon behaviour.

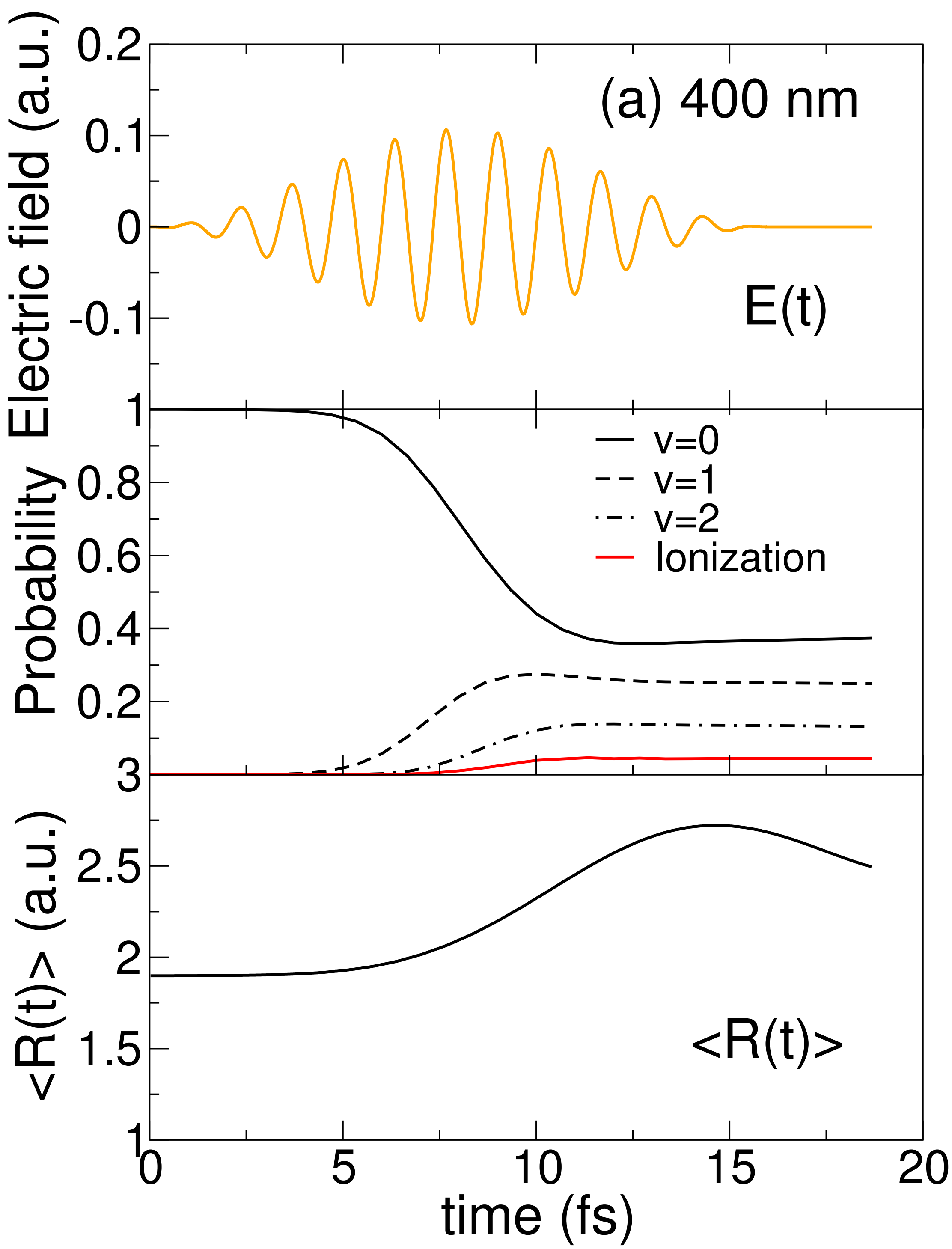

Figure 7.2.3.: Time evolution of different observables for the (400 nm, $4 \times 10^{14}$W/cm$^2$) pulse. Electric field (top), population of the lowest vibrational states and ionization probability (middle figure), and mean value of the internuclear distance $\langle R(t) \rangle$ (bottom).

# 8

# RESOLVENT OPERATOR METHOD ON A FULL DIMENSIONAL CALCULATION ON $H_2^+$

In this Chapter, we present the results obtained with the ROM in a 3D calculation of the $H_2^+$ molecule. A major disadvantage of the 1+1D calculations is the absence of angular distributions, which can be obtained in a full 3D calculation. The KER, CKE and $CAK_N$ spectra is calculated for pulses in the XUV domain. In particular, we have performed several calculations for three different frequencies, $\omega = 0.4,\ 0.6,\ 0.8$ a.u. and different pulse durations ranging from $0.5$ to $2.5$ fs. We have considered $T = 0.76$, $1.14$ and $2.5$ fs for $\omega = 0.8$ and $T = 0.5$, $1.0$ and $2.5$ fs for $\omega = 0.6$ and $0.4$. The intensity of the laser pulse is the same for all calculations ($10^{12}$ W/cm$^2$) which is a rather low intensity. Indeed, the Keldysh parameter is much larger than $1$ ($\gamma \gg 1$) and this indicate us that we are working in the multiphoton ionization regime.

We have solved the TDSE in a numerical box with $|z| < 100$, $\rho < 50$ and $R < 30$ with grid spacings of $\Delta z = 0.1$, $\Delta\rho = 0.075$ and $\Delta R = 0.05$ at the center of the grid. These grid spacings gradually increase as we go to the limits of the box. For the electronic propagation, we use a time step $\Delta t_{elec} = 0.011$ a.u., and for the nuclear propagation, we use a time step $10$ times larger ($\Delta t_{nuc} = 0.11$ a.u.). The convergence of these parameters were checked. The initial state is the groundstate of the $H_2^+$ molecule.

In regard to the ROM, we have chosen the values of $n_e = n_N = 2$, $\delta_e = 0.02$ and $\delta_N = 0.01$. For the angular resolution we have used $\Delta\theta = 4^\circ$. The ROM analysis is performed at the end of the pulse. We have checked that the results are converged by propagating $1$ fs after the end of the pulse. The results are practically the same.

The results that are presented in this Chapter were published in [29].

## 8.1 BORN-OPPENHEIMER CURVES

To study photoionization in a molecule we must first look at the BO diagrams (see Fig. 8.1.1). For the three different photon energies considered, one can expect that, in the monochromatic limit ($T \to \infty$), one-photon ionization is a forbidden process since the photon energy is smaller than the vertical ionization potential. In this limit, two-photon ionization is expected to be the dominant channel for $\omega = 0.8$ and $0.6$ a.u., and three-photon ionization is expected to be dominant for $\omega = 0.4$ a.u.

When the laser pulse duration is on the sub-fs scale, this picture no longer holds due to the large bandwidth of these pulses. When the bandwidth of the pulse is large enough new ionization channels are opened. Indeed, Fig. 8.1.1 shows that the bandwidth of the shortest pulses are of the order of the central frequency of the corresponding pulse $\omega$. Thus, for $\omega = 0.8$ a.u. [Fig. 8.1.1(a)] and a duration $T = 0.76$ fs, the one-photon absorption channel is also open in the Franck-Condon region since the electronic continuum can be reached in a vertical transition from the ground state by absorption of a photon lying in the large-$R$ region of the Franck-Condon zone. Similarly, for $\omega = 0.4$ a.u. [Fig. 8.1.1(c)] and a duration $T = 0.5$ fs, two-photon absorption is also possible. As the probability of absorbing $N - 1$ photons is much higher than that of absorbing $N$ photons (in perturbation theory), one can expect that the $(N - 1)$-photon ionization channel will be comparable to or even dominate over the $N$-photon ionization channel.

## 8.2 KINETIC-ENERGY-RELEASE SPECTRA

In Fig. 8.2.1, we show the KER spectra for all the cases considered in this Chapter. According to the Franck-Condon principle, we expect that, in all cases, the spectra will be centered at $E_N \approx 1/R_{eq} \approx 0.5$ a.u., which is the value of the repulsive Coulomb potential-energy curve associated with the ionization limit at the equilibrium internuclear distance $R_{eq}$. However, if one examines the results more closely, deviations from the expected results can be noticed.

In Fig. 8.2.1(b), for $\omega = 0.6$ a.u., we compare our results with those available in the literature [2]. The agreement is good. In this case, two-photon ionization is the dominant process for the three pulse durations considered in our calculations. For the shorter pulse, $T = 0.5$ fs, the distribution is very similar to that resulting from the Franck-Condon overlaps between the initial vibrational state and the final dissociative states (see [2]), which proves that, for such a short pulse duration, two-photon absorption is a near-vertical transition. One can see, however, that, as pulse duration increases, the maximum of the ionization probability shifts to higher nuclear energies, thus departing from the Franck-Condon behavior. This is a consequence of the variation of the one- and two-photon dipole transition amplitudes with internuclear distance and the fact that, as pulse duration increases, a resonant one-photon transition populates the $2p\sigma_u$ state at smaller $R$, thus generating a nuclear wave packet that can significantly move before the second photon is absorbed. The combination of these effects destroys the picture of a vertical two-photon vertical transition from the ground state. Similar effects explain the shift in the probability maximum for $\omega = 0.4$ a.u. [see Fig. 8.2.1(c)].

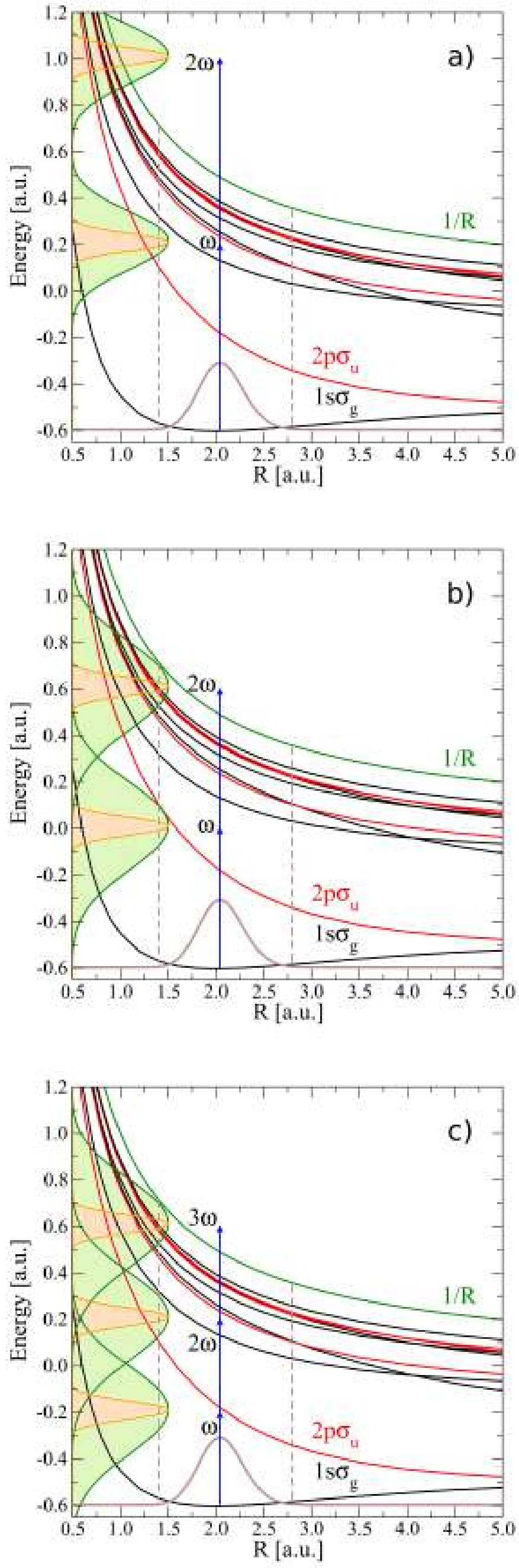

Figure 8.1.1.: Born-Oppenheimer potential energy curves for the $H_2^+$ molecule. The black curves correspond to states of $\sigma_g$ symmetry, and the red ones show those of $\sigma_u$ symmetry. The blue arrows represent a vertical transition from the ground state to the ionization continuum. (a) shows arrows corresponding to the photon energy $\omega = 0.8$ a.u. and the Fourier transforms of pulses with durations $T = 2.5$ fs and $T = 0.76$ fs, in green and orange, respectively, shifted by the energy of the photon. (b) and (c) are similar to (a) but are for a photon energy $\omega = 0.6$ a.u. and $\omega = 0.4$ a.u., respectively. In this case, the Fourier transform corresponds to pulses of duration $T = 2.5$ fs and $T = 0.5$ fs. The Franck-Condon region lies in between the vertical dashed lines.

The case shown in Fig. 8.2.1(a) for $\omega$ =0.8 a.u. is more interesting. First, we notice that the total probability is larger for the shorter than for the longer pulses, in contrast to the behavior observed in the two cases discussed in the previous paragraph. The reason for this behavior is that, for durations $T$ =0.76 fs and $T$ =1.14 fs, the one-photon ionization channel is open, and the corresponding ionization probability is much larger than that of the two-photon ionization channel. For the longest pulse duration, $T$ =2.5 fs, the one-photon ionization channel is closed, so that the total ionization yield follows a pattern closer to that discussed for $\omega = 0.6$ a.u. For the shortest pulse ($T = 0.76$ fs), the signature of the one-photon ionization process is the maximum at $E_N \approx 0.4$ a.u., while that of the two-photon ionization process is the shoulder at $E_N \approx 0.5$ a.u. The lower value of $E_N$ in the former case is due to the fact that, for this pulse duration, reaching the ionization continuum by absorption of a single photon is possible only at the larger values of $R$ within the Franck-Condon region. This is the only region where the ionization potential is smaller than the energy of the higher spectral components of the pulse [see Fig. 8.1.1(a)]. For the intermediate pulse duration ($T = 1.14$ fs), one- and two-photon ionization processes cannot be so easily identified. In this case, as we will see later, the analysis of the CKE and $\mathrm{CAK_N}$ spectra will provide a much more complete picture.

## 8.3 CORRELATED SPECTRA

In this section, we present our results for the CKE and the $\mathrm{CAK_N}$ spectra, which provide a more detailed information of the ionization process.

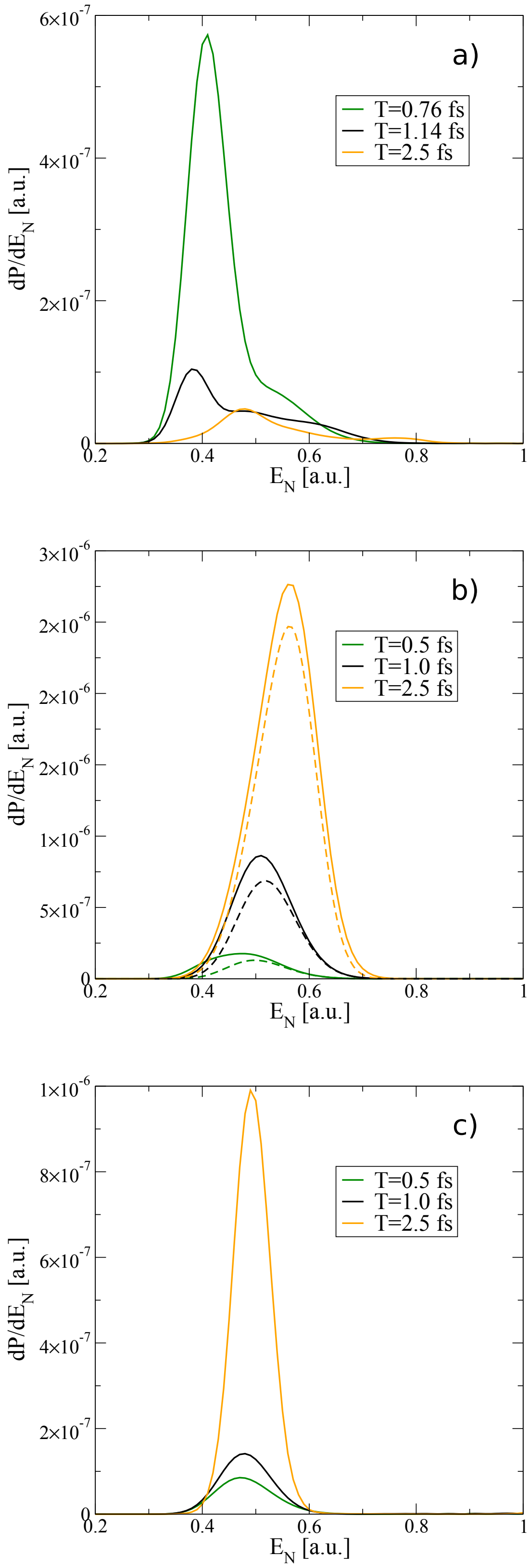

Figure 8.2.1.: KER spectra resulting from pulses with central frequencies (a) $\omega$ =0.8 a.u., (b) $\omega$ =0.6 a.u., and (c) $\omega$ =0.4 a.u. The pulse duration is indicated in each panel. In (b) we also show the results from [2] (dashed lines).

### 8.3.1 *Correlated kinetic-energy spectra*

We show in Fig. 8.3.1 the calculated CKE spectra for all the cases under study. Note that in the monochromatic limit (infinite pulse duration) we expect to see energy conservation lines [20, 28, 31, 73] satisfying the relationship $N\omega = E_e + E_N + D_{2H^+}$, where $D_{2H^+} = -E_0 = 0.597$ a.u. is the threshold energy required to produce two protons at infinite internuclear distance, $E_0$ is the ground-state energy, $E_e$ and $E_N$ are the electronic and nuclear energies, respectively, and $N$ is the number of absorbed photons. The expected energy-conservation lines are shown as dashed magenta lines in Fig. 8.3.1.

For the longest pulse, $T = 2.5$ fs, with central frequencies $\omega = 0.8$ and 0.6 a.u., one can clearly observe a strong signal along the energy-conservation line for $N = 2$ [right panels in Figs. 8.3.1(a) and 8.3.1(b)]. For $\omega$ =0.4 a.u. [right panel in Fig. 8.3.1(c)], the bright spot in the spectrum at $(E_N, E_e) \approx (0.5, 0.1)$ a.u. is explained by the same energy-conservation law with $N$ =3. In contrast, for the other pulse durations, it is harder to see a clear signature of energy-conservation lines. In particular, for a central frequency $\omega$ =0.4 a.u. and pulse durations $T = 0.5$ fs and $T = 1.0$ fs, the bright spot appearing at $E_N \approx 0.5$ a.u. is no longer present at the expected location, which is an indication of a two-photon process rather than a three-photon one. Also, for a central frequency $\omega = 0.8$ a.u. [Fig. 8.3.1(a)], one can clearly observe the transition from two-photon ionization to one-photon ionization as the pulse duration decreases, and for $\omega = 0.6$ a.u. [Fig. 8.3.1(b)], one can see the shift of the maximum to lower nuclear energy (see discussion in the previous section). Indeed, the region of low electron energies dominates the spectrum for $T = 0.76$ fs and $T = 1.14$ fs, which is an indication of one-photon ionization. For the longest pulse duration, however, the one-photon ionization channel is closed, and the ionization threshold can only be reached by absorption of two photons.

Finally, in Fig. 8.3.2 we present the comparison between the 3D and the 1+1D calculations. The results agree qualitatively but differ in the absolute value.

### 8.3.2 *Correlated angular and nuclear kinetic-energy spectra*

We show in Fig. 8.3.3 the correlated angular and nuclear kinetic-energy spectra ($\text{CAK}_\text{N}$) spectra. To interpret these results, one must take into account the fact that absorption of an odd (even) number of photons results in a combination of partial waves involving spherical harmonics $Y_l^{m=0} \propto P_l^{m=0}(\cos\theta)$ with odd (even) $l$. Thus, one can expect that the nodal structure of the corresponding Leg-endre polynomials will be imprinted in the $CAK_N$ spectra. Although $l$ is not a good quantum number for $H_2^+$ and therefore the photoionization selection rules are not the same as for atomic systems, one can still see reminiscences of the latter in the $CAK_N$ spectra because the ground state of $H_2^+$ has a predominant $l = 0$ character.

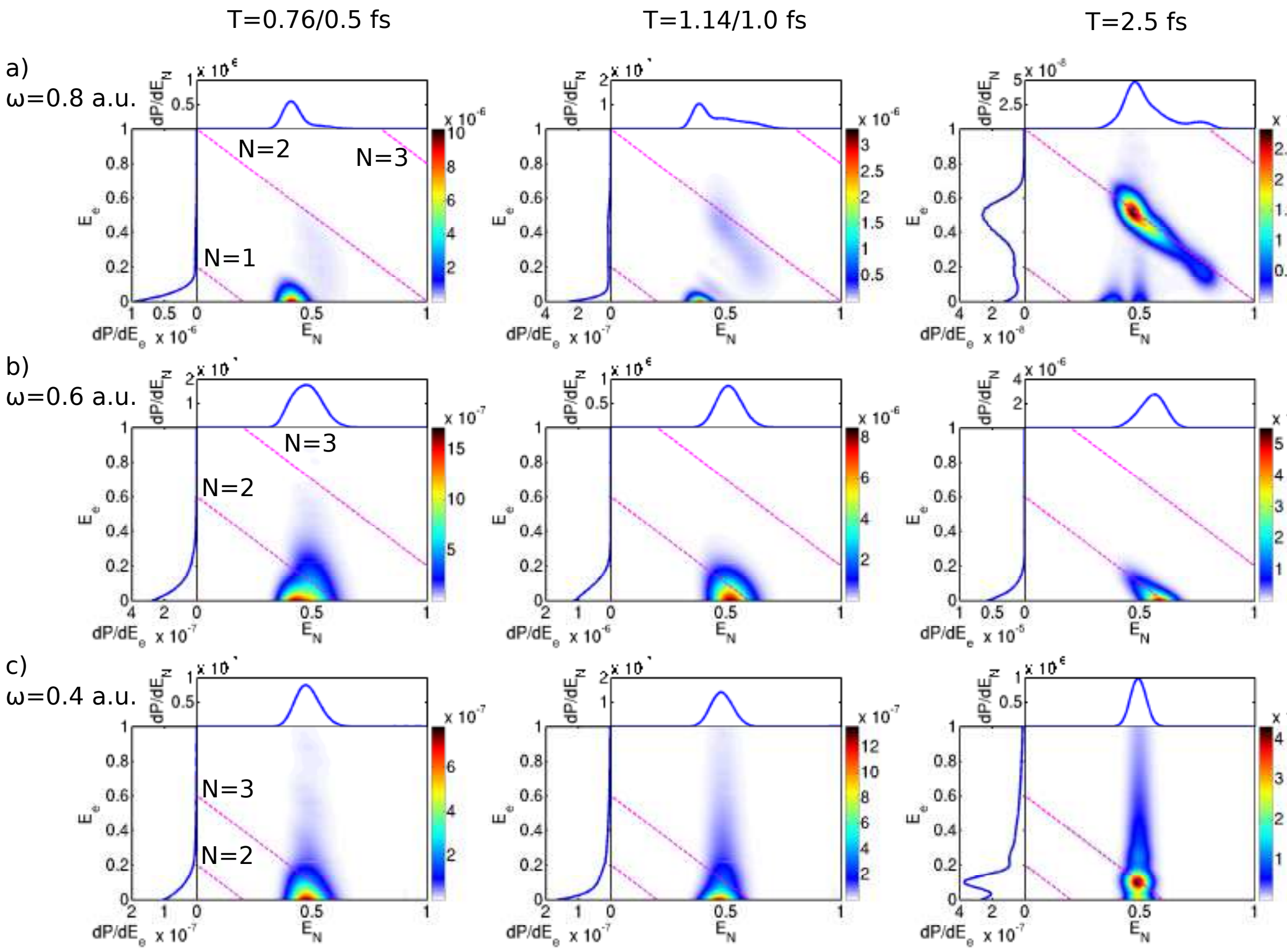

Figure 8.3.1.: CKE for different pulses. The projections (singly differential probabilities) are shown at the left and at the top of each CKE spectrum. (a) Central frequency $\omega = 0.8$ a.u. and pulse durations $T = 0.76$, 1.14, and 2.5 fs (left, middle, and right panels, respectively). (b) and (c) Central frequencies $\omega = 0.6$ a.u. and $\omega$ =0.4 a.u., respectively, and pulse durations $T = 0.5$, 1.0, and 2.5 fs (left, middle, and right panels, respectively). Energy-conservation lines for absorption of $N$ photons are indicated by dashed magenta lines. All the results and scales are in atomic units.

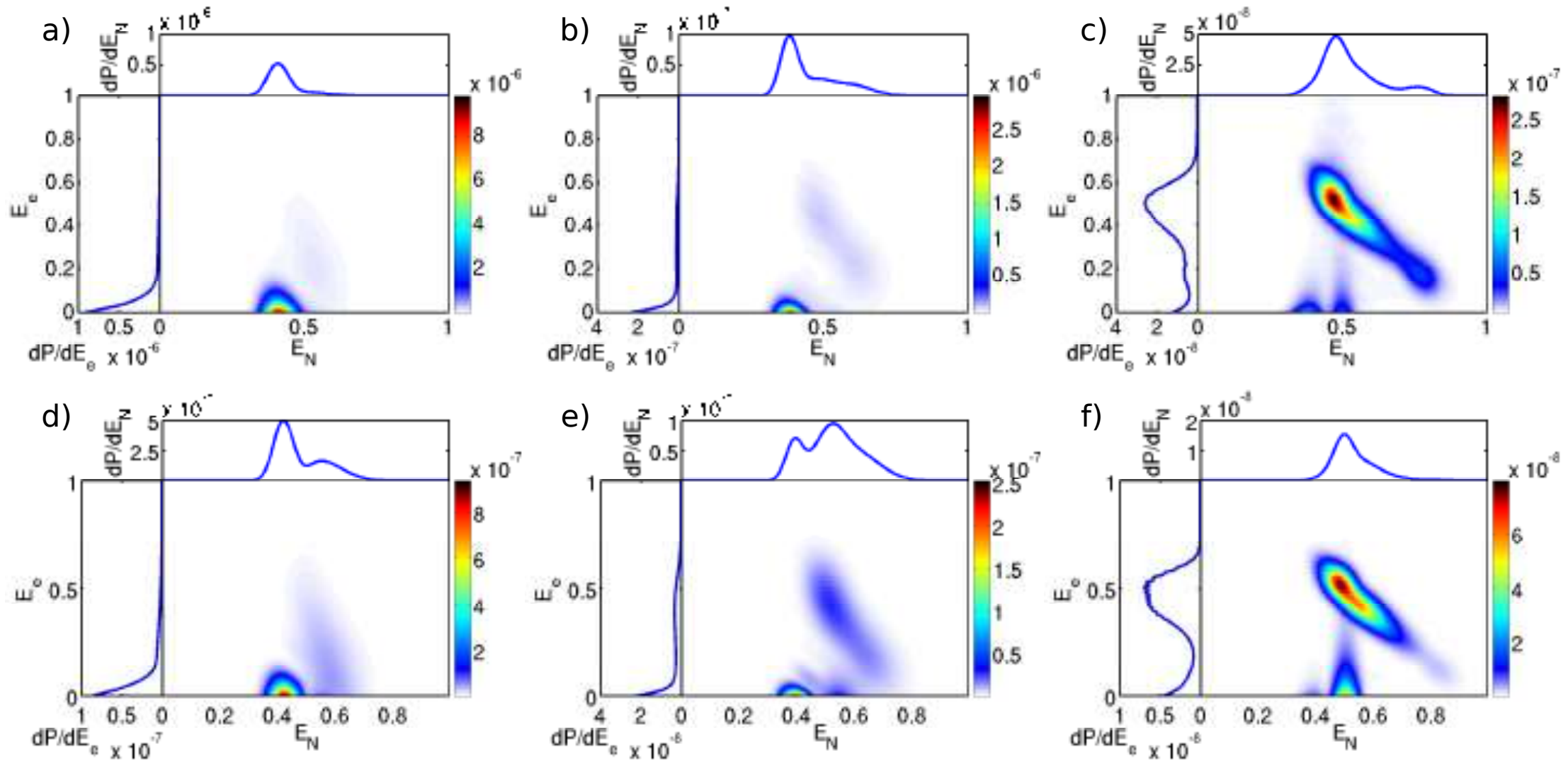

Figure 8.3.2.: CKE for different pulses. The projections (singly differential probabilities) are shown at the left and at the top of each CKE spectrum. (a,b,c) Central frequency $\omega = 0.8$ a.u. and pulse durations $T = 0.76$, 1.14, and 2.5 fs (left, middle, and right panels, respectively) for the 3D calculations. (d,e,f) Central frequency $\omega = 0.8$ a.u. and pulse durations $T = 0.76$, 1.14, and 2.5 fs (left, middle, and right panels, respectively) for the 1+1D calculations.

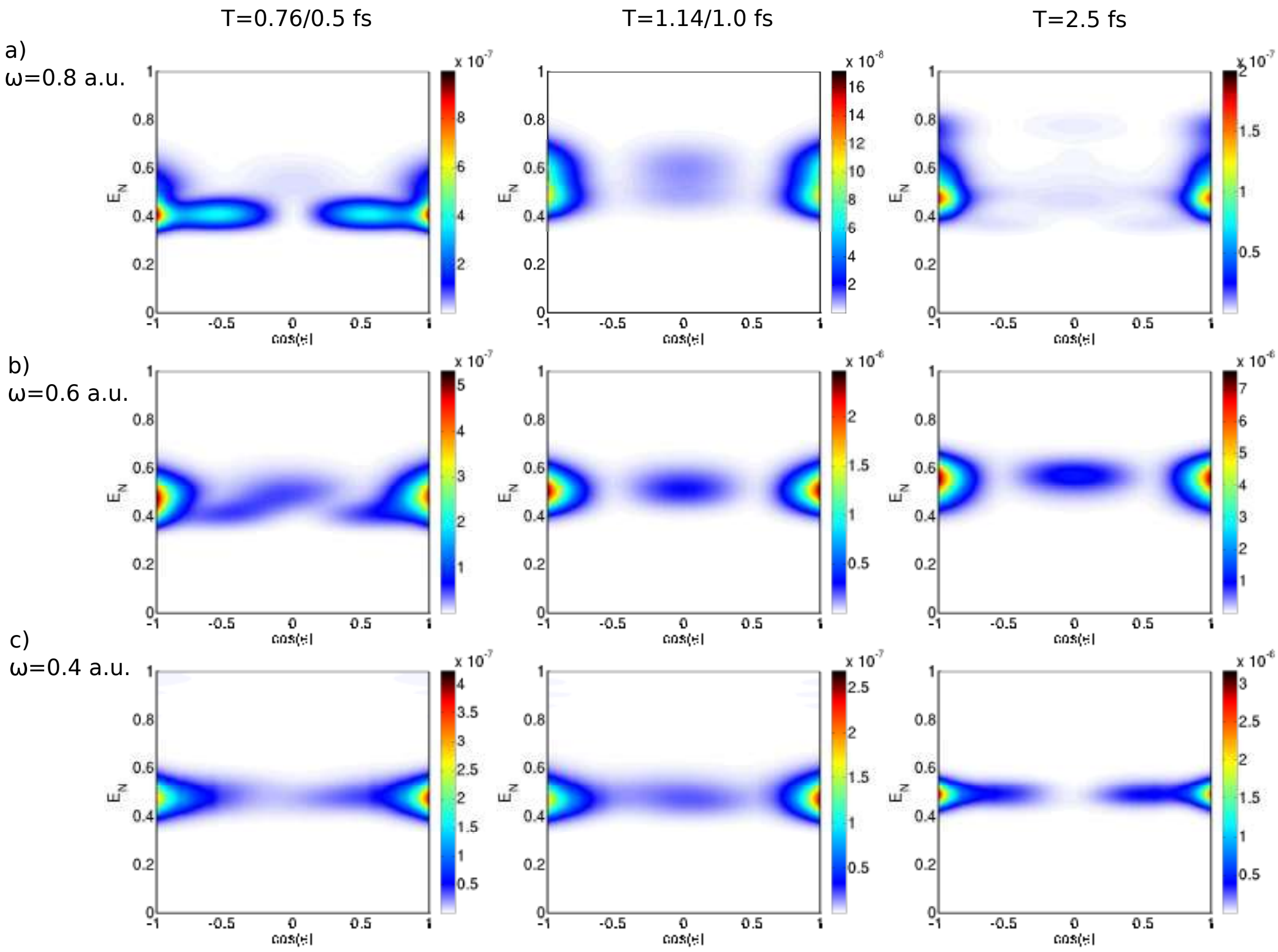

Figure 8.3.3.: Same as in Fig. 8.3.1, but for the $CAK_N$ spectra. All the results and scales are in atomic units.

Fig. 8.3.3(b) ($\omega$ =0.6 a.u.) shows that, for the pulse durations $T = 1.0$ fs and $T = 2.5$ fs, the $CAK_N$ spectra exhibit nodes at $\cos\theta \approx \pm 0.5$, which is the signature of a $d$ wave (the nodes of the $P_2^{m=0}$ Legendre polynomial strictly appear at $\cos\theta = \pm 0.57735$). For the shortest pulse ($T = 0.5$ fs), the angular distribution is not symmetric due to the fact that, for such a short duration, the effect of the carrier-envelope phase (CEP) is not negligible. In any case, the presence of the two nodes at $\cos\theta \approx \pm 0.5$ confirms that the spectra for a central frequency $\omega = 0.6$ a.u. are almost entirely due to a two-photon transition.

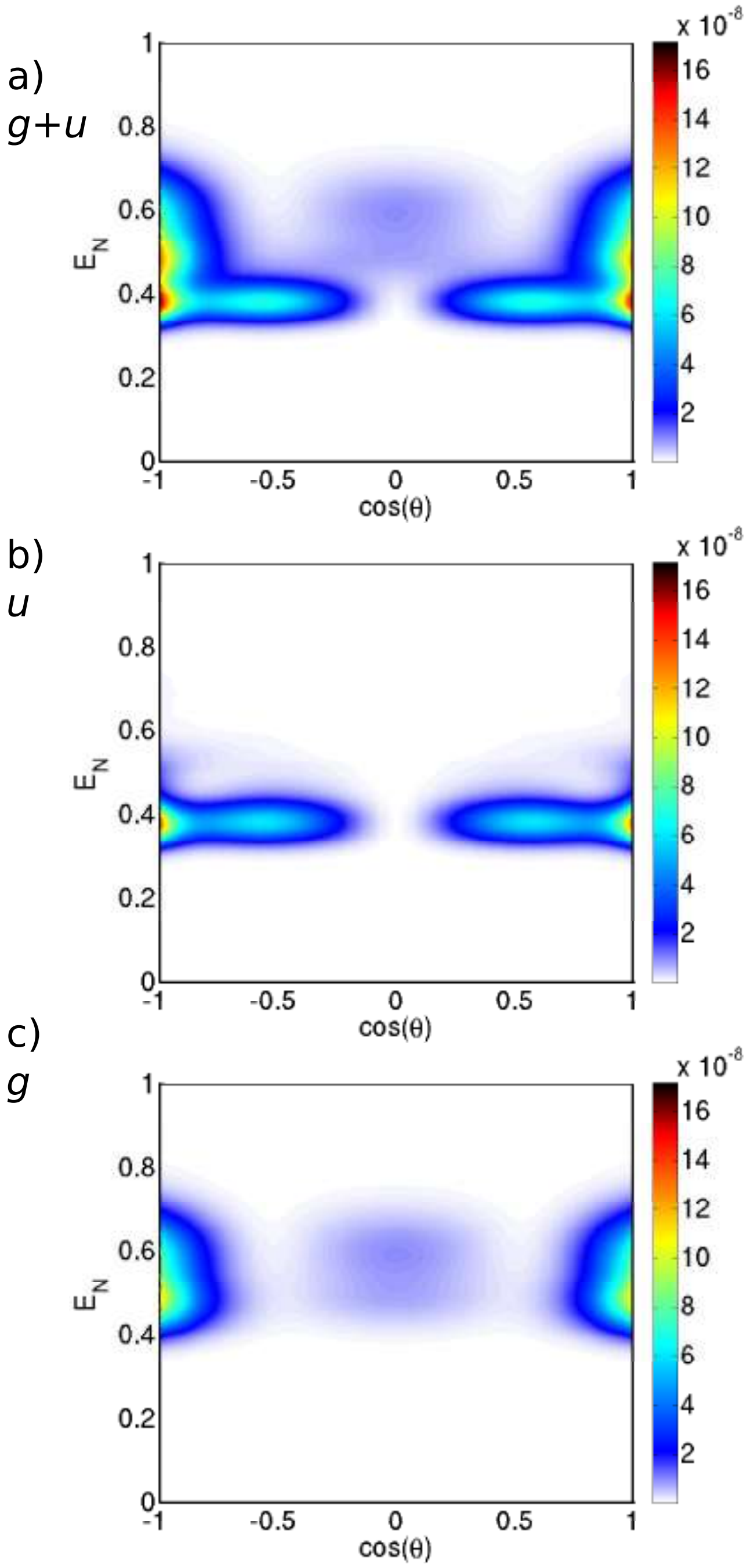

Figure 8.3.4.: Contributions to the $CAK_N$ spectrum from different molecular symmetries for a pulse with central frequency $\omega$ =0.8 a.u. and duration $T$ =1.14 fs. (a) The total $CAK_N$ spectrum and the (b) $u$ and (c) $g$ contributions. All the results and scales are in atomic units.

In Fig. 8.3.3(c) ($\omega = 0.4$ a.u.), the $\text{CAK}_\text{N}$ spectra look quite different depending on the pulse duration. For the shortest pulses, the angular distributions resemble those in Fig. 8.3.3(b) (signature of the $d$ wave). As discussed above, this is due to the fact that, as a result of the large bandwidth, two-photon ionization is possible and is the dominant process (see Fig. 8.1.1). We were already driven to this conclusion, although less clearly, by looking at the corresponding CKE spectra. However, such information could not be inferred at all by looking at the KER spectra. By increasing the pulse duration while keeping constant the central frequency [Fig. 8.3.3(c), right], we observe a nodal plane at $\cos\theta = 0$ and a little bump at $\cos\theta \approx 0.75$, which is the signature of an $f$ wave, thus indicating that absorption of an odd number of photons has occurred. Since one-photon ionization at these low frequencies is very unlikely [see Fig. 8.1.1(c)], we thus conclude that the spectra are dominated by three-photon ionization.

In Fig. 8.3.3(a) ($\omega = 0.8$ a.u.), one can also see a clear variation of the spectra with the pulse duration. We already know, from the analysis of the CKE spectra presented above, that as the pulse duration decreases, one passes from a dominant two-photon ionization regime to a different one in which the contribution from one-photon ionization becomes progressively more important. The $\text{CAK}_\text{N}$ spectra show this effect even more clearly. Indeed, Fig. 8.3.3(a) (left) shows the appearance of a nodal plane at $\cos\theta = 0$, thus indicating that absorption of one photon is the dominant process. As one moves from the left to the right panels in Fig. 8.3.3(a), i.e., as the pulse duration increases, one can see that the nodal plane at $\cos\theta = 0$ disappears and the overall shape of the spectrum becomes closer to that found for $\omega = 0.6$ a.u. [Fig. 8.3.3(b)]. In fact, for $T = 1.0$ fs and $T = 2.5$ fs, the $\text{CAK}_\text{N}$ spectra reflect contributions from both processes. To better visualize these contributions, we have performed a separate ROM analysis for the $g$ and $u$ symmetry components of the wave function. The results are shown in Fig. 8.3.4. One can clearly see that one-photon ionization, which, as explained above, leads to lower nuclear kinetic energies, appears in the $u$ part of the spectrum, where a node at $\cos\theta = 0$ is clearly visible. In contrast, two-photon ionization, which shows up at slightly higher nuclear kinetic energies, appears in the $g$ part of the spectrum, where the nodes at $\cos\theta \approx \pm 0.5$ are apparent.

# 9

# HIGH HARMONIC GENERATION FROM $H_2^+$ AND ITS ISOTOPES

In this Chapter, we present the HHG spectra obtained from 3D calculations in the $H_2^+$ molecule and its isotopes. The HHG spectra is calculated for pulses with $\lambda = 800$ nm. In particular, we have performed several calculations for different pulse durations, from 5 optical cycles to 20 optical cycles. The intensity of the laser pulse is the same for all calculations ($3 \times 10^{14}$ W/cm$^2$).

We have solved the TDSE in a numerical box with $|z| < 55$, $\rho < 50$ and $R < 30$ with grid spacings of $\Delta z = 0.1$, $\Delta \rho = 0.075$ and $\Delta R = 0.05$ at the center of the grid. These grid spacings gradually increase as we go to the limits of the box. For the electronic propagation, we use a time step $\Delta t_{elec} = 0.011$ a.u., and for the nuclear propagation, we use a time step $10$ times larger ($\Delta t_{nuc} = 0.11$ a.u.). The convergence of these parameters were checked. The initial state is the groundstate of the $H_2^+$ molecule. We have performed the calculation by putting the absorbers at $|z| > 35$, $\rho > 30$.

In principle, a symmetric homonuclear diatomic molecule subject to an intense IR field should generate only odd harmonics of the fundamental frequency. In the following, we will show that this is not always true. We will show that for light molecules and sufficiently long laser pulses, we can break this symmetry as a result of the nuclear motion and generate even harmonics.

## 9.1 HHG SPECTRA

In Figs. 9.1.1, 9.1.2, 9.1.3 and 9.1.4, we show the HHG spectra for the three different isotopes of the $H_2^+$ molecule for different pulse durations. For the shortest pulse, 5 optical cycles (Fig. 9.1.1), we observe that at the cutoff energies, the HHG spectrum of the heavier molecules is enhanced with respect to that of the $H_2^+$ molecule. This can be attributed to the fact that the nuclear motion is slower for the heavier molecules. Indeed, this effect has been explained earlier [74]: in this region, the vibrational autocorrelation function, the overlap between the nuclear wavepacket at the ionization time and the nuclear wavepacket at the recombination time, deviates more from the unity for the lighter molecules due to the fastest nuclear motion. The results for a 5 cycles pulse are pretty much the same as those obtained from the 1+1D calculations [21,75]. Also, we notice the presence of two localized minima in the HHG spectra below the ionization threshold. One minimum is present for all isotopes and it is present at the third harmonic. Another minimum appears between $13^{th}$ and $19^{th}$ harmonic, depending on the molecule. These minima are the result of the destructive interference between different vibrational states that contribute to the HHG process [75].

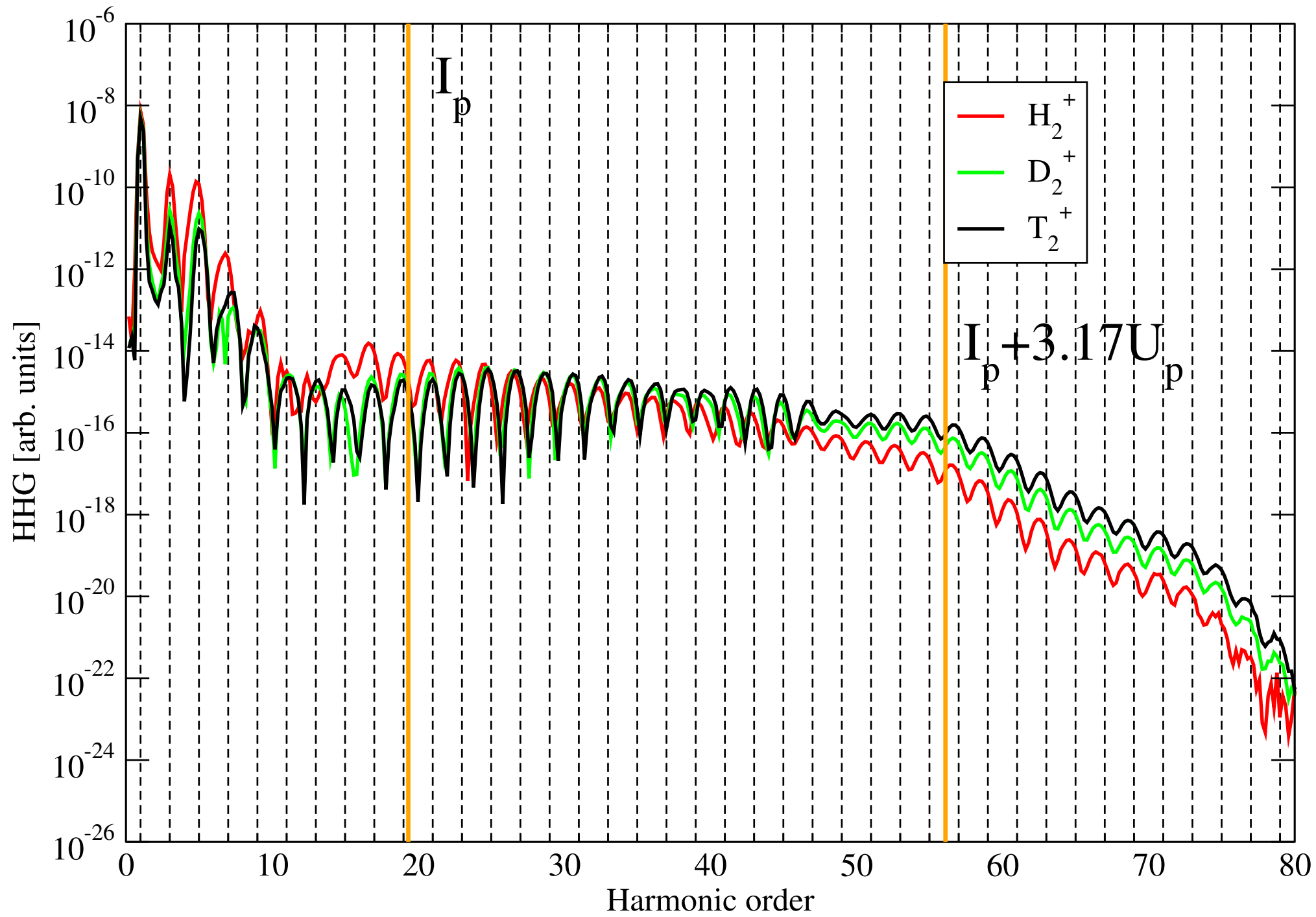

Figure 9.1.1.: HHG spectrum for a pulse with 800 nm, $I = 3 \times 10^{14}$W/cm$^2$ and 5 optical cycles for $H_2^+$, $D_2^+$ and $T_2^+$. The dashed vertical lines indicates odd harmonics.

For longer pulses (see Figs. 9.1.2, 9.1.3 and 9.1.4) the results differ significantly from the ones obtained for a 5 cycles pulse. First of all, we observe that the differences between the different isotopes are more pronounced than in the 5-cycles case. We see an enhancement of the HHG spectra in the $H_2^+$ for harmonics that are located at the plateau region. Furthermore, the location of the harmonic peaks in the $H_2^+$ molecule are displaced from that of the odd harmonics and we even see peaks that correspond to even harmonics. This result is not expected since we are working with a molecule with an inversion center and we would expect that only odd harmonics were generated. We can observe that even harmonics shift to higher harmonic orders when the pulse duration increases. This tendency can be more clearly seen in Fig. 9.1.5.

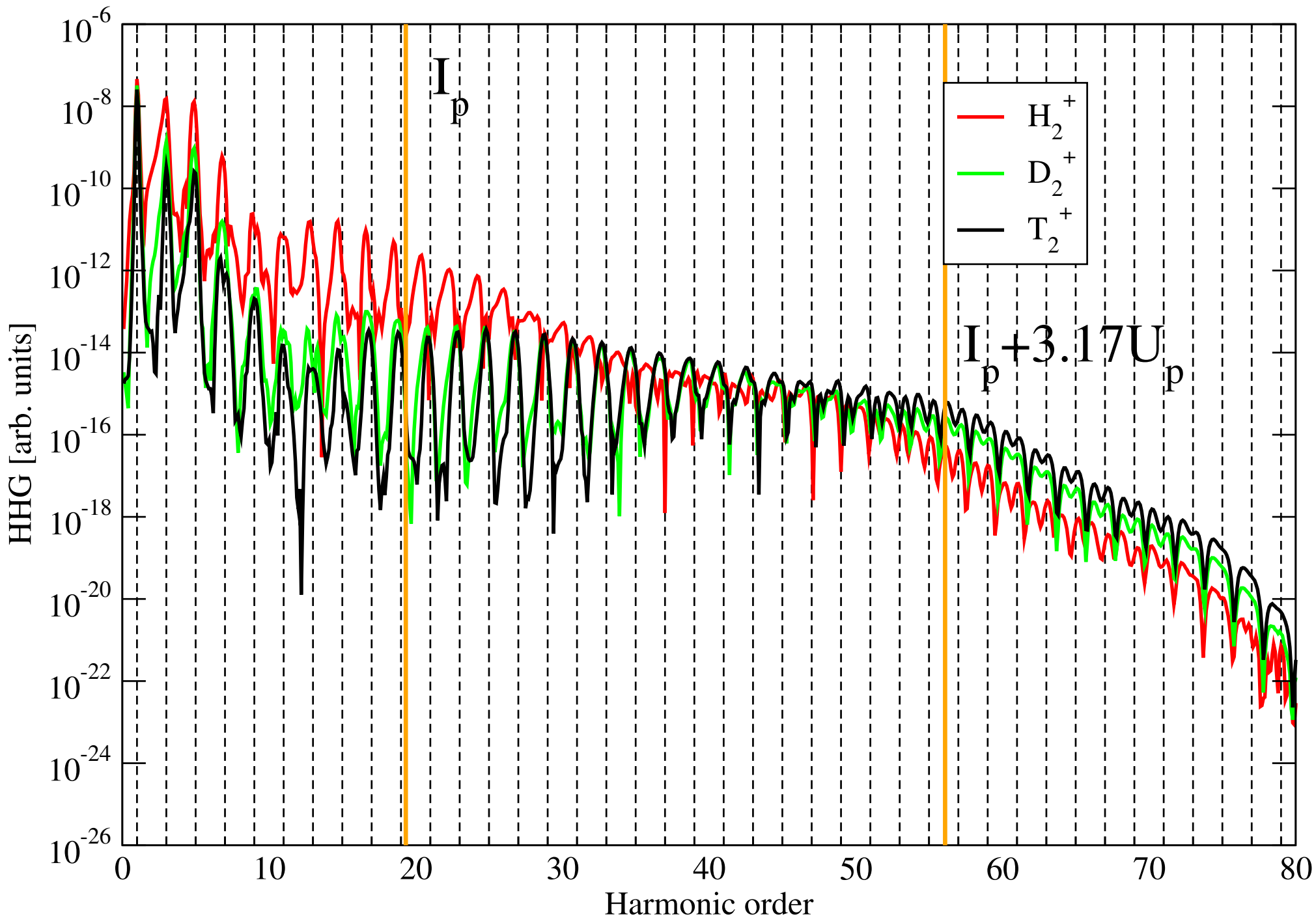

Figure 9.1.2.: Same as in Fig. 9.1.1 for 10 optical cycles.

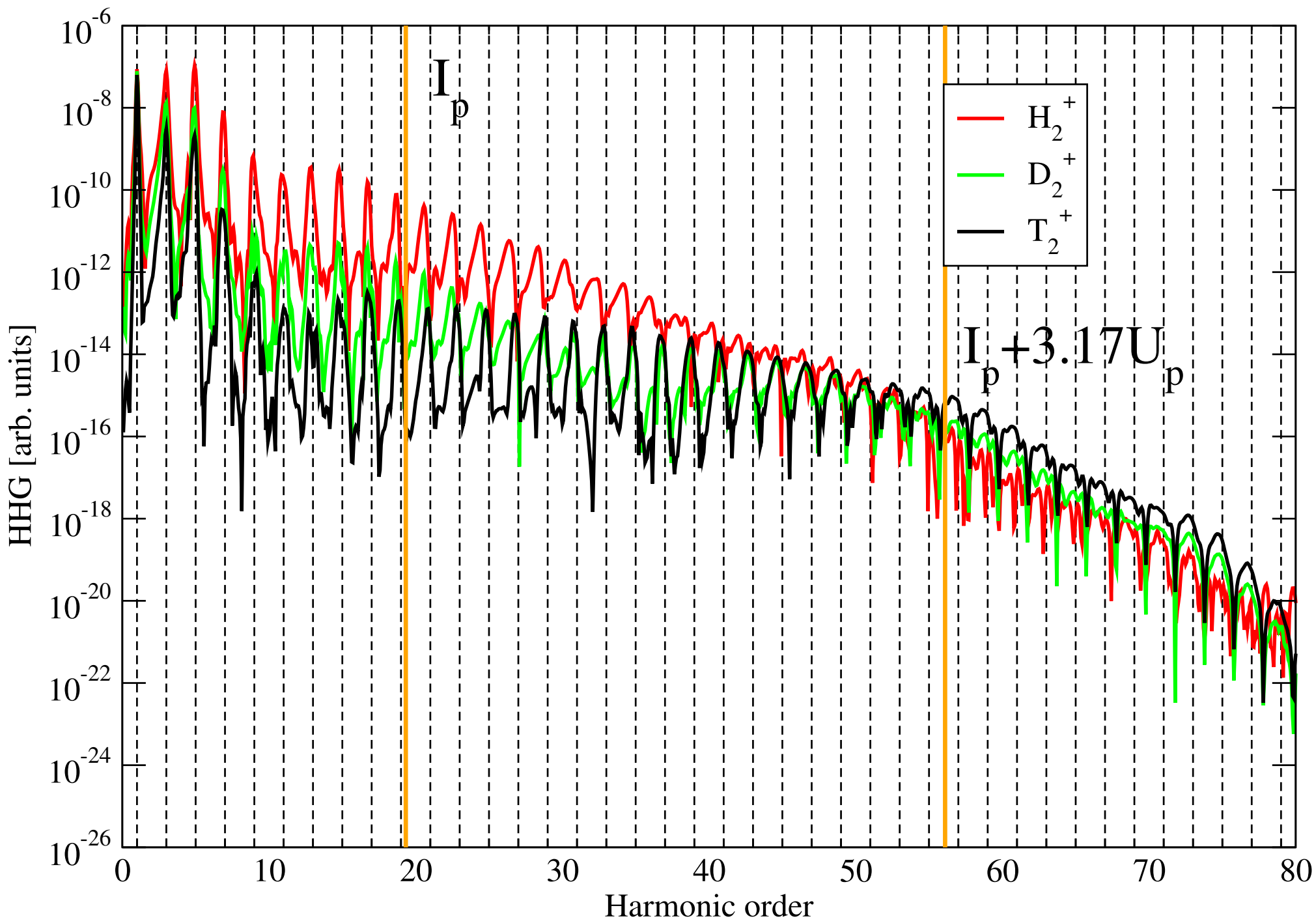

Figure 9.1.3.: Same as in Fig. 9.1.1 for 14 optical cycles.

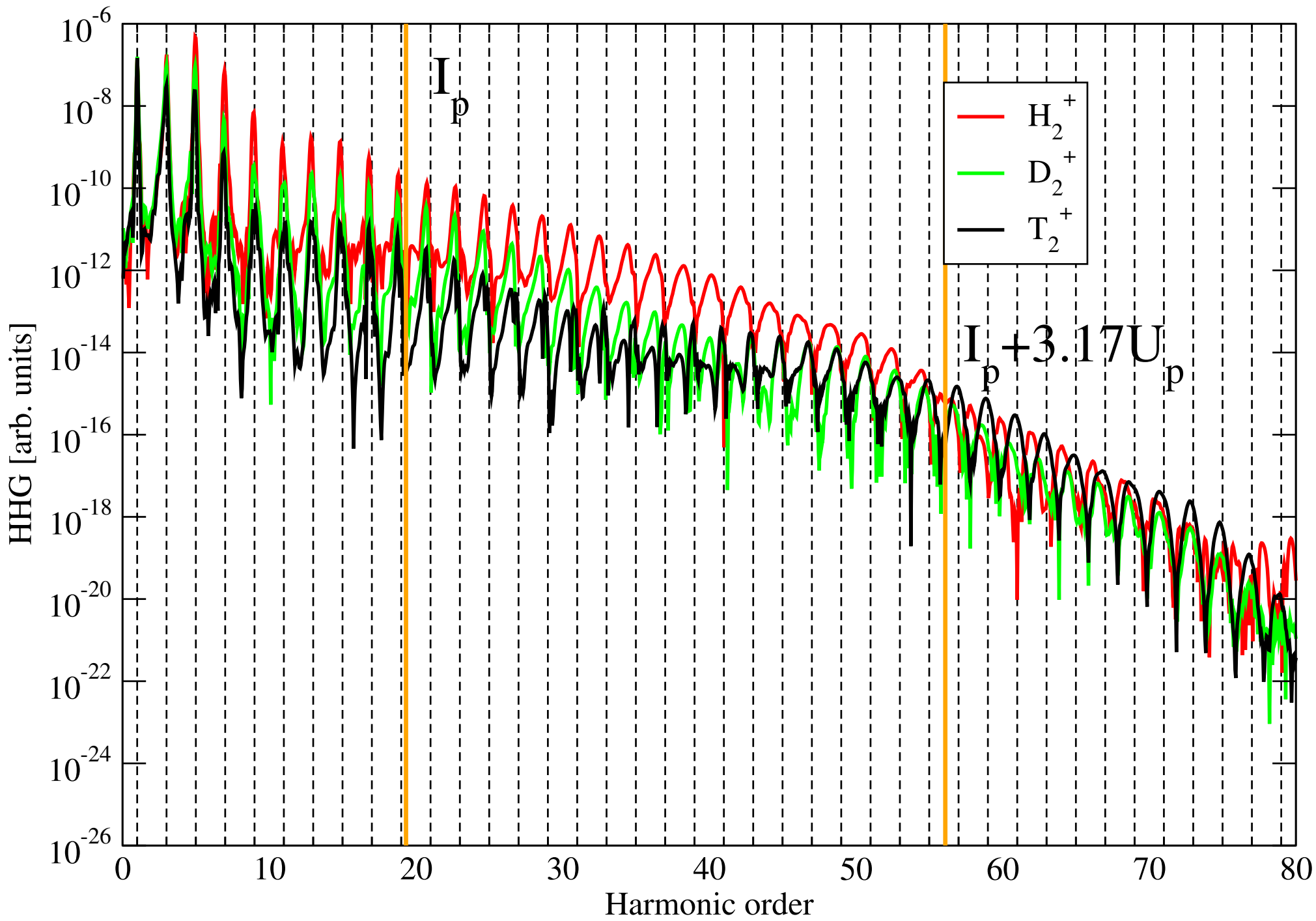

Figure 9.1.4.: Same as in Fig. 9.1.1 for 20 optical cycles.

The appearance of even harmonics in a homonuclear diatomic molecule was predicted in [21], from 1+1D calculations in the $H_2^+$ molecule. The main difference between those results and ours is that even harmonics are even more pronounced in the 3D calculations, up to the point that it becomes dominant over generation of odd harmonics. Recent results for lower harmonic orders have already pointed out a redshift of the odd harmonics peaks due to the asymmetric harmonic generation during the pulse [3]. However, as we will see below this cannot explain the appearance of even harmonics. We compare our numerical results with [3], and we observe that they agree perfectly, see Fig. 9.1.6. In the next section, we propose that electron localization is the ultimate explanation to the even harmonics.

## 9.2 ELECTRON LOCALIZATION

The odd harmonic rule is based on the fact that the dipole response obeys the following symmetry rule $d(t + T_{IR}/2) = -d(t)$, where $T_{IR}$ is the period of the infrared laser pulse. When this symmetry is broken, we expect to see the generation of both even and odd harmonics. One way of breaking this symmetry is by creating a localized state around one of the atomic centers of the molecule. In the $H_2^+$ molecule, a localized state can be created by the superposition of $1s\sigma_g$ and $2p\sigma_u$ states.

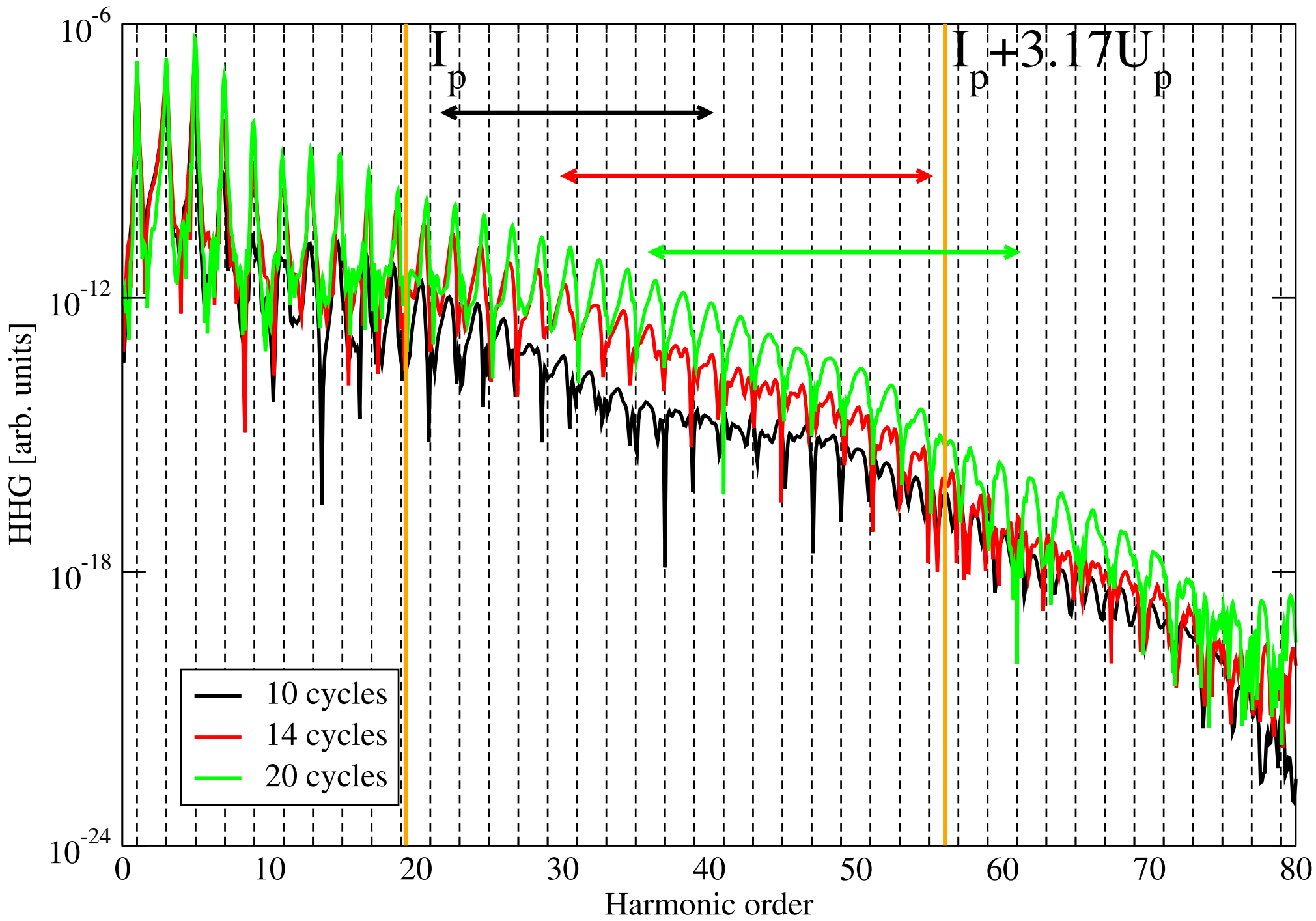

Figure 9.1.5.: HHG spectrum for a pulse with 800 nm, $I = 3 \times 10^{14}\text{W/cm}^2$ for $H_2^+$. Different pulse durations (10, 14 and 20 optical cycles) are shown in this figure. The horizontal arrows represent the even harmonic generation region for each pulse duration. The dashed vertical lines indicates odd harmonics.

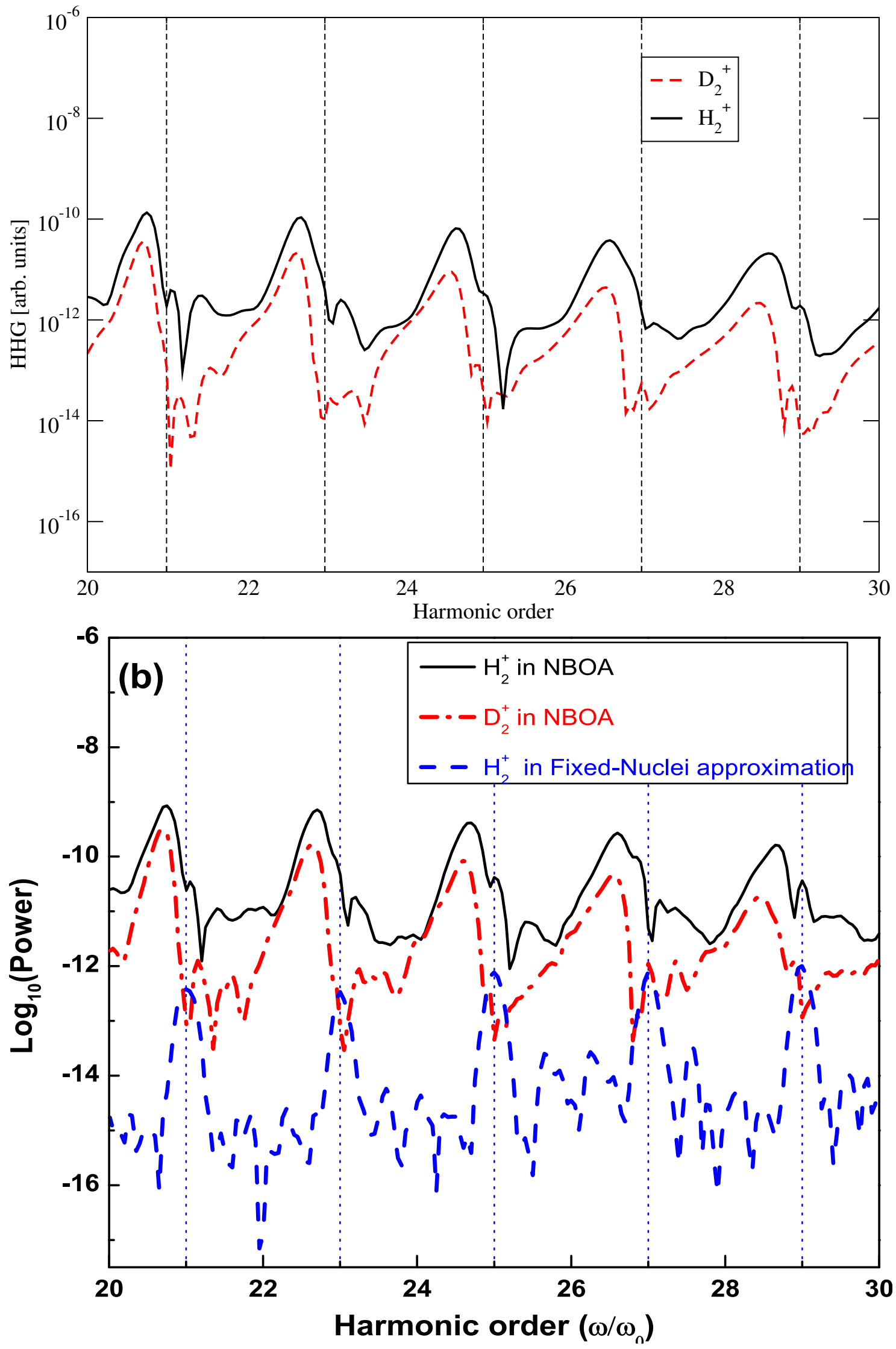

Figure 9.1.6.: HHG spectrum for a pulse with 800 nm, $I = 3 \times 10^{14} \mathrm{W/cm^2}$ and 20 optical cycles for $H_2^+$ and $D_2^+$. The top figure presents the results obtained in our calculations and the bottom figure presents results of [3].

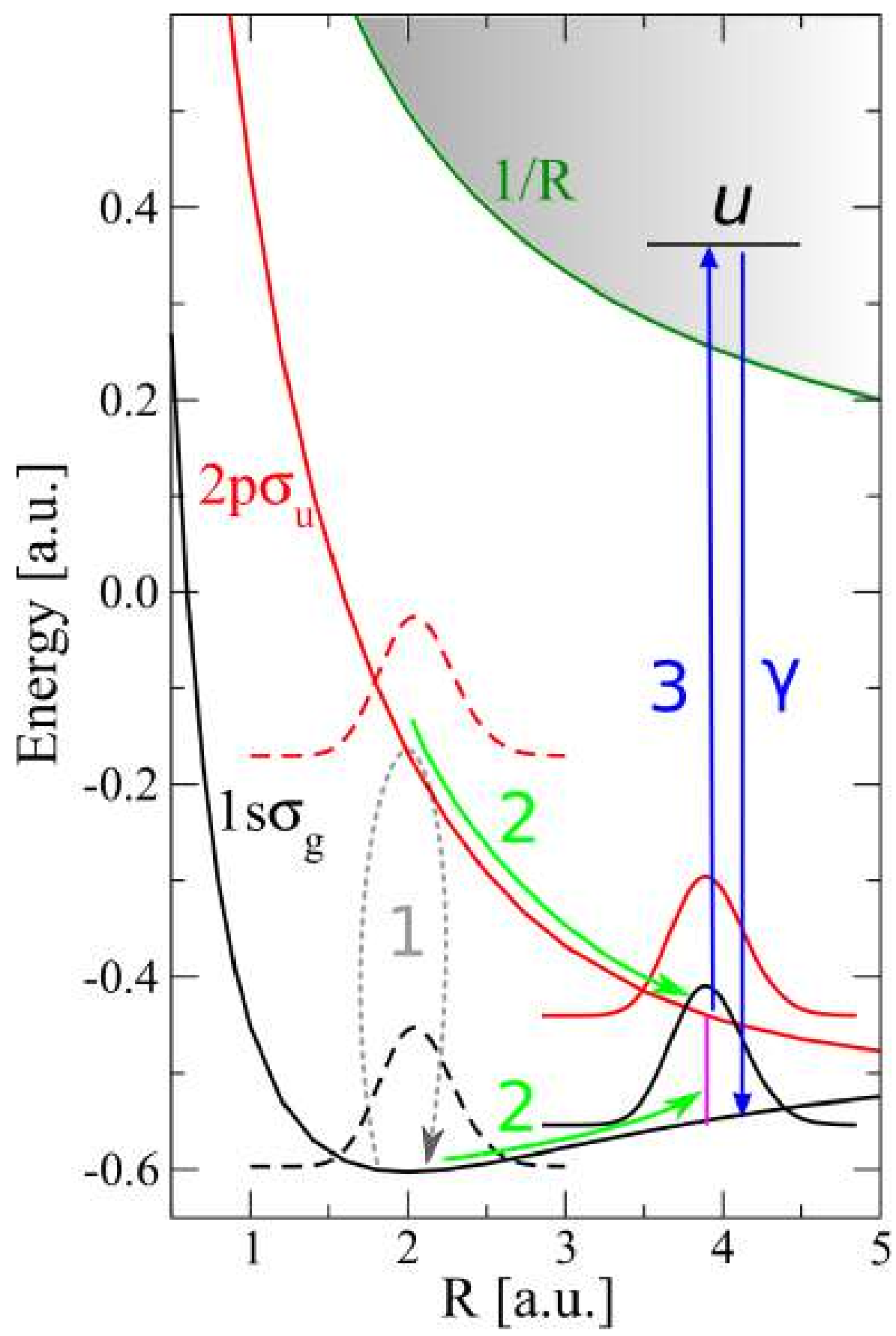

Figure 9.2.1.: Schematics of the even harmonic generation process.

In Fig. 9.2.1 we present a possible explanation for the even harmonic generation. Process 1 is the coupling between the first two electronic states of the $H_2^+$ molecule. As it was described in Chapter 7, this coupling leads to a nuclear wave packet that moves towards larger internuclear distances (process 2). Process 3 is the HHG at $R \approx 3.9$, where the $1s\sigma_g$ and $2p\sigma_u$ electronic states are separated by 2 IR photons.

To confirm this explanation, we have calculated a $R$-dependent HHG spectrum, see Fig. 9.2.2. The $R$-dependent HHG spectrum is calculated as the Fourier transform of $\ddot{d}(R,t)$ and

$$\ddot{d}(R,t) = \int \int \Psi^* (R,z,\rho)\, \hat{\ddot{d}} \Psi (R,z,\rho)\, \rho d\rho dz, \tag{9.2.1}$$

where $\ddot{d}$ is the dipole acceleration operator. We observe that the even harmonics are generated at internuclear distances higher than the internuclear equilibrium distance ($R \gtrsim 3.5$). Close to the equilibrium internuclear distance, $R_{eq} = 1.9$, we observe only odd harmonics. This would also explain the absence of even harmonics in the spectrum of heavier molecules, since heavier nuclei move slower and therefore would not have enough time to reach the region of large internuclear distances where localization occurs.

In Fig. 9.2.3, we can see the Gabor profile and the time evolution of the nuclear wavepacket in $H_2^+$. For each pulse duration we select the time, $t_C$, at which the density around the critical internuclear distance, $R_C = 3.9$, is largest. By looking at the Gabor profile at $t > t_C$ and at the harmonics that are emitted at $t_C$, we can predict the location of the even harmonics in the HHG spectra. For the 10 cycle pulse, $R_C$ is reached when the intensity of the laser pulse is decreasing. This leads to even harmonics in the lower region of the HHG spectrum. This can be seen at the corresponding Gabor profile. As long as we increase the pulse duration, we can see that $R_C$ is reached at a larger intensity of the field, so the even harmonics will appear at higher energies in the spectrum. This explains the shift of the even harmonic region to more energetic harmonics as the pulse duration increases.

Our results prove that it is possible to generate even harmonics from homonuclear diatomic molecules. Even harmonic generation depends both on the isotopic specie and on the pulse duration. The key point is electron localization which breaks the symmetry of our system, and is the consequence of the fast nuclear motion. So we can obtain even harmonics if we have a sufficiently fast nuclear motion and a sufficiently long laser pulse.

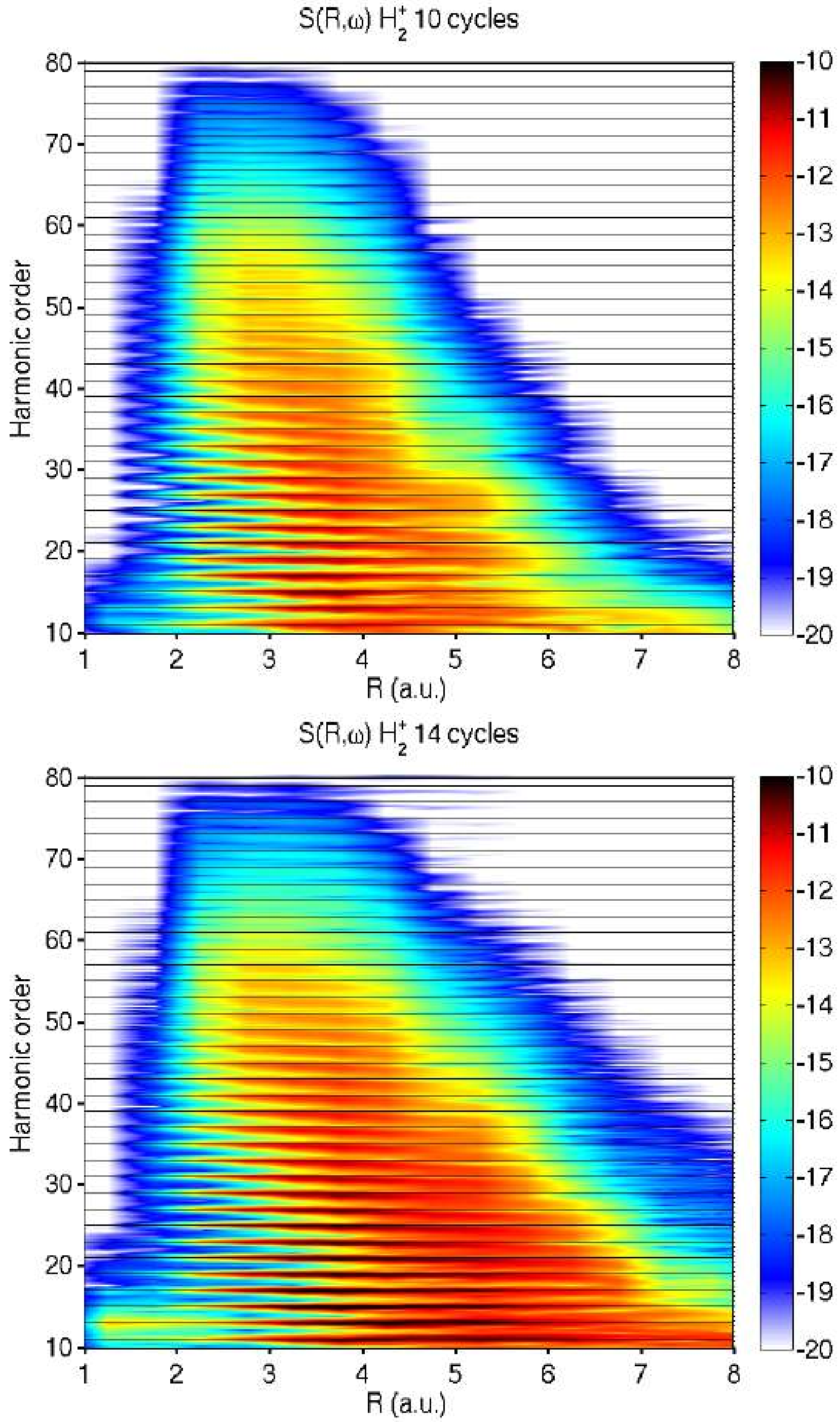

Figure 9.2.2.: $R$-dependent HHG spectrum for a pulse with 800 nm, $I = 3 \times 10^{14}$W/cm$^2$ and 10 (14) optical cycles for $H_2^+$, top (bottom) figure. The horizontal lines indicates odd harmonics.

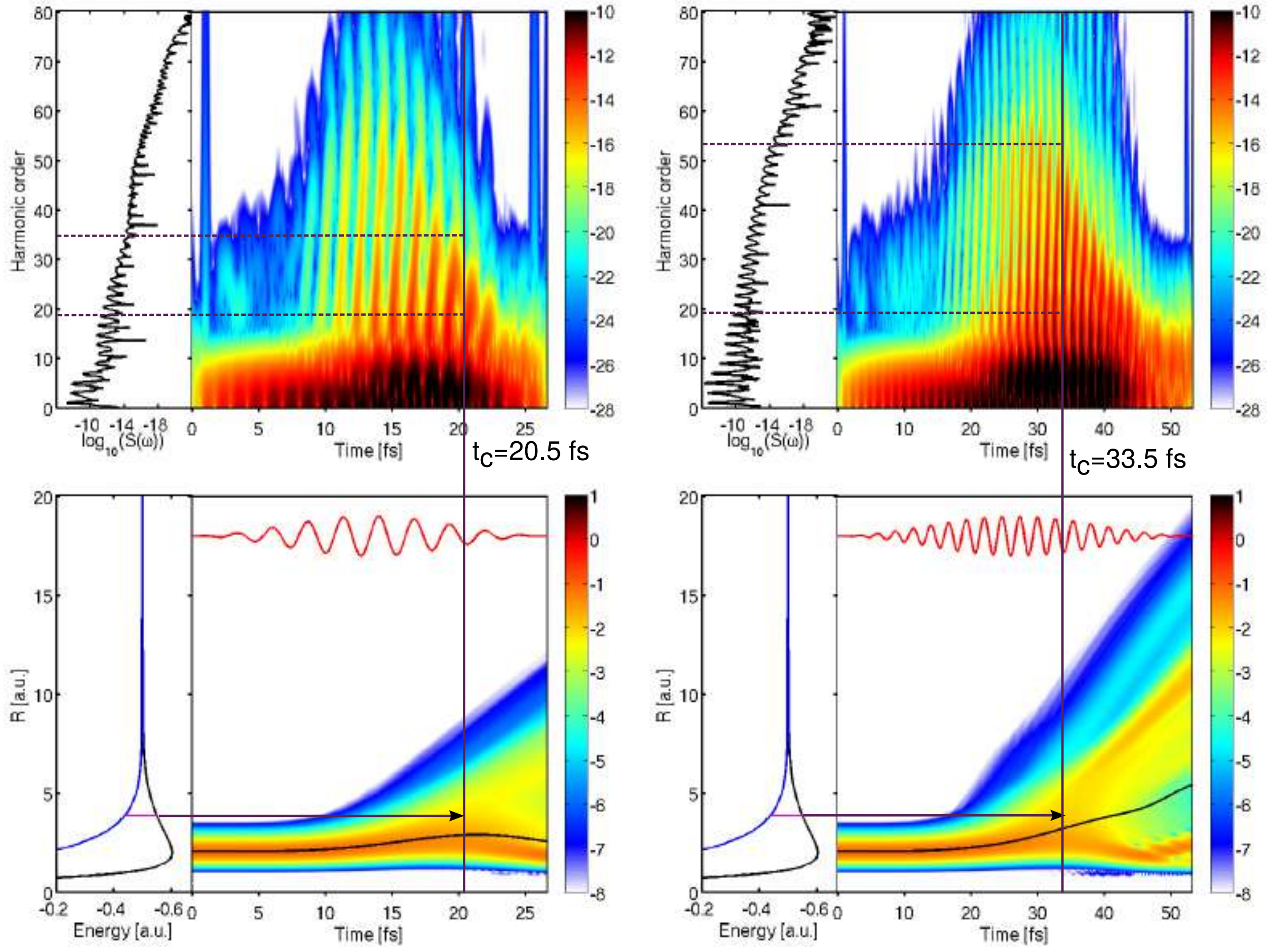

Figure 9.2.3.: Gabor profile and nuclear wavepacket distribution for a pulse with 800 nm, $I = 3 \times 10^{14}$W/cm$^2$ and 10 (left) and 20 (right) optical cycles for $H_2^+$.

# Part IV

# CONCLUSIONS

# CONCLUSIONS

We present in this thesis a new method to extract correlated photoelectron- and nuclear-kinetic energy spectra of molecules interacting with strong, ultrashort laser fields. We have focused on $H_2^+$ dissociation and photoionization by femtosecond laser pulses in the XUV-IR frequency range. We have applied this method to a 1+1D and a 3D model of the $H_2^+$ molecule.

We have firstly applied the resolvent operator method (ROM) to the 1+1D model of the $H_2^+$ molecule. We have shown results for few-photon absorption, for bound-bound electronic transitions, and for strong IR fields, in which multiphoton absorption is present. We have compared the results obtained by integrating the differential probabilities with those obtained by using the total ROM in which only one operator containing the total molecular Hamiltonian is used. The agreement between results obtained with both methods is excellent, showing that (i) the approximation made by projecting the wave function at finite time, in the differential ROM, is justified and (ii) the normalization used to obtain the probabilities is correct.

We have applied the ROM to the study of the correlation of electronic and nuclear dynamics in the strong field ionization of a 1+1D model of the $H_2^+$ molecule. The results show that electrons produced in multiphoton ionization share their energy with the nuclei, an effect that shows up in the correlated spectra in the form of energy-conservation fringes similar to those observed in weak-field ionization of diatomic molecules. In contrast, electrons resulting from tunneling ionization lead to fringes whose position does not depend on the proton kinetic energy; i.e., the molecular character is somewhat lost. At high intensity, the two processes coexist and the correlated spectra exhibit a complex structure, thus showing that the correlation between electron and nuclear dynamics in strong field ionization is involved as a result of the interplay between the electronic and the nuclear motion.

We have successfully applied the ROM to the 3D calculations of the $H_2^+$ molecule. The correlated kinetic-energy (CKE) and correlated angular and nuclear kinetic-energy ($CAK_N$) spectra have been evaluated and used to analyze the underlying mechanisms of the photoionization process. In particular, for pulses with a central energy $\hbar\omega = 0.8$ a.u., which is smaller than the vertical ionization potential of $H_2^+$ at the internuclear equilibrium distance, we have shown the opening of the one-photon ionization channel by decreasing the pulse duration down to the sub-fs time scale. This effect, namely, the opening of the $(N-1)$-photon ionization channel when the central energy is such that ionization requires $N$ such photons, is expected to occur for any pulse of sufficiently short duration. Our results for a central frequency $\hbar\omega = 0.4$ a.u. confirm this expectation. An inspection of the CKE and $CAK_N$ spectra clearly shows the variation of the relative contribution of $(N-1)$- and $N$-photon ionization processes with pulse duration. The latter information is difficult to obtain when only the kinetic energy release (KER) spectrum is measured. This points out the importance of performing multiple-coincidence measurements for better elucidation of competing ionization mechanisms, such as those arising when ultrashort pulses are used.

Finally, by studying HHG with the 3D model of the $H_2^+$ molecule, we were able to see the isotopic effects in the HHG spectra. The results obtained for the shortest pulse were in agreement with those obtained earlier in the literature from a 1+1D model [21, 75]. Our results have also been successfully compared with those previously obtained in the literature for lower harmonic orders [3]. Even harmonic generation was observed for light molecules and long laser pulses. We show that electron localization is the mechanism that breaks the symmetry of our medium and allows the generation of even harmonics.

# CONCLUSIONES

En esta tesis presentamos un nuevo método para extraer los espectros fotoelectrónicos y de energía cinética nuclear correlacionados de moléculas interaccionando con campos láser fuertes y ultra cortos. Hemos estudiado la dissociación y fotoionización de $H_2^+$ por pulsos láser de femtosegundos en el régimen de frecuencias desde el ultravioleta extremo hasta el infrarrojo. Hemos aplicado este método a modelos de diferente dimensionalidad del $H_2^+$.

Primeramente, hemos aplicado el método del operador resolvente (ROM) al modelo 1+1D de la molécula de $H_2^+$. Hemos enseñado resultados para la absorción de pocos fotones, para transiciones electrónicas entre estados ligados, y para campos IR fuertes, en que la absorción de varios fotones está presente. Hemos comparado los resultados obtenidos integrando las probabilidades diferenciales con las obtenidas usando el método del operador resolvente total en que solo un operador que contiene el Hamiltoniano molecular total es usado. El acuerdo entre los resultados obtenidos con ambos métodos es excelente, demostrando que (i) la aproximación hecha al proyectar la función de onda a tiempo finito, en el ROM diferencial está justificada y (ii) la normalización usada para obtener las probabilidades es correcta.

Hemos aplicado el ROM al estudio de la correlación entre la dinámica electrónica y nuclear en la ionización por campos fuertes en un modelo 1+1D de la molécula de $H_2^+$. Los resultados demuestran que los electrones producidos por ionización por varios fotones comparten su energía con los núcleos, un efecto que aparece en los espectros correlacionados en forma de líneas de conservación de energía semejantes a las observadas en la ionización por campos débiles de moléculas diatómicas. Por otro lado, electrones resultantes de ionización por túnel llevan a líneas cuya posición no depende de la energía cinética de los protones; i.e., el carácter molecular de alguna manera se pierde. A grandes intensidades, los dos procesos coexisten y los espectros correlacionados exhiben una compleja estructura, demostrando que la correlación entre la dinámica electrónica y nuclear en la ionización por campos fuertes interviene como resultado del intercambio entre el movimiento electrónico y nuclear.

Hemos aplicado con éxito el ROM a calculos en 3D en el $H_2^+$. Los espectros de energía cinética correlacionados (CKE) y los espectros de energía cinética nuclear angulares ($CAK_N$) fueran calculados y usados para el análisis de los mecanismos del proceso de fotoionización. En particular, para pulsos con energía central $\hbar\omega = 0{,}8$ a.u., que es menor que el potencial de ionización vertical del $H_2^+$ a la distancia internuclear de equilibrio, hemos demostrado la apertura del canal de ionización por un fotón disminuyendo la duración del pulso a la escala de los sub-fs. Este efecto, la apertura del canal de ionización por $(N-1)$ fotones cuando la energía central del pulso requiere la absorción de $N$ fotones, se espera que ocurra para cualquier pulso desde que se baje lo suficiente su duración. Nuestros resultados para una frecuencia central de $\hbar\omega = 0{,}4$ a.u. confirman esta hipótesis. Visualizando los espectros CKE y $CAK_N$ vemos claramente la variación de las contribuciones relativas de la ionización por $(N-1)$- y por $N$ fotones con la duración del pulso. Esta última información es difícil de obtener cuando medimos solamente el espectro de la energía cinética de los núcleos (KER). Esto sugiere la importancia de realizar medidas de múltiple coincidencia para una mejor elucidación de los distintos mecanismos de ionización, como los que surgen cuando se utilizan pulsos ultracortos.

Finalmente, al estudiar la generación de armónicos con el modelo 3D de la molécula de $H_2^+$, hemos sido capaces de ver los efectos isotópicos en los espectros de HHG. Los resultados obtenidos para el pulso más corto están en consonancia con otros obtenidos anteriormente en la literatura en un modelo 1+1D [21, 75]. Nuestros resultados han sido comparados con éxito a resultados obtenidos previamente en la literatura para armónicos de bajo orden [3]. La generación de armónicos pares ha sido observada para moléculas ligeras y pulsos laseres largos. Demostramos que la localización electrónica es el mecanismo responsable por la quiebra de simetría de nuestro medio y que permite la generación de armónicos pares.

# APPENDIX

# A

# PARTICLE UNDER A TIME-DEPENDENT ELECTRIC FIELD

Suppose that a point particle with a initial velocity $\boldsymbol{v}(t_0)$, mass $m$ and charge $q$, is exposed to a time-dependent electric field, without spatial dependence $\boldsymbol{E}(t)$. In this way we can define the vector potential that is given by

$$\boldsymbol{E}(t) = -\partial_t \boldsymbol{A}(t). \tag{A.1}$$

The second Newton equation is

$$m\dot{\boldsymbol{v}}(t) = q\boldsymbol{E}(t) \tag{A.2}$$

and integrating the equation of motion

$$\boldsymbol{v}(t) = \boldsymbol{v}(t_0) + \frac{q}{m}\int_{t_0}^{t} -\partial_t \boldsymbol{A}(t')\, dt' = \boldsymbol{v}(t_0) + \frac{q}{m}\left[-\boldsymbol{A}(t) + \boldsymbol{A}(t_0)\right]. \tag{A.3}$$

For an electron (in atomic units), $m = 1$ and $q = -1$, so

$$\boldsymbol{v}(t) = \boldsymbol{v}(t_0) + \left[\boldsymbol{A}(t) - \boldsymbol{A}(t_0)\right]. \tag{A.4}$$

We are interested in the final velocity of the electron. If $\boldsymbol{A}(\infty) = 0$ we have that

$$\boldsymbol{v}(\infty) = \boldsymbol{v}(t_0) - \boldsymbol{A}(t_0). \tag{A.5}$$

This correction must be done if we want to measure an electron's velocity when the electric pulse is still present.

# B

## DERIVATIVES IN AN INHOMOGENEOUS GRID

Here we derive a five-point formula for the derivative in a inhomogeneous grid. We will define the central point as $i$ and the four adjacent points as

$$i+\nu \qquad i+\beta \qquad i \qquad i+\alpha \qquad i+\mu. \tag{B.1}$$

The Taylor expansion at a point $x+\Delta x$ is

$$f\left(x+\Delta x\right)=f\left(x\right)+\Delta x f'\left(x\right)+\frac{\Delta x^2}{2!}f''\left(x\right)+\frac{\Delta x^3}{3!}f'''\left(x\right)+\mathcal{O}\left(\Delta x^4\right), \tag{B.2}$$

and if $f_\alpha\equiv f\left(x+\alpha\right)$ then we can express the expansions (B.2) at points (B.1) as a matrix,

$$\left[\begin{array}{c|cccc} - & f & f' & f'' & f''' \\ \hline f_\alpha & 1 & \alpha & \alpha^2/2 & \alpha^3/6 \\ f_\beta & 1 & \beta & \beta^2/2 & \beta^3/6 \\ f_\mu & 1 & \mu & \mu^2/2 & \mu^3/6 \\ f_\nu & 1 & \nu & \nu^2/2 & \nu^3/6 \end{array}\right] \tag{B.3}$$

After some algebra we arrive to

$$f'=\frac{\left(\beta^3-\alpha^3\right)\left(f_\nu-f_\mu\right)-\left(\nu^3-\mu^3\right)\left(f_\beta-f_\alpha\right)}{\left(\beta^3-\alpha^3\right)\left(\nu-\mu\right)-\left(\nu^3-\mu^3\right)\left(\beta-\alpha\right)}. \tag{B.4}$$

Defining $\gamma\equiv\left(\nu^3-\mu^3\right)/\left(\beta^3-\alpha^3\right)$ we will have

$$f' = \frac{\left(f_\nu - f_\mu\right) - \gamma\left(f_\beta - f_\alpha\right)}{(\nu - \mu) - \gamma\,(\beta - \alpha)}. \tag{B.5}$$

# PUBLICATIONS AND BIBLIOGRAPHY

# PUBLICATIONS

- Publications that are directly related to this thesis
  - **R. E. F. Silva**, F. Catoire, P. Rivière, H. Bachau and F. Martín. «Correlated Electron and Nuclear Dynamics in Strong Field Photoionization of $H_2^+$». *Phys. Rev. Lett.* **110** 113001 [2013].
    doi:10.1103/PhysRevLett.110.113001
  - F. Catoire, **R. E. F. Silva**, P. Rivière, H. Bachau and F. Martín. «Molecular resolvent-operator method: Electronic and nuclear dynamics in strong-field ionization». *Phys. Rev. A* **89** 023415 [2014].
    doi:10.1103/PhysRevA.89.023415
  - **R. E. F. Silva**, F. Catoire, P. Rivière, T. Niederhausen, H. Bachau and F. Martín. «Energy-and angle-resolved ionization of $H_2^+$ interacting with xuv subfemtosecond laser pulses». *Phys. Rev. A* **92**(1) 013426 [2015].
    doi:10.1103/PhysRevA.92.013426
  - **R. E. F. Silva**, P. Rivière, F. Morales, O. Smirnova, M. Ivanov and F. Martín. «High Harmonic generation from $H_2^+$ and its isotopes». *in preparation*
- Other publications that were produced during this thesis
  - **R. E. F. Silva**, P. Rivière and F. Martín. «Autoionizing decay of $H_2$ doubly excited states by using xuv-pump–infrared-probe schemes with trains of attosecond pulses». *Phys. Rev. A* **85** 063414 [2012].
    doi:10.1103/PhysRevA.85.063414
  - P. Rivière, **R. E. F. Silva** and F. Martín. «Pump-Probe Scheme To Study the Autoionization Decay of Optically-Forbidden $H_2$ Doubly Excited States». *The Journal of Physical Chemistry A* **46** 11304 [2012].
    doi:10.1021/jp3053136
  - M. Lara, **R. E. F. Silva**, A. Gubaydullin, P. Rivière, C. Meier and F. Martín. «Enhancing high-harmonic generation from light molecules with chirped pulses». *in preparation*

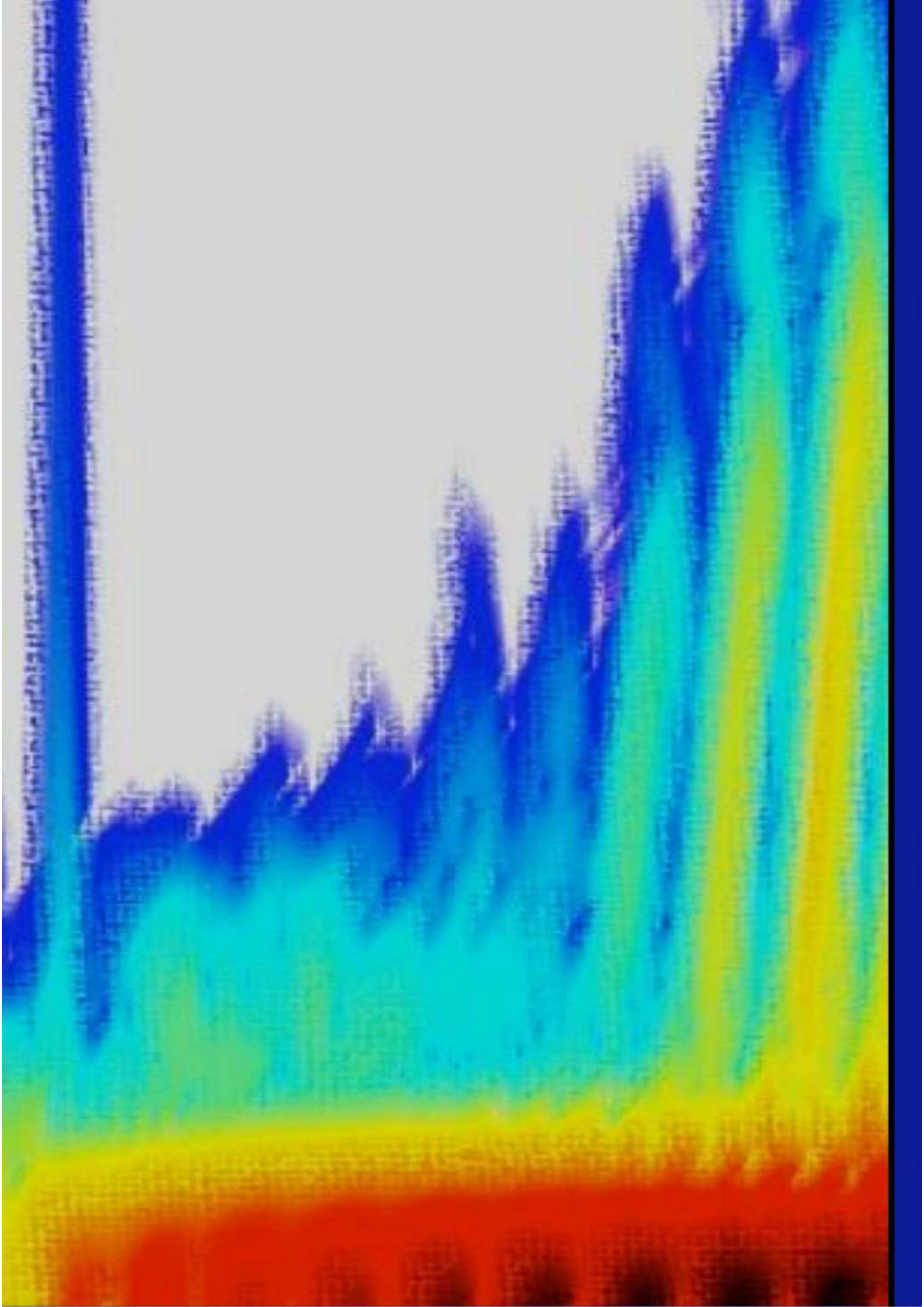